\documentclass[twocolumn]{revtex4-1}	
\usepackage{graphicx}
\usepackage{amsmath, amsfonts,amssymb}
\usepackage{color}
\usepackage{outlines}
\newcommand{\infint}{\int_{-\infty}^\infty\!\!\! }

\newcommand{\be}{\begin{equation}}
\newcommand{\ee}{\end{equation}}

\newcommand{\Tr}{{\hspace{0.5pt} \rm Tr}}
\newcommand{\llangle}{\langle \!\langle}
\newcommand{\rrangle}{\rangle \! \rangle}
\newcommand{\norm}[1]{\lVert#1 \rVert}

\usepackage[dvipsnames]{xcolor}
\newcommand{\Us}{\hat{\mathcal U}}
\newcommand{\Hs}{\hat{\mathcal H}}

\newcommand{\bs}[1]{\boldsymbol{#1}}

\definecolor{MyColor}{RGB}{0,0,240}

\renewcommand{\S}{{\mathcal S}}
\newcommand{\B}{{\mathcal B}}

\renewcommand{\Re}{{\rm Re}\,}
\renewcommand{\vec}[1]{{\bf #1}}

\newcommand{\sgn}{{\rm sgn}}
\newcommand{\vmax}{\Gamma_0}

\newcommand{\xiB}{\xi_{\hspace{0.5pt}\rm B}}

\newcommand{\xiM}{\xi_{\hspace{0.5pt}\rm M}}
\newcommand{\DeltaB}{\hat{\Delta}_{\hspace{0.25pt}\rm B}}
\begin{document}
\title{Universal Lindblad equation for open quantum systems}
\date{\today}
\author{Frederik Nathan and Mark S. Rudner}
\affiliation{Center for Quantum Devices, Niels Bohr Institute, University of Copenhagen, 2100 Copenhagen, Denmark}
\begin{abstract}

We develop a  Markovian master equation in the Lindblad form that enables the efficient study of a wide range of open quantum many-body systems that would be inaccessible with existing methods.
The validity of the master equation is  based entirely on properties of the bath and the system-bath coupling, without any requirements on the level structure within the system itself.
The master equation is derived using a  Markov approximation that is distinct from that used in earlier approaches.
We provide a rigorous bound for the error induced by this Markov approximation; the error  is controlled by a dimensionless combination of intrinsic correlation and relaxation timescales of the bath.
Our master equation is accurate on the same level of approximation as  the Bloch-Redfield equation.
In contrast to the Bloch-Redfield approach, our approach   ensures  preservation of the   positivity of    the density matrix.
As a result, our method  is robust, and  can be solved efficiently  using stochastic evolution of pure states (rather than density matrices). 
We discuss how our method can be applied to static or driven quantum many-body systems, and illustrate its power through numerical simulation of a spin chain that would be challenging to treat by existing methods.
\end{abstract}	

\maketitle
The theoretical description of a quantum system interacting with an environment is an important problem of both fundamental and practical interest. 
The problem arises in a diverse array of settings, from chemistry to atomic, molecular and optical physics, as well as condensed matter physics, high-energy physics, and quantum information processing~\cite{Scully_book,VanKampen_book,Nielsen_chuang,Feshbach_1958,Zanardi_1997,Bourennane_2004,Verstraete_2009,Diehl_2011}. 
Due to the  importance and long history of the problem,  there exists a wide range of well-established approaches for describing the dynamics of open quantum systems, see, e.g. Refs.~\cite{Breuer, GardinerZoller, Nakajima_1958,Zwanzig_1960, Wangsness_1953, Redfield_1965,Davies_1974, Majenz_2013, Mozgunov_2020}.

The Nakajima-Zwanzig (NZ) approach~\cite{Nakajima_1958,Zwanzig_1960} provides a systematic framework for describing the evolution of open quantum systems.
Although formally exact in its most general form, in practice there are many challenges associated with application of the NZ equation, even in approximate form.
For example, the Bloch-Redfield (BR) equation, which emerges as a lowest-order approximation to the time-convolutionless NZ equation, is not guaranteed to preserve positivity of the density matrix of the system and may therefore yield unphysical  solutions for long time evolution, with negative or diverging probabilities.
Moreover, solving these (NZ or BR) equations requires working with the density matrix of the system, whose dimension is the square of that of the system's Hilbert space.
This requirement may make their numerical solution prohibitively expensive, even for moderately-sized quantum systems~\cite{Breuer}. 

For Markovian systems where the correlation (or ``memory'') time of the bath is sufficiently short, Lindblad-form master equations provide an alternative to the NZ approach~\cite{Breuer}.
The Lindblad form is the most general form of   a time-local evolution equation that is guaranteed to preserve the trace and positivity of the density matrix~\cite{Lindblad_1976,Gorini_1976}.
Importantly, the Lindblad form also admits efficient numerical solution via stochastic evolution of {\it pure states}~\cite{Dalibard_1992,Dum_1992,Carmichael_1993,Breuer}, thus avoiding the computational cost of working with density matrices.
However, derivations of Lindbladian master equations, such as the quantum optical master equation~\cite{Davies_1974}, typically require stringent conditions on the level spacing of the system itself, thus limiting their applicability to specific classes of systems.
In particular, the quantum optical master  equation relies on the rotating wave approximation (RWA), and hence is only valid when the level-broadening arising from bath-induced transitions is small compared with the smallest level spacing in the system.
While this condition is well-satisfied in many important cases, for example in atomic physics and quantum optics, 
many types of systems (including many-body systems with dense spectra) and physical phenomena (such as Fano resonances) can not be described through this approach.

Our motivation in the present work is based on the following notion:
when the correlation time of the bath is much shorter than a characteristic timescale of system-bath interactions,  we heuristically expect that the evolution of the system should be generated by a Markovian master equation.
Hence, Markovianity should be a property of environment alone, independent of details of the system itself. 
Noting that a Markovian master equation for the density matrix must be in the Lindblad form, we thus seek to systematically derive a Lindbladian master equation without reference to any details of the system other than the operator(s) through which it couples to its environment.

The main result of this paper is the derivation of a ``universal Lindblad equation'' (ULE) 
that can be applied to  {\it any}  open quantum  system whose 
bath satisfies a particular Markovianity condition that is defined in terms of the bath spectral function and the system-bath coupling strength. 
In particular, the derivation of the ULE does not rely on the rotating wave approximation or any 
  other  assumption about the energy level spacings of the system.
We provide explicit expressions for the jump operators, and discuss their evaluation for static, Floquet, and arbitrarily driven many-body systems.
Importantly, the number of jump operators is equal to the number of independent terms (referred to as quantum noise channels) that 
couple the system and bath, independent of the details of the system.
As a result, for many cases,  the ULE features only  one or a few jump operators.
The jump operators are  straightforward to compute, either through exact diagonalization, or controlled expansions.

The principle underlying  our derivation of the ULE  is that there is no unique way of implementing a Markov approximation in the evolution of the density matrix. 
Instead there exists a continuous family of distinct approximations that 
result in Markovian dynamics of the system, all with error bounds of the same order in a  
dimensionless Markovianity parameter (see below). 
One particular choice out of this family of comparable Markov approximations leads to the Bloch-Redfield equation.
In this paper we 
employ a different Markov approximation from within this family which directly leads to a Lindblad-form master equation without any further assumptions about the nature of the system.

  We provide rigorous bounds on the relative error induced by making the Markovian approximation that results in the ULE. 
  The error is controlled by a dimensionless ``Markovianity'' parameter, defined from a combination of  correlation and relaxation timescales that we identify  from the bath and its coupling to the  system. 
  We  show that this error is of the same order as that incurred in deriving the Bloch-Redfield equation.
Unlike the BR equation, however, the ULE preserves the physicality (i.e., positivity and normalization) of the density matrix. 
Hence it is intrinsically robust and  amenable to solution using efficient stochastic methods~\cite{Dalibard_1992,Dum_1992,Carmichael_1993,Breuer}.

The universal Lindblad equation that we present here can be used for a wide range of physical situations.
In particular, it can be used to efficiently simulate the dynamics of open and noisy quantum many-body systems (i.e., systems with large Hilbert space dimension and small level spacing).
In addition, it provides a straightforward, general approach for describing the dynamics of driven systems coupled to  Markovian baths.

A master equation of the same form as we derive here {was} previously employed with phenomenological justification in Ref.~\cite{Kirsanskas_2018}. 
More recently, a similar master equation was also heuristically obtained in Ref.~\cite{Kleinherbers_2020}.
Here we provide a systematic, rigorous derivation of the universal Lindblad equation, and in particular show that it captures the dynamics of the system at the same level of error as the Bloch-Redfield equation.
(Some of our arguments appeared in a preliminary, heuristic derivation in the  PhD thesis of one the present authors~\cite{Phd_thesis}.)  
The ansatz in Ref.~\cite{Kirsanskas_2018} applies to systems coupled to independent bath observables, with static or weakly time-dependent Hamiltonians, such that the jump operators can be computed  within a quasistatic approximation for the system Hamiltonian.
Our approach covers systems with arbitrary time-dependence and system-bath couplings, and  in particular applies beyond the regime where the quasistatic approximation is valid.

Recently, another group of authors has also obtained a  Lindblad-form master equation for open quantum systems whose validity is independent of the details of the system~\cite{Mozgunov_2020}. 
The master equation of Ref.~\cite{Mozgunov_2020} is  distinct from the ULE that we obtain, and was
derived using a time-coarse graining approach that is of a fundamentally different nature from the novel Markov approximation that we employ here. 
Interestingly, the error bounds obtained by the authors of Ref.~\cite{Mozgunov_2020} were defined in terms of a closely related Markovianity parameter  to the one we identify here (see Appendix~\ref{app:time_scale_relationship}). 
The ULE we derive is thus valid on an equivalent level of approximation as the master equation of Ref.~\cite{Mozgunov_2020}.
The simultaneous validity  of these two distinct master equations reflects the  non-uniqueness of the Markov approximation discussed above.

{The rest of this paper is organized as follows.
In the main text we discuss the essential ideas of our work, while we provide technical details and  derivations  in several appendices. 
In Sec.~\ref{sec:ady_section} we provide a summary of our main results. 
In Sec.~\ref{sec:model} we formally introduce the general model of open quantum systems that we study,  
  review existing approaches to analyzing the dynamics of this class of systems, and present  important auxiliary results that are  used to derive the ULE.
In Sec.~\ref{sec:derivation} we derive the ULE, allowing for multiple baths and arbitrarily time-dependent system Hamiltonians. 
In Sec.~\ref{sec:application}, we discuss how to calculate and implement the jump operators of the ULE for a range of relevant special cases, including systems with time-independent Hamiltonians, periodically driven systems, and systems where exact diagonalization of the Hamiltonian is not feasible.  
In Sec.~\ref{sec:numerics} we demonstrate our approach via numerical simulations of a spin chain 
coupled to two baths at different temperatures. 
We conclude with a discussion in Sec.~\ref{sec:discussion}.

\section{Summary of results}
\label{sec:ady_section}
In this section, we summarize the main ideas and  results   of this paper. 
We  investigate the  dynamics of a   quantum system $\S$   
 connected to an external environment (bath) $\B$. 
For simplicity, in this section we illustrate  our results for the    case  where the system's Hamiltonian $H_\S$ is time-independent, and  the system and bath are connected through a single term $H_{\rm int} = \sqrt{\gamma }X B$ in the combined system-environment Hamiltonian. 
Here $X$ and $B$ are observables of the system and bath, respectively, and 
the energy $\gamma $ denotes the system-bath coupling strength, normalized such that  $X$ has unit spectral norm~\cite{fn:spectral_norm_def,fn:sb_hamiltonian}.
The results we present below for this system  are derived and discussed in detail  in Secs.~\ref{sec:model}-\ref{sec:derivation}, 
where we also  extend our results  to  general  system-bath couplings  and time-dependent system Hamiltonians $H_\S(t)$.

In Sec.~\ref{sec:derivation}, we   seek  conditions  on the bath  under which the time-evolution of the reduced density matrix of the system  $\rho$  takes the Lindblad form~\cite{Gorini_1976,Lindblad_1976,fn:general_lindblad_form}: 
 \be 
\partial _t \rho = -i [H_\S+\Lambda ,\rho] -\frac{1 }{2}\{  L^\dagger   L, \rho \} +     L\rho     L^\dagger .
\label{eq:lindblad_1}
\ee
In the above, $L$ is known as the jump operator, and determines  the dissipative component of the system's evolution, while the Hermitian Lamb shift  $\Lambda$ accounts for the renormalization of the Hamiltonian due to the  system-bath coupling.
Note that we set $\hbar = 1$ throughout.

Importantly, the conditions for 
 the Lindblad-form master equation that we identify in Sec.~\ref{sec:derivation} are formulated purely in terms of properties of the {bath} and the   system-bath coupling strength.
This situation stands in contrast to the quantum optical master equation, which is also a Lindblad-form master equation but is only valid under additional stringent requirements on the level spacing of the {\it system} itself~\cite{Breuer}. 

In a wide range of situations, all information of the bath $\B$ required to determine the evolution of $\rho$ is contained in the  bath spectral function $J(\omega)$~\cite{Breuer} (see Sec.~\ref{sec:born_markov} for definition). 
The conditions we obtain in Sec.~\ref{sec:derivation} for the Lindblad-form master equation  are also expressed in terms of this function:  from $J(\omega)$ we identify   an energy scale $\Gamma$ and timescale $\tau$ whose dimensionless product $\Gamma \tau$   serves as a  measure of Markovianity. 
As the main result of this paper, we show that, when $\Gamma \tau \ll 1$,    the time-evolution of $\rho$ is  accurately described by a (Markovian)    master equation in the Lindblad form [Eq.~\eqref{eq:lindblad_1}], with the single  jump operator 
\be 
L  = \sum_{mn}\sqrt{2\pi \gamma  J(E_n-E_m)} X_{mn} |m\rangle \langle n|.
\label{eq:ady_jump_operator}
\ee
Here $\{|n\rangle\}$ and $\{E_n\}$ denote the eigenstates and energies of the system Hamiltonian $H_\S$ (not including the Lamb shift), respectively, while $X_{mn}\equiv \langle m|X|n\rangle$. 
The Lamb shift $\Lambda$  is proportional to $\gamma $ and is defined from $X$ and the bath spectral function in Eq.~\eqref{eq:lamb_shift_static} below. 
Stated more precisely, as we show in Sec.~\ref{sec:derivation}, the time-derivative of $\rho$ is given by Eqs.~\eqref{eq:lindblad_1}-\eqref{eq:ady_jump_operator}, up to a correction of order $\Gamma^2 \tau$. 
In comparison,  the magnitude of the right-hand side of Eq.~\eqref{eq:lindblad_1} is typically well-estimated by $\Gamma$ (hence the correction is smaller by a factor $\Gamma \tau$). 
Due to its system-independent applicability, we refer to the master equation in Eqs.~\eqref{eq:lindblad_1}-\eqref{eq:ady_jump_operator}, along with its generalizations  in Sec.~IV, as the {\it  universal Lindblad equation}. 

The two quantities $\Gamma$ and $\tau$ that determine the accuracy of the universal Lindblad equation [Eqs.~\eqref{eq:lindblad_1}-\eqref{eq:ady_jump_operator}] are associated with the bath spectral function $J(\omega)$  and system-bath coupling $\gamma $. 
Specifically, $\Gamma$ and $\tau$ are derived from a related function, $g(t)$, that we call the ``jump correlator.''
 The jump correlator is defined via its Fourier transform, $g(\omega)$, as the square root of the spectral function: $J(\omega) = 2 \pi [g(\omega)]^2$. 
 In time domain, this gives
\be 
g(t) =\frac{1}{\sqrt{2\pi}} \infint {\rm d}\omega \, \sqrt{ J(\omega)}e^{-i\omega t}.
\label{eq:jump_correlator}
\ee
From this  jump correlator, $\Gamma$ and $\tau$ are given by 
\be 
\Gamma = 4 \gamma   \left[\int_{-\infty}^\infty \!\!\!{\rm d}t\, |g(t)|\right]^2, \quad \tau = \frac{\int_{-\infty}^\infty\!{\rm d}t\, |g(t) t|}{\int_{-\infty}^\infty\!{\rm  d}t\, |g(t)|},
\label{eq:time_scales}
\ee
where  $\gamma $ denotes the system-bath coupling strength.
As we explain  in Sec.~\ref{sec:bath_time_scales}, the timescale $\tau$ can be seen as a measure of the characteristic correlation time of the bath observable $B$, while  $\Gamma$  sets an upper bound for  the  rate of bath-induced evolution of the system, independent of any approximation.
In the limit $\Gamma \tau \ll 1$, where the {ULE} 
 is valid, the correlations of the bath decay  rapidly on  the characteristic timescale of  system-bath interactions $\Gamma^{-1}$.
 In this case, the standard heuristic arguments behind the Markov-Born approximation suggest that the dynamics of the system should be effectively Markovian~\cite{Breuer}. 
The  results we obtain here
hence    put this intuition on rigorous footing, independent of properties of the system itself.

We note that the bath-induced terms in Eq.~\eqref{eq:lindblad_1} scale linearly with the system-bath coupling strength $\gamma $.  
In contrast, the correction to Eqs.~[\eqref{eq:lindblad_1}-\eqref{eq:ady_jump_operator}] we identified above is of order $\Gamma^2 \tau$, and thus scales as $\gamma ^2$.
Hence, when the coupling-independent quantities  $\Gamma/\gamma $ and $\tau$ are  finite, a small enough value of $\gamma $ can in principle always be found such the system's dynamics are Markovian and well-described by the universal Lindblad equation in Eq.~\eqref{eq:lindblad_1}. 
In this way,  the condition $\Gamma \tau \ll 1$ gives a well-defined notion of the weak-coupling limit.

We demonstrate in Sec.~\ref{sec:bloch_redfield_conditions} that the Bloch-Redfield equation is also valid up to a correction of order $\Gamma^2 \tau$.
In this sense, the ULE is valid on an equivalent level of approximation as the  Bloch-Redfield equation.

The universal Lindblad equation in Eqs.~\eqref{eq:lindblad_1}-\eqref{eq:ady_jump_operator} is consistent with existing results for open quantum systems~\cite{Breuer}. 
In particular, the ULE naturally reduces to   the quantum optical master equation in the limit where the latter is valid, namely when the  rate of system-bath interactions  $\Gamma$ is much smaller than any energy level spacing of the Hamiltonian $H_\S$  (i.e., when the rotating wave approximation is valid). 
However, in contrast to the quantum optical master equation, the derivation of the ULE does not rely on the rotating wave approximation (or any other assumptions about the energy levels of the system); hence  it can also be  applied beyond the regime where the quantum optical master equation is valid. 
In addition to being consistent with the quantum optical master equation as explained above, 
Eqs.~\eqref{eq:lindblad_1}-\eqref{eq:ady_jump_operator}  reproduce   Fermi's golden rule:
Fermi's golden rule  states that the transition rate between two  energy levels  of the Hamiltonian, $m$ and $n$,  is given by $\Gamma_{n\to m}  =2 \pi |X_{mn}|^2 J({E_n-E_m})$. 
This  result follows from Eqs.~\eqref{eq:lindblad_1}~and~\eqref{eq:ady_jump_operator} by identifying $\Gamma_{n\to m }=\langle m| \partial _t \rho(t)|m\rangle|_{t=0}$ when  taking $\rho(0)= |n\rangle\langle n|$. 


We note that the expression for the jump operator $L$ in Eq.~\eqref{eq:ady_jump_operator}  was previously
 hypothesized in Ref.~\cite{Kirsanskas_2018}. 
Ref.~\cite{Kirsanskas_2018} showed that the master equation in Eqs.~\eqref{eq:lindblad_1}-\eqref{eq:ady_jump_operator} was consistent with   Fermi's golden rule, and reproduced the quantum optical master equation in the regime  $\gamma  \to 0$ where the latter is valid.
Based on these results and numerical demonstrations, Ref.~\cite{Kirsanskas_2018} conjectured that the master equation in Eqs.~\eqref{eq:lindblad_1}-\eqref{eq:ady_jump_operator} could accurately describe the evolution of open quantum systems. 
In this work, by rigorous derivation we recover the hypothesis of Ref.~\cite{Kirsanskas_2018}, and identify the precise conditions under which the universal Lindblad equation [Eqs.~\eqref{eq:lindblad_1}-\eqref{eq:ady_jump_operator}] holds.
Crucially, the conditions we identify  rely solely on the properties of the bath and system-bath coupling, and hold well beyond the regime where the quantum optical master equation is valid.
 In addition, our results generalize the hypothesized master equation from Ref.~\cite{Kirsanskas_2018} to arbitrary system-bath couplings and time-dependent  Hamiltonians.

\section{Open system dynamics: formulation and characteristic timescales}
\label{sec:model}
We now set out to derive  the universal Lindblad equation by rigorous means, for general open quantum systems. 
As a first step, in this section we define the model we study and review standard theory for open quantum systems. 
We moreover present two auxiliary results that   play an important role for the derivation of the ULE: 
we establish a rigorous upper bound for the correction to the Bloch-Redfield equation (Sec.~\ref{sec:bloch_redfield_conditions}, see also Ref.~\cite{Mozgunov_2020}), and obtain an upper bound for the rate of bath-induced quantum evolution (a so-called ``quantum speed limit'') [Eq.~\eqref{eq:speed_limit_m}].
These results, which may also be of interest on their own, are derived in Appendix~\ref{app:br_correction}.
The concepts and basic assumptions described 
in this section will form the foundation for the derivation of the {ULE} in the  next section.

The system we consider in this paper consists of a quantum (sub)system $\S$ which is connected to an  external system, referred to as the  bath  $\B$.
The subsystem  $\S$  may  be anything from a two-level spin to a many-body  system, while the   bath $\B$ is typically a large system with  a dense energy spectrum, such as a phononic or electromagnetic environment, or the fermionic modes in an electronic lead. 
The bath $\B$ can also consist of  several  ``sub-baths'' with distinct physical origins and properties. 
Without loss of generality, the Hamiltonian $H$ of the  full system  $\S\B$ (including the bath)
  takes the form 
\be
 H =  H_\S +  H_\B  +H_{\rm int},  \label{eq:SystemBathHamiltonian}
\ee
where $ H_\S$   and $H_{\B}$ are  the  Hamiltonians of the subsystem  and bath,  respectively, while  $H_{\rm int}$ contains all terms in the Hamiltonian that couple the two. 
In the following, we allow $H_\S$ to depend  on time, while we assume $H_\B$ and $H_{\rm int}$ to be time-independent.

It is useful to decompose  $H_{\rm int}$ as follows:
\be 
H_{\rm int} = \sqrt{\gamma }\sum_{\alpha } X_\alpha  B_\alpha ,
\label{eq:h_sb_def}
\ee
where, for each $\alpha $, $X_\alpha $ is a dimensionless Hermitian operator on the subsystem $\S$,  $B_\alpha $ is a Hermitian operator acting on the bath $\B$, with units of $[{\rm Energy}]^{1/2}$, and the energy  $\gamma $   parametrizes  the  system-bath coupling strength  (see footnote~\onlinecite{fn:sb_hamiltonian}).
We  normalize $\gamma $  and $B_\alpha $ such that $X_\alpha $ has unit spectral norm  for each $\alpha $~\cite{fn:spectral_norm_def}}. 
While  $\gamma $ can still be absorbed into the operators $\{B_\alpha\} $, and thus  in  principle remains arbitrary,  we include it in Eq.~\eqref{eq:h_sb_def} to highlight the  scaling of various quantities  with respect to the system-bath coupling in the discussion below. 
We  note that the decomposition above is always possible with a sufficiently high, but finite, number of terms in the sum $N$.
We refer to each such term as a (quantum) noise channel in the following. 

For simplicity, in the remainder of this section, and in the derivation of the {ULE} in Sec.~\ref{sec:single_channel},  we consider the case where the sum in Eq.~\eqref{eq:h_sb_def} consists of a single term, and refer to the  system and bath operators as $X$ and $B$, respectively. 
In Sec.~\ref{sec:multiple_noise_channels}, we generalize our results to the  case where the sum in Eq.~\eqref{eq:h_sb_def} contains multiple terms.  

To describe the dynamics of observables in the system $\S$, it is sufficient to know the evolution of the reduced density matrix of $\S$, 
\be 
 \rho(t)\equiv \Tr_\B \left[\rho_{\S\B}(t)\right].
\ee
Here $\Tr_\B$ traces out all the  degrees of freedom in $\B$, and  $\rho_{\S\B}(t) $ denotes the density matrix of the combined system $\S\B$.
Crucially, it is possible to obtain  an equation of motion for $ \rho(t)$  which depends only on  $H_\S$, $X$, and the statistical properties of the bath. 
Such an equation of motion is known as a master equation.
There exists several approximation schemes for obtaining master equations for $\rho$ (see, for example, Refs.~\onlinecite{Breuer,GardinerZoller,Nakajima_1958,Zwanzig_1960,Wangsness_1953,Redfield_1965,Davies_1974}). 
While useful in their respective regimes of applicability, each of these methods has its limitations on which cases they  may be applied (either due to physical limitations on the regime of applicability, or practical issues associated with numerical implementation).
The goal of our paper is  to derive a new Markovian master equation that can be applied to a wider range of cases, which unifies and extends  some of these previous approaches.

\subsection{Born-Markov approximation}
\label{sec:rho_dynamics}
\label{sec:born_markov}
Before deriving the universal Lindblad equation, we review one of the  existing approaches to obtaining  a master equation for $\rho$, namely the Born-Markov approximation. 
This standard approach leads to a master equation for $\rho$ known as  the Bloch-Redfield {(BR)} equation.
The concepts introduced here will be used in the derivation of  the {Universal Lindblad equation} in Sec.~\ref{sec:derivation}.

\subsubsection{Derivation of Bloch-Redfield equation}

To derive the {BR} equation,  we assume that   the bath was in a steady state  at some arbitrary time $t_0$ in the remote past. 
Specifically, we assume that $\rho_{\S \B}(t_0) = \rho_\S(t_0) \otimes \rho_\B$, where $\rho_\B$ describes a steady state of the bath: $[H_\B , \rho_\B] = 0$.
The bath state $\rho_\B$ can for example describe a thermal equilibrium state with a specific temperature and chemical potential. 
If $\B$ consists of several  sub-baths, $\rho_\B$ can also be a direct product of thermal states out of equilibrium with each other. 
Due to  its macroscopic size, the state of the bath remains practically  unaffected  by the  system $\S$ at  later times, except for short-lived fluctuations arising from the system's evolution in the recent past. 
Without loss of generality, we may assume that each bath operator $B_\alpha $ has vanishing expectation value in the bath state $\rho_\B$: $\Tr_\B(B_\alpha  \rho_\B) = 0$, since nonzero expectation values can be eliminated by appropriate redefinition of $H_\S$ and $B_\alpha $ in Eqs.~\eqref{eq:SystemBathHamiltonian}~and~\eqref{eq:h_sb_def}.

We note that, due to the finite memory and relaxation times of the bath and of the system, respectively, the evolution of $\rho(t)$ should be independent of the details of the initialization in the remote past. 
Supporting this, in Appendix~\ref{seca:transient_correction} we show that the evolution of the system is independent of  the details of the state $\rho_\S(t_0)$ and the exact value of  $t_0$, when $t_0$ is sufficiently far in the past. 

The {BR} equation is most  easily  derived in the interaction picture. 
We transform the problem to the interaction picture by  applying a rotating frame transformation  generated by the Hamiltonian $H_\S(t)+H_\B$. 
After this transformation, the Hamiltonian of  the combined system $\S\B$ in the interaction picture  is  given by:
\be
\tilde H(t) = \sqrt{\gamma}  \tilde X (t)\tilde B (t).
\label{eq:interaction_picture_hamiltonian}
\ee
Here 
$
\tilde X (t) \equiv U^\dagger(t) X  U(t)$, and 
$
\tilde B (t) \equiv e^{iH_\B t}B e^{-iH_\B t},
$
where $U(t) \equiv \mathcal T e^{-i \int_{0}^t dt'H_\S(t') }$  is the time-evolution operator of the subsystem $\S$ relative to an arbitrary origin of time, and  $\mathcal T$ is the time-ordering operation. 
We let $\tilde \rho(t)$  denote  the reduced density matrix of $\S$ in the interaction picture. 
Specifically, $\tilde \rho(t) \equiv \Tr_\B[\tilde \rho_{\S\B}(t)]$, where $\tilde \rho_{\S\B}(t)$ denotes the state of the combined system  $\S\B$ when  time-evolved with   $\tilde H(t)$ from the state $\tilde \rho_{\S\B}(0)=  \rho_{\S\B}(0)$.
From $\tilde \rho(t)$, one  can  straightforwardly obtain the time-evolution of the system in the Schr\"odinger picture  through  the relation   $\rho(t) = U(t)  \tilde\rho(t)U^\dagger(t)\label{eq:interaction_to_schrodinger}
$.

After transforming to the interaction picture, the system's dynamics   occur on a timescale which is set by the  system-bath coupling $\gamma $.
When this  coupling  is sufficiently weak, $\tilde \rho(t)$ can be assumed static  on the  intrinsic correlation timescale   of the bath (see Sec.~\ref{sec:ady_section}). 
This   so-called weak-coupling limit forms the basis for the derivation of the   {BR} equation, using the   Born-Markov approximation~\cite{Breuer}.
To  employ the Born-Markov approximation, we integrate the von Neumann equation $\partial _t  \tilde \rho_{\S\B}(t) = - i [\tilde H(t) , \tilde \rho_{\S\B}(t)]$ once, obtaining $\partial _t \tilde \rho_{\S\B}(t)=-\intop_{t_0}^t {\rm d}t'\, [\tilde H(t),[\tilde H(t'),\tilde \rho_{\S\B}(t')]]$. 
(Here we exploited the fact that $\Tr_\B (\tilde B(t_0) \rho_{\S\B}) = \Tr_\B (B \rho_{\S\B}) $ vanishes by assumption as described above, to eliminate the term arising from the boundary term of the integration).
The Born approximation amounts to setting $\tilde \rho_{\S\B}(t') \approx  \tilde \rho(t') \otimes \rho_\B$ inside the integral. 
The next step is to take the partial trace over the bath, to obtain an equation of motion for the reduced density matrix of the system, $\tilde{\rho}(t)$. 
Using the fact that  $\tilde{X}(t)$ acts only on the system, while $\tilde{B}(t)$ acts only on the bath, we obtain
$\partial _t \tilde \rho(t) \approx     - \gamma \int_{t_0}^{t} {\rm d}t'    J(t-t') \big[\tilde X(t) , \tilde X(t')\tilde \rho (t')\big] + H.c.$, where we introduced the (two-point) bath correlation function 
\be
J(t-t')\equiv \Tr_\B\left(  \tilde B(t) \tilde B(t')\rho_\B\right).
\label{eq:correlation_function_def}
\ee

Finally, the Markov approximation is implemented by assuming that $\tilde \rho(t')$ is stationary over the characteristic decay time of the bath correlation function $J(t)$ (see below for  discussion). 
By making the replacement $\tilde{\rho}(t')\approx \tilde \rho(t)$ inside the integral over the history of the system $(t')$, and taking  the limit $t_0 \to -\infty$, we obtain~\cite{Breuer}: 
\be 
\partial _t \tilde \rho(t)  = \mathcal D _{\hspace{0.5pt}{\rm R}}(t)[\tilde\rho(t)] + \xi(t), 
\label{eq:redfield_dissipator}
\ee
where
\begin{align}
 \mathcal D_{\hspace{0.5pt}{\rm R}} (t)[\rho] \equiv - \gamma\! \int_{-\infty}^{t} \!\!\!\!\!{\rm d}t'     J(t-t') \big[\tilde X(t) ,  \tilde X(t')\rho\big] +\! H.c., \! 
\label{eq:redfield_superoperator}
\end{align}
and $\xi(t)$  denotes the correction  arising from the Born and Markov approximations above.
The Bloch-Redfield equation is obtained by assuming the error induced by the Born-Markov approximation, $\xi(t)$, to be negligible in Eq.~\eqref{eq:redfield_dissipator}. 
In Appendix~\ref{app:br_correction}, we derive an upper bound for this correction,  thus obtaining rigorous conditions for the validity of the BR equation.
 See Sec.~\ref{sec:bloch_redfield_conditions} for a further discussion. 

Note that the last approximation in the  above derivation resulted in an equation of motion for $\tilde{\rho}(t)$  which is {\it Markovian}: in Eq.~\eqref{eq:redfield_dissipator}, the time derivative $\partial_t \tilde{\rho}(t)$ depends only on the value of $\tilde{\rho}(t)$ at the the same time, $t$.
As we demonstrate in Sec.~\ref{sec:derivation} (see Sec.~\ref{sec:markov_family} for discussion), the Born-Markov approximation above is not the only way of approximating $\partial _t \tilde \rho$ by a Markovian master equation in the weak-coupling limit. 
In Sec.~\ref{sec:derivation}, we will develop a different Markovian approximation for $\partial _t \tilde \rho$ which  is valid under the same conditions  as the  standard Markov-Born approximation 
 above, but, unlike the former, leads to a master equation in the Lindblad form.

The bath correlation function in Eq.~\eqref{eq:correlation_function_def}, or equivalently its Fourier transform $ J(\omega) \equiv \frac{1}{2\pi}\int_{-\infty}^\infty {\rm d}t\, J(t) e^{i\omega t}$, known as the bath spectral function,  plays a crucial role for describing the dynamics of the system $\S$:
in the  {BR} equation 
 [Eq.~\eqref{eq:redfield_dissipator}], 
 $J(t)$ contains all information of the bath required to determine the evolution of $\tilde\rho$. 
Importantly, even without   the Born-Markov approximation, a wide class of baths (so-called Gaussian baths) are fully characterized by  the two-point correlation function
 $J(t)$. 
This situation for instance arises if the  bath consists of a large collection of decoupled subsystems, such as continua of independent  fermionic or bosonic modes.  
While we note that our approach below can also be applied to cases where higher-order bath correlations  are relevant, in this paper, we assume for simplicity that the bath is Gaussian.

The spectral function $ J(\omega)$, which is real and non-negative,  can in  many cases be  computed or phenomenologically assumed~\cite{Breuer} (see for example  Sec.~V, where we  calculate $J(\omega)$ for a  bath of bosonic modes). 
While in the above $\tilde B (t)$ emerged as a time-evolved observable in a quantum mechanical bath,  the results in this paper also apply to the case where $\tilde B (t)$ is a classical noise signal, and  the bath trace is  replaced by the statistical average over noise realizations. 
In this case, $J(\omega)$ is symmetric in $\omega$ and gives the spectral density of the classical noise signal.

\subsubsection{Correction to Bloch-Redfield equation}
\label{sec:bloch_redfield_conditions}
As an important secondary result, in this paper we derive a rigorous upper bound for the correction to the Bloch-Redfield equation,  $\xi(t)$,  which is independent of the details of the system Hamiltonian, $H_\S$.
This error bound is used in the next section, where we   derive   the ULE.

To derive the error bound, we  assume that  the bath is Gaussian, and that the bath and system were decoupled at some point in the remote past (see beginning of this subsection). 
Using these assumptions, in Appendix~\ref{app:br_correction} we systematically  expand the time-derivative of $\tilde \rho$   in powers of the  dimensionless ``Markovianity parameter''
 $\Gamma \tau $, where the bath timescales $\Gamma^{-1}$ and $\tau$ were defined from the bath spectral function and the system-bath coupling strength, $\gamma $, in Eqs.~\eqref{eq:jump_correlator}~and~\eqref{eq:time_scales} 
(in Sec.~\ref{sec:bath_time_scales}, we further discuss the physical meaning of these timescales).
As we show in Appendix~\ref{app:br_correction},  truncation of the expansion  of $\partial _t \tilde \rho$ to leading order in $\Gamma \tau$  yields $\mathcal D_{\rm R}[\tilde \rho(t)]$. 
This truncation is thus  is equivalent to making the Born and Markov approximations, while  the correction $\xi(t)$  corresponds to the  sum of all subleading terms in the expansion. 
In Appendix~\ref{app:br_correction} we obtain a bound for  this subleading correction:
\be
\norm{\xi(t)}\leq \Gamma^2 \tau,
\label{eq:br_bound}
\ee
where (here and in the following)
$\norm{\cdot}$ refers to the spectral norm (see footnote~\cite{fn:spectral_norm_def}).

{Note that consistent correction bounds for the BR equation were recently  obtained elsewhere~\cite{fn:correction_bounds,Mozgunov_2020}.}
 Our derivation of Eq.~\eqref{eq:br_bound} holds in the general case where the system and bath are connected through multiple noise channels, 
 with  the generalized definitions $\Gamma$ and $\tau$  given     in Eq.~\eqref{eq:time_scales_m} below (see Ref.~\cite{Breuer} or  Appendix~\ref{app:br_correction} for the multi-channel generalization of the BR equation).

%


In Eq.~\eqref{eq:speed_limit_m} below, 
 we also show that the spectral norm of $\partial _t\tilde  \rho$ on the left-hand side of Eq.~\eqref{eq:redfield_dissipator} is bounded by $\Gamma/2$.
 Comparing this bound with Eq.~\eqref{eq:br_bound} above,  we conclude  that $\Gamma\tau\ll 1$ is a necessary condition for the Born-Markov approximation to be justified by our arguments.

%

\subsection{Characteristic  timescales of the bath}
\label{sec:bath_time_scales}
\begin{figure}
\includegraphics[width=0.99\columnwidth]{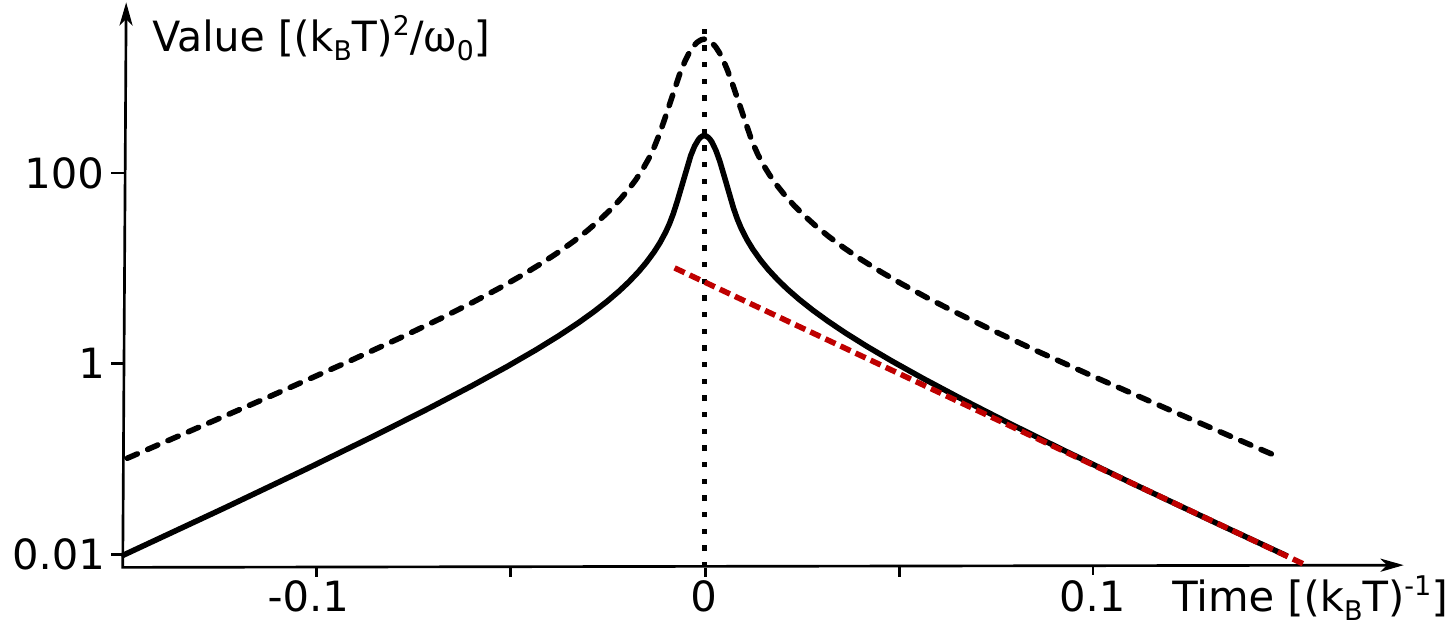}
\caption{ 
Absolute value of the jump correlator $g(t)$  (solid) and bath correlation function $J(t)$ (black dashed line) for the Ohmic bath studied  in Sec.~\ref{sec:bath_time_scales}. 
Red line indicates the fit used to obtain the    exponential decay constant for $g$ (see Sec.~\ref{sec:bath_time_scales}).
}\label{fig:jump_correlator}
\end{figure}

Above we found that the  validity of the Born-Markov approximation is 
 determined from the 
 characteristic 
  timescales
   $\Gamma^{-1}$  and $\tau$, which are intrinsic to the bath (and its coupling to the system). 
In  Sec.~\ref{sec:derivation},  we show that  $\Gamma $ and $\tau$  also  determine the accuracy of the universal Lindblad equation.
Here, we briefly discuss the nature of these quantities. 
As a demonstration, we  moreover explicitly calculate  the jump correlator $g(t)$ and the correlation time $\tau$ for the case of an Ohmic bath.

To highlight the physical meaning of the time scale $\Gamma^{-1}$, in Appendix~\ref{app:speed_limit} we show that $\Gamma$  provides a strict bound on the rate of change of $\tilde\rho$,
\be 
\norm{\partial _t \tilde \rho} \leq \Gamma /2.
\label{eq:speed_limit_m}
\ee  
Note that Eq.~\eqref{eq:speed_limit_m} is exact, and is derived without any approximations, other than the assumption of  a Gaussian bath. 
In this way, the rate $\Gamma/2$ can   be seen as a ``quantum speed limit'' for dissipative quantum evolution~\cite{DelCampo_2013,Deffner_2017}.  
Heuristically,
$\Gamma^{-1}$ thus
 characterizes the ({shortest}) typical interval between real or virtual system-bath interaction events, such as, e.g., photon emission or absorption.
Note that the inequality in Eq.~\eqref{eq:speed_limit_m} extends to  the case of multiple noise channels, with $\Gamma$ as defined in Eq.~\eqref{eq:time_scales_m} below (see Appendix~\ref{app:speed_limit}).

The timescale $\tau$ captures the characteristic decay time of correlations in the bath.
To see this,    note from  Eq.~\eqref{eq:time_scales} that $\tau$ gives the mean value of $|t|$ associated with the normalized distribution $|g(t)|/C$, where $C\equiv {\int_{-\infty}^\infty\!{\rm  d}t\, |g(t)|}$. 
The existence of a finite value of $\tau$ requires that $g(t)$    effectively decays faster than $C \tau /t^2$ for $|t|\gg \tau$~\cite{fn:g_decay}.
Noting  from Eq.~\eqref{eq:jump_correlator}
that the bath correlation function $J(t)$ is given by the convolution of the jump correlator with itself, $J(t-t') = \int_{-\infty}^{\infty}{\rm d}s\, g(t-s)g(s-t')$,  the bath correlation function  hence must also decay on a timescale of magnitude $\tau$. 
The conditions described above hold under the assumption that $\tau$ takes a finite value; a divergent value of $\tau$ indicates that long-term memory is present in the bath; in this case, the system cannot be well-described by a Markovian master equation.

To illustrate the above relationship between  $J(t)$, $g(t)$, and the correlation time $\tau$, we explicitly compute 
$J(t)$,  $g(t)$, and $\tau$, for an Ohmic bath. 
The  Ohmic bath  consists of a  continuum of bosonic modes with Hamiltonian $H_\B =\intop_0^\infty \!\! {\rm d}\omega\, \omega b ^\dagger(\omega)b (\omega)$, 
where $b(\omega)$ denotes the  annihilation operator of modes with frequency $\omega$, satisfying $[b(\omega),b^\dagger(\omega')]=\delta(\omega-\omega')$ and  $\delta(\omega)$ denotes the Dirac delta function. 
In the framework of Eqs.~(\ref{eq:SystemBathHamiltonian})~and~(\ref{eq:h_sb_def}),  the bath operator $B$ is given by the bosonic field operator $\intop_0^\infty\!\! {\rm d}\omega\,\sqrt{S(\omega)} [b (\omega)+b ^\dagger(\omega)]$, where $S(\omega)$ denotes the effective spectral density of the bath, including the frequency-dependence of the system-bath coupling.
The class of models above is commonly used in the literature~\cite{Breuer,GardinerZoller}, and  can for  example describe a phononic or electromagnetic environment of an electronic system.

We consider the Ohmic   spectral density $S(\omega)  = \omega  e^{-\omega^2 /2\Lambda^2 }/\omega_0$, 
with an ultraviolet energy cutoff  set by the scale $\Lambda$. 
The energy scale $\omega_0$ is  introduced  to keep $S(\omega)$ dimensionless.
Assuming the bath is in equilibrium at temperature $T$,
a straightforward  calculation~\cite{Breuer,GardinerZoller} 
yields the bath spectral function 
\be 
J (\omega) =  \frac{1 }{\omega_0}\frac{\omega e^{-\frac{\omega^2}{2\Lambda^2}}}{1-e^{- \omega/ T }},
\label{eq:bosonic_sf}
\ee
where we work in units where $k_{\rm B}=1$.  

Using Eq.~\eqref{eq:jump_correlator}, we numerically compute the jump correlator  $g(t)$ from the  spectral function  in Eq.~\eqref{eq:bosonic_sf}, for the case where  $\Lambda = 50\,  T$. 
By explicit computation [see Eq.~\eqref{eq:time_scales}], we find for this case $\tau\approx0.007\, T^{-1}$.
In Fig.~\ref{fig:jump_correlator}, we plot $|g(t)|$  on a logarithmic scale (solid line),
along with the bath correlation function $|J(t)|$ (dashed line). 
{As Fig.~\ref{fig:jump_correlator} shows, both $J(t)$ and $g(t)$  decay exponentially, at approximately the same rate, after a sharp initial drop at short times. 
We confirm numerically (data not shown here) that the short-time peak arises from high-energy modes in the bath, and is controlled by the cutoff, $\Lambda$.
By linear regression  (red dashed line in Fig.~\ref{fig:jump_correlator}) we find the slope of  $\log g(t)$ outside this initial decrease to be given by approximately $0.023\,  T^{-1}$.   
The difference between the exponential decay constant from $\tau$  is caused by the short-time peak of $g(t)$, and is  thus controlled by $\Lambda$.   

\section{Universal Lindblad equation}
\label{sec:derivation}

While useful, the standard Born-Markov approximation  discussed in Sec.~\ref{sec:rho_dynamics} has some shortcomings. 
In particular, the  Bloch-Redfield equation in Eq.~\eqref{eq:redfield_dissipator} is not in the Lindblad form.
As a result, as was explained in the introduction, integration of the {BR} equation may   yield  negative or diverging probabilities, and can be impractical to implement numerically  even for moderately sized quantum systems. 
In this section, we  derive a  master equation for $\tilde \rho$ which is valid under the same conditions  as the {BR} equation, but will be in the Lindblad form and thus  free of the limitations above.
Specifically, our new master equation  is accurate up to a  correction of the same magnitude   as the correction bound  $\Gamma ^2 \tau$  we identified for the {BR} equation in Sec.~\ref{sec:bloch_redfield_conditions}. 
The new master equation, which we term the universal Lindblad equation {(ULE)}, constitutes the  main  result of our paper. 
Crucially,  the {ULE}  does  not
 require any special conditions on the system to be valid; rather, its validity  relies solely on the properties of the {\it bath} itself (along with its coupling to the system).

To make the physical basis for the {ULE} most transparent,  in Sec.~\ref{sec:single_channel}, we derive the ULE on an intuitive level of argumentation, focusing on the case where the system and bath are coupled through a single noise channel.
In Appendix~\ref{app:ule_derivation} we  provide a rigorous derivation of these results that also holds for general system-bath couplings.
In Sec.~\ref{sec:markov_family}, we comment on the principle underlying our derivation; namely, the existence of distinct, but equivalently valid, Markov approximations.
The results for general system-bath couplings are given in Sec.~\ref{sec:multiple_noise_channels}, and expressed in the Schr\"odinger picture in Sec.~\ref{sec:schrodinger_picture}.

\subsection{Single noise channel}
\label{sec:single_channel}

In this subsection we heuristically derive the universal Lindblad equation for the case of a single quantum noise channel.
As a first step in our derivation, we identify an alternative form of Markov approximation, which is valid at the same level of approximation as  the standard Born-Markov approximation   (i.e., up to a correction of order $\Gamma^2\tau$). 
Subsequently, we demonstrate that this Markov approximation results in a master equation in the Lindblad form (in contrast, the standard Born-Markov approximation does not lead to a Lindblad-form master equation).

The starting point for our derivation is the BR  equation [Eq.~\eqref{eq:redfield_dissipator}], whose error bounds we obtained in Sec.~\ref{sec:bloch_redfield_conditions} above.
Below, we apply additional manipulations to the BR equation, which induce errors of the same magnitude as those inherent in the Born-Markov approximation used in deriving Eq.~(\ref{eq:redfield_dissipator}). 
These additional steps hence lead to a new  master equation that is different from the BR equation,  but  is valid on the same level of approximation. 
Unlike the BR equation, our new master equation is crucially in the Lindblad form. 
Our modification procedure  is equivalent to  employing a distinct Markovian approximation from the standard 
Markov approximation (reviewed in Sec.~\ref{sec:born_markov}) that is used to obtain the {BR} equation [note that our derivation still makes use of the ``conventional''  Born approximation, as described in the paragraph above Eq.~\eqref{eq:correlation_function_def}]. 
In Sec.~\ref{sec:markov_family} below, we discuss the diversity of possible Markov approximations in more detail.

As the first step of our derivation, we decompose the bath correlation function $J(t-t')$ [Eq.~(\ref{eq:correlation_function_def})] as a convolution using ``jump correlator'' $g(t)$ defined in Eq.~\eqref{eq:jump_correlator}: $J(t-t') = \int_{-\infty}^{\infty}{\rm d}s\, g(t-s)g(s-t')$. 
 Using this decomposition, the BR equation  [Eq.~\eqref{eq:redfield_dissipator}] can be (exactly) rewritten as:   
\be 
\partial _t \tilde \rho (t) \approx  \int_{-\infty}^\infty \!\!\! {\rm d}t'\int_{-\infty}^{\infty}\!\!\!{\rm d}s\, \mathcal F(t,s,t')[ \tilde \rho(t) ],
\label{eq:f_superoperator}
\ee
where 
\be
\mathcal F(t,s,t')  [\tilde \rho]  = \gamma   \theta(t-t') g(t-s)g(s-t') [\tilde X(t),\tilde  \rho \tilde X(t')]  +H.c..
\label{eq:f_def}
\ee
The approximate equality in Eqs.~\eqref{eq:redfield_dissipator} and \eqref{eq:f_superoperator} captures the correction   to the {BR} equation,  $\xi(t)$, whose bound (with respect to the spectral norm) we identified in Sec.~\ref{sec:bloch_redfield_conditions}.
For brevity, we do not   include this correction in the derivation below.
Note that  $\mathcal F(t,s,t')$ is a linear operator acting  on system operators.

Next, we  integrate Eq.~\eqref{eq:f_superoperator}  with respect to $t$   to compute the   change of $\tilde \rho$ over a finite time interval from $t_1$ to $t_2$, that we will choose much longer than $\tau$: 
\be 
\tilde \rho(t_2)- \tilde \rho(t_1) 
\approx \int_{t_1}^{t_2}\!\! {\rm d}t  \int_{-\infty}^\infty \!\!\! {\rm d}t'\int_{-\infty}^{\infty}\!\!\!{\rm d}s\, \mathcal F(t,s,t') [\tilde \rho(t)].
\label{eq:discrete_change_1}
\ee

We now argue that the weak-coupling limit $\Gamma \tau \ll 1$ allows us to  apply two approximations to the right-hand side above, which yield a new expression that 
is valid on an equivalent level of approximation as the standard Bloch-Redfield equation in Eq.~\eqref{eq:redfield_dissipator}.

To make the first approximation, we note  that, in the limit $\Gamma \tau \ll 1$, the condition $\norm{\partial _t \tilde \rho} \leq \Gamma /2$ in Eq.~\eqref{eq:speed_limit_m} ensures that $\tilde{\rho}(t) = \tilde{\rho}(s) + \mathcal{O}(\Gamma \tau)$ for $|t-s|\lesssim \tau$. 
Additionally, since  $g(t)$ decays  on the  timescale $\tau$ (see Sec.~\ref{sec:bath_time_scales}),  
 $\mathcal F(t,s,t')$ is  suppressed when the difference between any two of its time-arguments is much larger than $\tau$ [see Eq.~\eqref{eq:f_def}] (in particular, note that  $g(t-s)g(s-t')$, and thus $\mathcal F(t,s,t')$, must be small when $t-t'\gg \tau$).
These two  results suggest that we may replace  $\tilde \rho(t)$ by $\tilde \rho(s)$ in the right-hand side of  Eq.~\eqref{eq:discrete_change_1} when $\Gamma \tau \ll 1$. 
In Appendix~\ref{app:ule_derivation}, we implement this substitution in a systematic way, and prove that replacing $\tilde{\rho}(t)$ by $\tilde{\rho}(s)$ in Eqs.~\eqref{eq:f_superoperator}~and~\eqref{eq:discrete_change_1} results in a correction to $\partial _t \tilde{\rho}$ of order $\Gamma^2 \tau$.


%

To make our second approximation, we again use the fact that $\mathcal F(t,s,t')$ decays when the difference between any of its time-arguments exceeds $\tau$. 
Thus,  since $\tau\ll t_2-t_1$ by assumption, most of the contribution to the integral in  Eq.~\eqref{eq:discrete_change_1} comes from 
the region where all three integration variables $t,s,t'$ are located in the interval $[t_1,t_2]$. 
As a result, the right-hand side of Eq.~\eqref{eq:discrete_change_1} is approximately unaffected if we change the integration domain from $-\infty < (s ,t') < \infty , \  t_1\leq t\leq t_2$ to the domain $-\infty < (t ,t') < \infty ,\  t_1\leq s\leq t_2$. 
Indeed, in Appendix~\ref{app:ule_derivation}, we show that this change of  integration domain results in a  correction to $\tilde \rho$ which is bounded by $\Gamma \tau$. 

After making the two approximations described above [i.e., setting $\rho(t)\approx \rho(s)$ in Eq.~\eqref{eq:discrete_change_1}, and subsequently changing the domain of integration], we obtain
\be 
\tilde \rho(t_2) -\tilde \rho(t_1) \approx 
 \int_{t_1}^{t_2}\!\!\! {\rm d}s  \int_{-\infty}^\infty \!\!\! {\rm d}t \int_{-\infty}^\infty \!\!\!{\rm d}t'\, \mathcal F(t,s,t') [\tilde \rho(s)].
\ee
By taking the derivative with respect to $t_2$, and renaming the variables of integration,  we obtain the (time-local) master equation
\be 
\partial _t \tilde \rho(t) \approx  \mathcal L (t) [\tilde \rho(t)], 
\ \ 
\mathcal L(t)   = \!\int_{-\infty}^\infty \!\!\!\! {\rm d}s\! \int_{-\infty}^\infty \!\!\!\!{\rm d}s'\, \mathcal F(s,t,s') .
\label{eq:l_def_0}
\ee

In Appendix~\ref{app:ule_derivation} we put the above line of arguments on rigorous footing:  we  identify  a slightly modified density matrix $ \rho'(t)$ whose norm-distance to $\tilde \rho(t)$ remains bounded by $\Gamma \tau$ at all times. 
Assuming that the bath is Gaussian, and that the bath and system were decoupled at some point in the remote past, we show that $\rho'$   evolves according to  the master equation
\be
\partial _t \rho'(t) = \mathcal L (t)[\rho'(t)] + \xi'(t),
\label{eq:accurate_master_eq}
\ee 
where $\mathcal{L}(t)$ is defined by Eqs.~(\ref{eq:f_def}) and (\ref{eq:l_def_0}), and $\norm{\xi'(t)}\leq 2 \Gamma^2 \tau$ for 
 all times, $t$. 
In the Markovian limit $\Gamma \tau \ll 1$, $\rho'(t)$ is nearly identical to $\tilde \rho(t)$, and  the   evolution
 of the system is thus well-described by  $\rho'(t)$.
The same condition $\Gamma \tau\ll 1 $ is already required for the BR equation to be valid by our arguments (see Sec.~\ref{sec:bloch_redfield_conditions}) and hence  does  not impose additional constraints on the system. 
Consistent with Eq.~\eqref{eq:speed_limit_m}, we show in Appendix~\ref{app:ule_derivation} that  $\norm{\mathcal L(t)[\rho]}\leq \Gamma/2$.
Hence,  by the same arguments as in Sec.~\ref{sec:bloch_redfield_conditions}, $\Gamma \tau \ll 1$ is also a necessary condition for  error $\xi'$ above  to  be  negligible.

As a final step, we verify  that the master equation in Eq.~\eqref{eq:l_def_0} is in the  Lindblad form. 
By  decomposing the step function $\theta(t-t')$ in Eq.~\eqref{eq:f_def} into its symmetric and antisymmetric components, $\theta(t) = \frac{1}{2} + \frac{1}{2} \sgn(t)$, we find through a straightforward  computation  (see  Appendix~\ref{sec:lindblad_form} for details) that 
\be 
\mathcal L (t) [\tilde\rho] =-i  [\tilde \Lambda(t),\tilde \rho] -   \frac{1}{2}\{\tilde L^\dagger(t)\tilde  L(t),\tilde \rho \}   +\tilde L(t)\tilde \rho \tilde  L^\dagger(t) ,
\label{eq:lindblad_form_1}
\ee
where the jump operator $\tilde L(t)$ is given by 
\be
\tilde L (t) = \sqrt{\gamma } \infint {\rm d}s \,g(t-s)\tilde X (s)
\label{eq:jump_operator_def}
\ee
and 
\be 
\tilde \Lambda(t) =   \frac{\gamma}{2i}  \infint {\rm d}s{\rm d}s' \tilde X(s) g(s-t)g(t-s')\tilde X(s')\, \sgn(s-s').
\label{eq:lamb_shift}
\ee
The Lamb shift $\tilde \Lambda(t)$ is by construction Hermitian, due to   the  symmetry of the jump correlator  $g(t) = g^*(-t)$, which results from  its definition  in Eq.~\eqref{eq:jump_correlator}.

Comparing with the BR equation [Eq.~\eqref{eq:redfield_dissipator}], we see that the universal Lindblad equation [Eqs.~\eqref{eq:accurate_master_eq}-\eqref{eq:lamb_shift} with $\xi'(t)$ neglected] yields an expression for $\partial _t\rho'(t)$ which is  accurate up to a  correction bounded by the same value as the correction  for the BR equation (up to a factor of $2$). 
In this sense, the ULE  and the BR equation are valid on an equivalent level of approximation~\cite{fn:error_accumulation}.

To summarize this section,  we showed that, in the weak-coupling limit $\Gamma \tau \ll 1$,  the interaction picture density matrix of the subsystem $\S$, $\tilde \rho(t)$, evolves according to  the Lindblad-form master equation in Eqs.~\eqref{eq:l_def_0}-\eqref{eq:lindblad_form_1}.
At each time $t$, the error in $\partial _t \tilde \rho$ is of the same magnitude as that of the Bloch-Redfield  equation.
Thus the ULE is valid over the same regimes as previously developed Markovian master equations, while offering important gains in usability and applicability.

\subsection{Equivalence of  Markov approximations}
\label{sec:markov_family}
Above, we derived a time-local master equation for $\tilde \rho(t)$ that is distinct from the Bloch-Redfield equation, but is valid on an equivalent level of approximation. 
As we explain here, the existence of distinct but equivalently valid time-local master equations reflects the existence of a  class of distinct 
but equivalently valid Markov approximations.
We refer to two approximations as being ``equivalently valid'' if they are both valid up to an error of the same order in the Markovianity parameter $\Gamma\tau$.

We demonstrate the existence of equivalently valid Markov approximations by means of  a simple example.
Consider the master equation for   $\tilde{\rho}(t)$ that  results from the Born   approximation [see text above Eq.~\eqref{eq:correlation_function_def}]:}
$\partial _t \tilde \rho(t) =  - \intop_{t_0}^t {\rm d}t'\,J(t-t')[X(t),[X(t'),\tilde \rho(t')]]+H.c.$.
As explained in Sec.~\ref{sec:born_markov}, the standard Markov approximation amounts to approximating $\tilde \rho(t')\approx \rho(t)$ in this expression. 
This approximation is justified when the bath correlation time $\tau$ is much shorter than the characteristic timescale  of system-bath interactions, $\Gamma^{-1}$. 
By the same arguments, however,  instead of  setting  $\tilde\rho(t')\approx \tilde\rho(t)$, we just as well could  have approximated $\tilde \rho(t')$ by any weighted average of $\tilde \rho(s)$ within a window  of times  $s$ near $s=t'$, as long as the width of the time-window is much smaller than $\Gamma^{-1}$.
These different choices of weight functions     result in distinct, but equivalently valid, time-local master equations. 
The infinite family  of suitable  weight-functions can thus be seen as generating a class  of distinct Markov approximations. 

As we show in  Appendix~\ref{app:ule_derivation}, there are also other classes of equivalent Markov approximations of more subtle origin than the  simple  example above.
These  other classes of equivalent approximations can be identified using similar approaches as above.
The approximations in Eq.~\eqref{eq:f_superoperator}-\eqref{eq:l_def_0} constitute such an alternative Markov approximation. 
A rigorous definition and discussion of this  approximation is given in Appendix~\ref{app:ule_derivation}.

\subsection{General system-bath couplings}
\label{sec:multiple_noise_channels}

In  Sec.~\ref{sec:single_channel}, we derived  the universal Lindblad equation for the case where the system-bath coupling $H_{\rm int}$ in Eq.~\eqref{eq:h_sb_def} holds a single noise channel.
In this subsection we extend our results to   the most  general case of   system-bath couplings, namely the case where $H_{\rm int}$ contains an arbitrarily high  (but finite) number of noise channels $N$:
$
H_{\rm int} = \sqrt{\gamma } \sum_{\alpha =1}^N X_\alpha  B_\alpha.
$
Here,  for  each $\alpha $,  $X_\alpha $ and $B_\alpha $ are observables of the system $\S$ and bath $\B$, respectively.
These are normalized such that  $\norm{X_\alpha } =1$, while the bath operators $\{B_\alpha \}$ may have different scales of magnitude.

For general system-bath coupling, the {ULE} can be derived through straightforward generalization of the 
single-channel case  in Sec.~\ref{sec:single_channel}. 
Because of this, the  derivation of the {ULE} in Appendix~\ref{app:ule_derivation} that we quoted in Sec.~\ref{sec:single_channel} considers  the case of multiple noise channels. 
Here, we present the results from  Appendix~\ref{app:ule_derivation}. 

As for the single-channel case, the validity of the {ULE} is determined solely by properties of the bath correlation (or, equivalently, spectral) functions. 
In the case where  the system and bath are connected through $N$  quantum noise channels,  the bath correlation function introduced  in Eq.~\eqref{eq:correlation_function_def}  takes values as an $N \times N$ matrix $\boldsymbol{J}(t)$ with matrix elements
\be 
J_{\alpha \beta }(t-s)\equiv \Tr_\B [  \tilde B_\alpha (t) \tilde B_\beta (s)\rho_\B].
\label{eq:correlation_function_m}
\ee
Here $\tilde B_\alpha (t) \equiv e^{iH_\B t }B_\alpha e^{-iH_\B t}$ denotes the interaction picture version of the bath operator $B_\alpha $, and the indices $\alpha $ and $\beta $ label the  noise channels,  taking values $1\ldots N$. 
Using the definition above, one can verify that the bath spectral function  $\boldsymbol{J}(\omega) \equiv \frac{1}{2\pi} \int_{-\infty}^\infty\! {\rm d}t\, \boldsymbol J(t) e^{i\omega t}$ forms a positive-semidefinite matrix for all $\omega$. 
The fact that  $\boldsymbol{J}$ is positive-semidefinite  generalizes the single-channel result  that the scalar-valued bath spectral function $J(\omega)$ is  non-negative (see Sec.~\ref{sec:born_markov}).

To  establish the conditions under which the {ULE} holds,  we generalize the jump correlator $g(t)$ from Eq.~\eqref{eq:jump_correlator} to the multiple-channel case.
Using the fact  that $\boldsymbol J(\omega)$ is positive-semidefinite, we define the matrix-valued jump correlator $\boldsymbol g(t)$ as follows: 
\be 
\boldsymbol g(t) = \int_{-\infty}^\infty\!\!\!{\rm d\omega}\, \boldsymbol g(\omega)e^{-i\omega t}, \quad 
\boldsymbol g  ( \omega)=\sqrt{ \boldsymbol J(\omega) /2\pi}.
\label{eq:jump_correlator_m}
\ee
Here the square root in the second equation  denotes the matrix square root; i.e., 
$
J_{\alpha \beta }(\omega) = \frac{1}{2\pi} \sum_\lambda g_{\alpha \lambda}(\omega)g_{\lambda \beta } (\omega),
$
where $\{g_{\alpha \beta }(\omega)\}$ denote the matrix-elements of $\boldsymbol g(\omega)$.
Since $\boldsymbol J(\omega)$ is positive-semidefinite, the Fourier transform of  the jump correlator
 $\boldsymbol g(\omega)$ is itself a well-defined  positive-semidefinite matrix for all values of $\omega$.

From the multi-channel jump-correlator $\boldsymbol g(t)$, we
 define the  quantities $\Gamma$ and $\tau$ from the multi-channel jump correlator $\boldsymbol g(t)$   as follows: 
 \be 
\Gamma = 4 \gamma   \left[\int_{-\infty}^\infty dt \norm{\boldsymbol g(t)}_{2,1}\right]^2,\quad \tau = \frac{\int_{-\infty}^\infty dt\norm{\boldsymbol g(t) t}_{2,1}}{\int_{-\infty}^\infty dt \norm{\boldsymbol g(t)}_{2,1}}.
\label{eq:time_scales_m}
\ee
Here for   any matrix $\boldsymbol M$ with  elements $M_{\alpha \lambda }$, 
$\norm{\boldsymbol M}_{2,1}  \equiv \sum_\lambda   (\sum_{\alpha   }|M_{\alpha \lambda }|^{2})^{1/2}$. 
The matrix  norm $\norm{\cdot}_{2,1}$ is also known as the $L_{2,1}$ matrix norm~\cite{Ding_2006}, and is identical  to the trace norm for diagonal matrices. 
As we require, for the special case of a single noise channel ($N=1$), the definitions  of $\Gamma$ and $\tau$ above are identical to the definitions in  Eq.~\eqref{eq:time_scales}.

In Appendix~\ref{app:ule_derivation}, we show  that  when $\Gamma \tau\ll 1$ [with $\Gamma$ and $\tau$ as defined  in Eq.~\eqref{eq:time_scales_m}], the system's dynamics  are effectively Markovian, and $\tilde \rho$ evolves according to the following Lindblad-form master equation:
\begin{widetext}
\begin{align}
\partial _t \tilde \rho(t) =      - i[\tilde\Lambda(t),\tilde \rho(t)] + \sum_{\lambda=1}^N&\left( \tilde  L_\lambda(t)\tilde \rho  (t)\tilde L^\dagger_\lambda(t)-\frac{1}{2}\{\tilde L^\dagger_\lambda(t)\tilde L_\lambda(t), \tilde \rho (t)\}\right)   + \xi'(t), 
\label{eq:general_ip_lindblad}
\end{align}
\end{widetext}
{where $\norm{\xi'(t)}\leq 2\Gamma^2 \tau$, and} $\tilde L_{\lambda}(t)$   is the jump operator associated with the emergent noise  channel $\lambda$ (which may involve operators from multiple baths). 
Explicitly, $\tilde L_\lambda(t)$ is given by 
\be 
\tilde L_\lambda(t)  = \sqrt{\gamma } \sum_\alpha \infint {\rm d}s \,    g_{\lambda\alpha }(t-s)\tilde X_\alpha(s),
\label{eq:jump_operator_m}
\ee
while the multiple-channel Lamb shift $\tilde \Lambda(t)$ is given by 
\begin{align}
\tilde \Lambda(t)  =   \frac{\gamma}{2i}  \infint\!\! {\rm d}s\! \infint\!\! {\rm d}s' & \sum_{\alpha \beta } \tilde X_\alpha (s)\tilde X_\beta (s')\phi_{\alpha \beta }(s-t,s'-t).
\label{eq:lamb_shift_multi}
\end{align}
Here
$
\{\phi_{\alpha \beta }(t,s)\}$ denote the matrix elements of the $N\times N$ matrix $\bs \phi(t,s)\equiv \bs g(t)\bs g(-s) \sgn(t-s).
$
The  above expressions for $\tilde L_\lambda(t)$ and $\tilde \Lambda(t)$ simplify further  in the case of independent noise channels (i.e., when the bath spectral function $\boldsymbol  J(\omega)$ is diagonal), since $\boldsymbol g(t)$ in this case is diagonal.
%
%

Analogous to the single-channel case,  Eqs.~\eqref{eq:general_ip_lindblad}-\eqref{eq:lamb_shift_multi}  hold up to a correction of order $2\Gamma^2 \tau$ for a density matrix $\rho'$ whose norm-distance to $\tilde \rho$ remains bounded by $\Gamma \tau$ at all times.
Hence, we expect Eq.~\eqref{eq:general_ip_lindblad} to accurately describe the evolution of $\tilde \rho$ in the weak-coupling limit $\Gamma \tau \ll 1$ (see discussion in Sec.~\ref{sec:single_channel}).

\subsection{Schr\"odinger picture}
\label{sec:schrodinger_picture}

We conclude this section by expressing the universal Lindblad equation [Eqs.~\eqref{eq:general_ip_lindblad}-\eqref{eq:lamb_shift_multi}] in  the  Schrodinger picture. 
Using the transformation between the interaction and Schr\"odinger pictures below Eq.~\eqref{eq:interaction_picture_hamiltonian}, we obtain the  master equation for the reduced density matrix of the system in the Schr\"odinger picture, $\rho$:
\be
\partial _t \rho (t)= - i[H_\S(t) + \Lambda(t),\rho(t)] +   \sum_{\lambda=1}^N\mathcal D_\lambda [\rho(t),t], 
\label{eq:sp_lindblad}
\ee
where we suppressed the correction of order $\Gamma^2 \tau$ from Eq.~\eqref{eq:general_ip_lindblad}.
In the above, $\mathcal D_\lambda$ denotes the dissipator associated with emergent noise channel $\lambda$, and is given by 
\be
\mathcal D_\lambda[\rho,t]  \equiv  L_\lambda(t) \rho   L^\dagger_\lambda(t) -  \frac{1}{2}\{ L^\dagger_\lambda (t)L_\lambda(t),  \rho \}.
\label{eq:sp_dissipator_def}
\ee
Here the Schr\"odinger picture jump operators and Lamb shift are  given by  $L_\lambda(t)=U(t) \tilde L_\lambda(t) U^\dagger(t)$ and $\Lambda(t) = U(t) \tilde \Lambda(t) U^\dagger(t)$, where $U(t)$ denotes the unitary evolution operator of the system $\S$, while 
 $\tilde \Lambda(t)$ and $\tilde L_\lambda(t)$ were given in Sec.~\ref{sec:multiple_noise_channels} above. 
By direct computation, we find, in particular  
\be 
 L_\lambda(t)  = \sqrt{\gamma }\sum_\alpha \infint {\rm d}s \,    g_{\lambda\alpha }(t-s) U(t,s) X_\alpha U^\dagger(t, s), 
 \label{eq:sp_l}
\ee
where $U(t,s)\equiv \mathcal T e^{-i\int_{s}^t {\rm d}t'\, H_\S(t')}$ denotes the time-evolution operator of the system from time $s$ to time $t$, defined such that $U(s,t)= U^\dagger(t,s)$.
An analogous expression can be obtained for $\Lambda(t)$. 
As we show in Sec.~\ref{sec:time_independent}, the jump operators $\{L_\lambda(t)\}$ and Lamb shift  $\Lambda(t)$ above are  time-independent when the system Hamiltonian $H_\S(t)$ is time-independent.

\section{Practical implementation}
\label{sec:application}

In this section, we  discuss how to implement the universal Lindblad equation [Eq.~\eqref{eq:sp_l}]  in practice. 
We separate our discussion into   three often-arising cases: 
in Sec.~\ref{sec:time_independent},   we consider systems with time-independent  Hamiltonians,
in Sec.~\ref{sec:time_dependent}, we  consider systems with time-dependent Hamiltonians,
and in  Sec.~\ref{sec:many_body}, we demonstrate  how  the {ULE} can be implemented in cases where exact diagonalization of the system Hamiltonian is not feasible (such as, e.g., quantum  many-body  systems). 

\subsection{Systems with time-independent Hamiltonians}
\label{sec:time_independent}

We first consider the case of time-independent Hamiltonians. 
We moreover assume that the system's Hamiltonian can be efficiently diagonalized, either analytically or numerically. 
In this case   the jump operators and Lamb shift   can be easily computed from  the Hamiltonian's eigenstates and energies.
In Sec.~\ref{sec:many_body} we  discuss an efficient approximate implementation for cases where exact diagonalization is not practically possible.

When the system $\S$ has a time-independent Hamiltonian $H_\S$, the time-evolution operator  of the system is given by $U(t,s)= \sum_{n}|n\rangle\langle n|e^{-iE_n (t-s)}$ where  $\{|n\rangle\}$ and $\{E_n\}$ denote the  eigenstates and energy spectrum of $H_\S$. 
Inserting this result into Eq.~\eqref{eq:sp_l}, we obtain the following simple expression for the system's jump operators:
\be  
L_\lambda  = 2 \pi \sqrt{\gamma }\sum_{m,n,\alpha } g_{\lambda \alpha } (E_n-E_m)  X^{(\alpha)}_{mn}|m\rangle  \langle n|  ,
\label{eq:static_jump_operators}
\ee
where $X^{(\alpha )}_{mn} \equiv\langle m|X_\alpha |n\rangle$.
The result above holds for an arbitrary number of quantum noise channels, and  was quoted in Sec.~\ref{sec:ady_section} for the single-channel case. 
Note that  the jump operators are time-independent, 
 as required by  the time-translation symmetry in this case of the problem. 

The Lamb shift $\Lambda$ can be expressed in similar terms as above, and also inherits the time-independence of $H_\S$:
in  Appendix~\ref{app:lamb_shift_static}, we find 
 \be 
\Lambda=    \sum_{l,m,n}f_{\alpha \beta }(E_m-E_l,E_n-E_l) X^{(\alpha )}_{ml}X^{(\beta )}_{ln}  |m\rangle\langle n|.
\label{eq:lamb_shift_static}
\ee
Here the functions
$\{f_{\alpha \beta }(E_{1},E_{2})\}$ denote the  elements of the 
 matrix $\boldsymbol f(E_1,E_2)  \equiv  2\pi \gamma  \, \mathcal P\!  \int_{-\infty}^\infty \!\! {\rm d} \omega\, \omega^{-1}\boldsymbol g(\omega-E_{1})\boldsymbol g(\omega+E_{2})$,  with $\mathcal P\int $ denoting the Cauchy principal value 
  integral.

We confirm that 
the master equation above reproduces previous results for open quantum systems in the limit  of large level spacing  and weak $\gamma $ (i.e., in the regime  where the quantum optical master equation is valid)~\cite{Breuer}: in this limit, 
standard arguments~\cite{Breuer} show that the rotating wave approximation can be applied to Eq.~\eqref{eq:sp_dissipator_def}.
Using  the jump operator in Eq.~\eqref{eq:static_jump_operators}, this approximation reduces Eq.~\eqref{eq:sp_lindblad}  to the quantum optical master equation.
Similarly,  to first order in $\gamma $, the Lamb shift $\Lambda$ 
  renormalizes    each energy level of the system $E_n$     by the amount $\delta E_n = \langle n| \Lambda|n\rangle$. 
Using $\boldsymbol g(\omega )^2 = 2\pi\boldsymbol  J(\omega)$ in Eq.~\eqref{eq:lamb_shift_static}, one can verify that $\delta E_n$ is identical to previous expressions for the Lamb-shift renormalization of energy levels  in the small-$\gamma$ limit~\cite{Breuer}.

\subsection{Systems with time-dependent Hamiltonians}
\label{sec:time_dependent}

We now  consider the situation where the system's Hamiltonian $H_\S(t)$ varies with time. 
While in this case one can always obtain the jump operators from  Eq.~\eqref{eq:sp_l},   here we obtain more convenient expressions 
in two important situations of wide applicability.

The first case  we consider arises when the time-dependence of $H_\S$ is slow on the bath correlation timescale, $\tau$.
In this case,  $H_{\S}(t)$ may be assumed constant in Eq.~\eqref{eq:sp_l}, and  the jump operators $\{L_\lambda(t)\}$ and Lamb shift $\Lambda(t)$  can be calculated from the  energies and eigenstates of the instantaneous Hamiltonian $H_\S(t)$.
Specifically, as we show in Appendix~\ref{app:time_independent_approx}, calculating the  jump operator from the instantaneous Hamiltonian as above yields a  correction of order  up to $ \sqrt{\Gamma} \tau^2 \norm{\partial _t H_\S}$ (recall  that $L_\lambda(t)$ has units of $[{\rm Energy}]^{1/2}$).
See  Appendix~\ref{app:time_independent_approx} for   further details.
The above result confirms that $L_\lambda(t)$ may be calculated from the eigenstates and energies of the instantaneous Hamiltonian when its time-derivative $\partial _t H_\S$ is sufficiently small compared to $1/\tau^2$. 
The approach in Appendix~\ref{app:time_independent_approx} can also be used to identify similar corrections  for the Lamb shift. 

The second situation where the universal Lindblad equation simplifies  is the  special   case of  periodically driven systems, where  $H_\S(t) = H_\S (t+T)$ for some driving period $T$. 
In this case, $L_\lambda(t)$ and $\Lambda(t)$ can be exactly computed from the time-periodic Floquet states  $|\phi_n(t)\rangle = |\phi_n(t+T)\rangle$ and quasienergies of the system~\cite{fn:floquet_intro}, $\varepsilon _n$:  
for periodically driven systems, the evolution operator of the system is given by $U(t,s) = \sum_n |\phi_n(t)\rangle\langle \phi_n(s)| e^{-i\varepsilon _n (t-s)}$. 
Using this  in Eq.~\eqref{eq:sp_l},  {one can verify by straightforward computation that} 
\begin{align}
\nonumber L_\lambda(t) &= \sum_{m,n} \sum_{z=-\infty}^{\infty}  L _{mn;z}^{(\lambda)} |\phi_m(t)\rangle\langle\phi_n(t)|e^{-i\Omega z t} ,
\end{align}
where
\begin{align}
 L _{mn;z}^{(\lambda)}  &\equiv \sum_\alpha  \int_0^T\!\frac{{\rm d}t}{T} \langle \phi_m(t)|X_\alpha  |\phi_n(t)\rangle e^{i\Omega z t}g_{\alpha \lambda}(\varepsilon _{nm}^z).
\label{eq:floquet_jump_operator}
\end{align}
Here $\Omega \equiv {2\pi}/{T}$,  while  $\varepsilon _{nm}^z \equiv \varepsilon _n- \varepsilon_m +z\Omega$. 
Note that the jump operators inherit the time-periodicity of the Hamiltonian, as required by discrete time-translation  symmetry: $L_\lambda(t)=  L_\lambda(t+T)$. 
The Lamb shift $\Lambda(t)$ has a similar expression in terms of the Floquet states  and also satisfies   $\Lambda(t+ T) = \Lambda(t)$.  
Interestingly, when the time-dependence of $H_\S(t)$ is slow compared to the bath correlation time $\tau$ (which may be very short), the  results above show that the jump operators in Eqs.~\eqref{eq:floquet_jump_operator} are equivalent to the jump operators generated from the instantaneous eigenstate basis of the  Hamiltonian $H_\S(t)$ through Eq.~\eqref{eq:static_jump_operators}. 
The above form of the jump operators was used  by the one of the  authors to numerically simulate the dynamics of a driven-dissipative quantum cavity in Ref.~\cite{Nathan_2019}.

The results above  reproduce  the generalization 
of the quantum optical master equation to  periodically driven systems, for example derived in Refs.~\cite{Blumel_1991,Kohler_2004,Hone_2009}.
In these works,  a Lindblad-form master equation is obtained for the system using a   rotating wave approximation (RWA) which assumes the relaxation rate ($\Gamma$) much smaller than the smallest possible level spacing  in the system's quasienergy spectrum $\delta\varepsilon _{\rm min} = {\rm min}_{m\neq n}(\varepsilon _n - \varepsilon _m + z\Omega)$. 
The  approaches we present above do not rely on such a rotating wave approximation, and hence are valid for a wider class of systems. 
We note that both approaches presented here are equivalent to the above RWA master equations  in the limit $\Gamma \ll \delta \varepsilon $ where the latter are valid. 
In this limit, one can verify that the steady-state of the system   is diagonal in the Floquet state basis.

\subsection{Obtaining jump operators without diagonalization}
\label{sec:many_body}
We finally show  how the universal Lindblad equation can be implemented in cases where
  diagonalization of the system Hamiltonian  $H_\S$ is not feasible, such as for large quantum many-body systems.
In this case, the jump operators and Lamb shift of the {ULE} cannot be obtained from
 the eigenstate decompositions  presented above.
Instead, as we show here,  these operators can be easily    obtained  through    a systematic, convergent expansion of Eq.~\eqref{eq:sp_l} in powers of $\tau/\tau_X$, where $\tau$ is the bath correlation time, and  $\tau_X$ denotes the characteristic time-scale 
for the dynamics of the system observables $\{X_\alpha \}$ (see below for definition).

For simplicity we consider here  the time-independent single-channel case.
The approach we present  below generalizes straightforwardly to  time-dependent Hamiltonians and multiple noise channels. 
We moreover focus on computing the jump operator $L$ (we suppress $\lambda$ since we  consider the single-channel case); 
the Lamb shift can be obtained through a similar approach.

To compute  the jump operator $L$, we note that, for  a time-independent Hamiltonian  $H_\S$,    $U(t,s) = e^{-i(t-s)H_\S}$.
Next, we note that
$ 
e^{iH_\S t}X e^{-iH_\S t } = \sum_{n=0}^\infty {(it )^n}({\rm ad}_{H_\S})^n [  X]/{n!}
$
where $ {\rm ad}_{H_\S} $ denotes the commutation operation by $H_\S$, i.e., 
$ {\rm ad}_{H_\S} [\mathcal O] = [H_\S , \mathcal O]$.
Using these results in Eq.~\eqref{eq:sp_l}, we obtain
\be 
L = \sqrt{\gamma } \sum_{n=0}^\infty  c_n   \, ({\rm ad}_{H_\S})^n [ X],\quad c_n\equiv \frac{i^n}{n!} \int_{-\infty}^\infty \!\!\!{\rm d}t\, g(t) t^n.
\label{eq:l_many_body_series}
\ee
Crucially, the 
 coefficients  $\{c_n\}$ are  system-independent, and  can be  easily computed from the jump correlator. 

The convergence of the series in Eq.~\eqref{eq:l_many_body_series}  can be ensured  by introducing a temporal cutoff $\tau_{\rm max}$, such that  $g(t)$ is set to zero for $|t|\geq \tau_{\rm max}$. 
The error in $L$ resulting from this approximation is bounded by $2\int_{ \tau_{\rm max}}^\infty\!{\rm dt} |g(t)|$ [see Eq.~\eqref{eq:sp_l}],
and can thus be made arbitrarily small by choosing $\tau_{\rm max}$ sufficiently large~\cite{fn:gamma_finite}.
In particular, we expect the error to be negligible when $\tau_{\rm max} \gg \tau$. 
With the temporal cutoff imposed, $c_n \sim \tau_{\rm max}^n/n!$ for large $n$. 
As a result, the series in Eq.~\eqref{eq:l_many_body_series} converges at order $ \tau_{\rm max}/\tau_X$, where $\tau_X$ denotes the typical  timescale associated with the dynamics of $X$, such that $\norm{({\rm ad}_{H_\S})^n [ X]} \sim 1/\tau_X^{n}$. 

The expansion of the jump operator simplifies further when the Hamiltonian $H_\S$ is composed of an easily diagonalizable term $H_0$   (such as a quadratic term)  and a weak non-integrable perturbation $V$ (such as an interaction term): $H_\S = H_0 + V$. 
By transforming to the   interaction picture with respect to $H_0$, we find $e^{-iH_\S t } = e^{-iH_0t} U'(t)$, where 
$U'(t)\equiv \mathcal T e^{-i\int_0^t \!\!{\rm d}t' \tilde V(t')}$, and $\tilde V(t) = e^{iH_0t} V e^{-iH_0 t}$. 
Using this result in Eq.~\eqref{eq:sp_l}, and 
 expanding 
$U'(t) X [U'(t)]^\dagger$  in powers of $\tilde V(t)$ as above Eq.~\eqref{eq:l_many_body_series}, we obtain a series expansion, where the $n$th order term is bounded by $(\tau_{\rm max}/\tilde \tau_X)^n/n! $, where $\tilde\tau_X$ denotes the time-scale for the dynamics of $X$ induced by the perturbation $\tilde V$,  such that $\norm{{\rm ad}_{\tilde V}^n X} \lesssim\tilde \tau_X^{-n}$. 
 As a result, only terms up to order $\tau_{\rm max}/\tilde \tau_X$ contribute in the expansion of the jump operator. 
Importantly, $\tilde\tau_{X}$ is inversely proportional to the strength of the perturbation $V$.
Thus, in the limit of weak  perturbations, where $\tau_{\rm max} \ll \tilde\tau_X$, the expansion above can be truncated at order zero. 
In this case, we may thus ignore the non-integrable perturbation in the calculation of the jump operator, and obtain $L$ from the spectrum and eigenstates of the integrable Hamiltonian $H_0$. 
In the case where $V$ is small, but not completely negligible, the jump operator may still be efficiently approximated by including the first few terms of the  expansion discussed above.

\section{Numerical demonstration: Heisenberg spin chain}
\label{sec:numerics}
\begin{figure*}
\includegraphics[width=1.95\columnwidth]{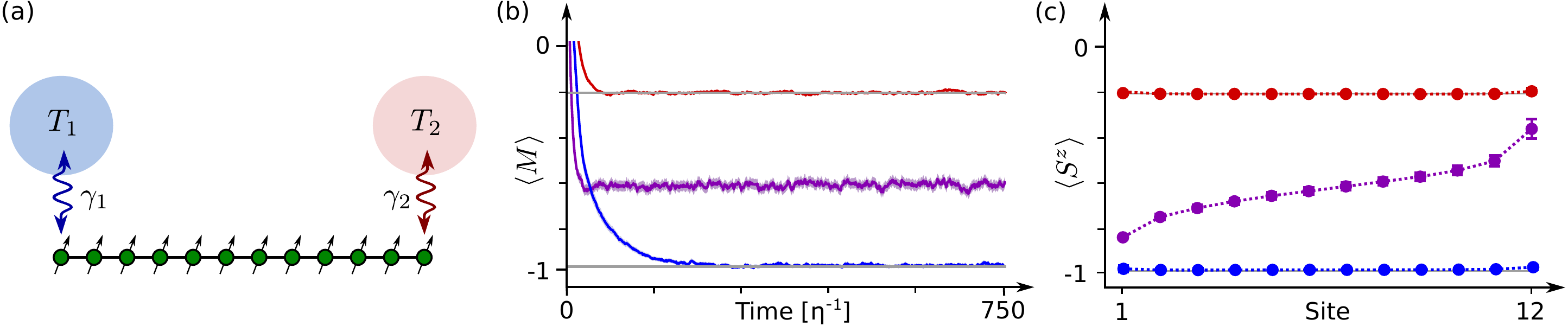}
\caption{ 
Simulation of the open Heisenberg spin chain model in Sec.~\ref{sec:numerics} (see main text for further details): 
(a)  Schematic depiction of the  system. 
 (b) Average $z$-magnetization in the chain, 
 as a function of time,  
  for the cases where the chain is connected  to bath $2$ (red), to bath $1$ (blue), and to both baths (purple). 
Grey lines:  expectation values of the $z$-magnetization in the  Gibbs states  at the temperatures   of baths $1$ (lower) and  $2$  (upper). 
Shaded areas surrounding curves (only visible for purple curve) indicates the uncertainty due to the finite  number of sampling states. 
(c) Average $z$-magnetization in the chain $\langle S_n^z\rangle$ for the final duration $150\eta^{-1}$ of the simulation, as a function of site index $n$ (using same coloring scheme as in  panel (b)).
Error bars indicate the uncertainty due to the finite number of sampling states in the simulation.
}
\label{fig:simulation}
\end{figure*}

In this section, we demonstrate how the  universal Lindblad equation can be used  in a numerical simulation. 
We consider a ferromagnetic spin-$1/2$  Heisenberg  chain
  coupled to two Ohmic  baths that are out of equilibrium with each other, as  schematically depicted in Fig.~\ref{fig:simulation}a.
By numerically solving the {ULE}, we 
obtain the  nontrivial  steady states and transport properties of the spin chain, along with its  transient relaxation dynamics. 


The system we consider  cannot be easily simulated by current master equation techniques, and hence 
our demonstration  highlights the utility of the {ULE}. 
For instance, the quantum optical master equation can only be employed when the system's relaxation rate is small compared to the  level spacing of the system  Hamiltonian. 
For  the  spin chain we consider, this level spacing is exponentially suppressed in the number of spins $N$, and hence, even for moderately-sized chains, the quantum optical master equation  only works for extremely weak system-bath couplings. 
In contrast,   the validity of the {ULE} is independent of the  level spacing in the system. 
Thus the ULE is valid for system-bath couplings many orders of magnitude larger than  allowed by the quantum optical master equation.

Another common master equation approach, the Bloch Redfield equation, is also ill-suited for the spin chain  we consider:
 the {BR} equation is often not stable, and may yield unphysical results, as discussed in the beginning of Sec.~\ref{sec:derivation}. 
 In contrast, the {ULE} is in the Lindblad form, and thus inherently robust.
Even without  instabilities, integration of the {BR} equation is numerically expensive,  since it requires evolving  the $D\times D$   density matrix  of the system $\rho$,  where  $D=2^N$ is the Hilbert space dimension of the system.
{On the other hand}, 
Lindblad-form master equations can be integrated  with the stochastic-Schr\"odinger equation, which only requires evolving a $D$-component state vector.
This  significantly reduces the  computational cost, with  the relative gain scaling  
exponentially   with the size of the system. 

The spin chain Hamiltonian is given by
\be
H_\S=    -B_z \sum_{n=1}^{N}S_{n}^z  - \eta \sum_{n=1}^{N-1}\vec S_n \cdot \vec S_{n+1},
\ee
where $B_z$ denotes the strength of a  uniform Zeeman field, $\eta$ is the nearest-neighbor coupling strength, and $\vec S_n = (S^x_n,S^y_n,S^z_n)$, where $S^\mu _n$  denotes the   spin-$\mu $ operator on site $n$ in the chain.

For the simulations below, we take $N=12$ sites.
The system is connected to two baths, $\B_1$ and $\B_2$, via spins $\vec S_1$ and $\vec S_N$ at the opposite ends of the chain, as schematically depicted in  Fig.~\ref{fig:simulation}a.
For demonstration, we couple the baths to the spins through their $x$-components, $S_1^x$ and $S_N^x$, with coupling strengths $\gamma_1$ and $\gamma_2$.
The  baths $\B_1$ and $\B_2$  are  modeled as Ohmic  baths in thermal equilibrium,  with spectral functions 
given in Eq.~\eqref{eq:bosonic_sf}. 
For the simulations below we take the baths to have the same values of the cutoff $\Lambda$ and  $\omega_0$, but distinct temperatures, $T_1$ and $T_2$.
The system-bath coupling of the system is hence given by
\be 
H_{\rm int} =  \sqrt{\gamma _1 } S_1^x  B'_1  +  \sqrt{\gamma _2} S_N^x B'_2,
\label{eq:sb_coupling_spin_chain}
\ee
where, for $\alpha =1,2$, $B'_\alpha$ is a bosonic field operator in the Ohmic bath $\B_\alpha $  with spectral function $J_\alpha (\omega)$ given by Eq.~\eqref{eq:bosonic_sf} with  $T=T_\alpha $.

To obtain the master equation
for the system, we cast the above   system-bath coupling  into the form given in  Eq.~\eqref{eq:h_sb_def}.
Within this framework, the bath {consists of two noise channels} with $X_1=S_1^x$ and $X_2 = S_N^x$.
The  corresponding bath operators are given by $B_\alpha  =\sqrt{\tilde \gamma _\alpha } B'_\alpha $ for $\alpha = 1, 2$, where $\tilde \gamma _\alpha  \equiv \gamma _\alpha /\gamma $   denotes the relative system-bath coupling strength,
and  $\gamma $ denotes the redundant  energy scale  we introduced in Eq.~\eqref{eq:h_sb_def} to parameterize the overall system-bath coupling [see discussion below Eq.~\eqref{eq:h_sb_def}]. 
Straightforward calculations~\cite{Breuer}  show that the  elements of the  $2\times 2$ spectral function matrix $\boldsymbol J(\omega)$  are given by  $J_{\alpha\beta } (\omega) =\delta_{\alpha \beta }  \tilde \gamma _\alpha  J_{\alpha }(\omega)$, where $\delta_{\alpha \beta }$ denotes the Kronecker symbol and $J_\alpha (\omega)$ denotes the spectral function of bath operator $B_\alpha '$ (see above).
Note that the coefficients $\tilde \gamma _\alpha $ appear in the spectral function due the parametrization of $H_{\rm int}$ in Eq.~\eqref{eq:h_sb_def}, in which a single common coupling scale $\gamma$ is factored out of the coupling Hamiltonian.

\subsection{Relaxation to thermal steady state}

 We first seek to verify that, when connected only to bath $2$, the system relaxes to a steady-state in thermal equilibrium with the bath, as we expect from basic thermodynamics. 
 In our simulation, we therefore set $\gamma _1 = 0,\gamma _2 = 0.02 \eta$. 
 The remaining parameters are set to $B_z = 8 \eta$,  $\Lambda =100 \eta$, $\omega_0 =2\eta$ (for both baths), while $T_2 = 20\eta$, and $T_1=2\eta$.
All parameters except for $\gamma _1 $ and $\gamma _2$  are given by the  same values throughout this section. 

As a first step, we compute the characteristic bath timescales $\Gamma^{-1}$ and $\tau$, 
which define the regime of applicability of the universal Lindblad equation.
Using Eq.~\eqref{eq:time_scales_m},  we find   $\Gamma = 0.41 \eta$ and $\tau = 0.00013 /\eta$, resulting in $\Gamma \tau =0.00053$.
Thus, following the discussion in Sec.~\ref{sec:derivation}, we expect the {ULE} to be valid.
In particular, the {ULE}  correctly describes the rate of change of the system's density matrix, $\partial _t\rho$,  up to a correction bounded by $2\Gamma^2\tau = 0.0043 \eta$.
This error bound is several orders of magnitude smaller than the other energy scales of the model, and we thus expect the {ULE} to faithfully capture the system’s evolution and steady states~\cite{fn:error_accumulation}.

To solve the {ULE}, we computed the system's jump operators $\{L_\lambda\}$ by exact diagonalization of  $H_\S$, using  Eq.~\eqref{eq:static_jump_operators}. 
Note that Eq.~\eqref{eq:l_many_body_series} can be used if diagonalization is not feasible.
We excluded the  Lamb shift from the simulation,  since this term only  weakly perturbs $H_\S$; thus  we do not expect it to affect the system's dynamics significantly~\cite{Breuer}. 
In contrast, the jump operators,  no matter how weak,  break the unitarity of time-evolution, and hence  cannot be neglected in the master equation.
We initialized  the system in the state with all spins  aligned against the uniform field $B_z$, and integrated the {ULE} numerically using the stochastic Schr\"odinger equation, with  an ensemble of $100$ states~\cite{Dalibard_1992,Dum_1992,Carmichael_1993}.

In Fig.~\ref{fig:simulation}b, we plot the expectation value of the  average $z$-magnetization  in the chain, $M= \frac{1}{N}\sum_{n=1}^N S^z_n$, as a function of time (red line). 
The  uncertainty of the expectation value $\langle M\rangle \equiv \Tr[\rho(t)M]$ arising  from the finite number of ensemble states 
is smaller than the thickness of the line. 
As Fig.~\ref{fig:simulation}b shows,   $ \langle M\rangle$ reaches a stationary value after a transient relaxation period of approximate duration $ 50 \eta^{-1}$. 
The steady-state value of $\langle M\rangle$ is identical to the  expectation  value of $M$ in a  Gibbs state at temperature $T_2$ (upper grey line), up to the accuracy of the simulation. 
A similar result arises in the case where the chain is  connected  only to bath $\B_1$: $\gamma _1 = 0.1 \eta$ and $\gamma _2 =0$ (blue curve in Fig.~\ref{fig:simulation}b).
Thus, we confirm that the universal Lindblad equation reproduces the expected equilibrium steady-states, further supporting its validity. 




\subsection{Nonequilibrium steady state with two baths}
We now consider the case where the spin chain is simultaneously connected to both baths, $\B_1$ and $\B_2$, with 
$\gamma _1 = 0.1\eta$ and $\gamma _2 =0.02\eta$.
In this case, due to the temperature difference between the baths,   we expect the system to reach a non-equilibrium steady state characterized by nonzero transport of energy and magnetization between the baths. 
With the parameters above, the characteristic  timescales as defined in Eqs.~\eqref{eq:time_scales} evaluate to $\Gamma \approx 3.6 \eta$ and $\tau \approx 0.0032 \eta^{-1}$.
Thus,  $\Gamma \tau \approx 0.011$ and $2\Gamma^2 \tau \approx 0.079 \eta$, indicating that the  universal Lindblad equation   should accurately describe the system's dynamics. 

In Fig.~\ref{fig:simulation}b, we plot the  the  magnetization   in the chain, $\langle M\rangle$, as a function of time (purple), obtained with the universal Lindblad equation. 
Similar to the two equilibrium cases, the magnetization  settles to a steady-state value after a transient relaxation period of duration $\sim 50 \eta^{-1}$. 
However, the relaxed system is not in  a  Gibbs state, but rather   a more complicated non-equilibrium steady state:
to demonstrate this, in Fig.~\ref{fig:simulation}c we show  the site-resolved magnetization  $\langle S_n^z(t)\rangle$, averaged over a time-window of length $150\eta^{-1}$ at the end of the  simulation.
As Fig.~\ref{fig:simulation}c clearly shows,  the local magnetization of the system is not uniform, but   gradually increases from the left to the right end of the chain, indicative of  a  non-equilibrium steady state.
In contrast, for the two  cases where  only a single bath is connected to the chain (red and blue), the local magnetization is uniform throughout the chain, consistent with a thermal Gibbs state at  temperatures $T_2$ and $T_1$ of the connected baths (horizontal grey lines in Fig.~\ref{fig:simulation}c). 
The skewed magnetization profile in the non-equilibrium case above reflects 
 a nonzero transport of heat and magnetization (magnons) between the two baths through the chain. 
By direct computation (see Appendix~\ref{app:transport} for details), we compute the average rate of heat transfer $\bar I_E$ and magnetization transfer $\bar I_M$ from bath $2$ to bath $1$ over a time-window of duration $75\eta^{-1}$ at the end of the simulation, finding  $\bar I_E = 2.1 \eta^2\pm 0.2\eta^2 $ and  $\bar I_M =0.33\eta  \pm 0.02\eta$.

\section{Discussion}
\label{sec:discussion}
In this paper we derived a Lindblad-form master equation for open quantum many-body systems:
the universal Lindblad equation (ULE). 
We identified rigorous upper bounds for the correction to the  ULE, expressed in terms of the intrinsic timescales of the bath and the system-bath coupling.
Crucially, the correction bounds we obtained for the ULE  are independent of the details of the system, and are of the same magnitude as the error bounds we obtained for the Bloch-Redfield (BR) equation, which is not in the Lindblad form. 
In this sense, the ULE is valid on an equivalent level of approximation as the BR equation. 

The universal Lindblad equation opens up new possibilities for  systematically studying a wide class of open quantum systems. 
These classes of systems include quantum many-body systems, and  general driven quantum systems with dense energy spectra, for which the stringent conditions of the quantum optical master equation are not met. 
In addition, the ULE can be implemented with lower computational cost and greater stability than the BR equation, since by construction it preserves the positivity and trace of the reduced density matrix of the system.

We have demonstrated the utility of the ULE in numerical simulations of an open Heisenberg spin chain, where we used it to extract the transport characteristic of the system's steady state in a nonequilibrium setting. 
We expect the ULE can be used to easily infer other non-equilibrium characteristics of the chain, such as, e.g., the correlations  of magnetization or heat current fluctuations, without adding any additional cost in the simulation. 
In addition to the spin chain model we considered here for demonstration, the universal Lindblad was {recently} used by one of the authors to simulate the dynamics of  a periodically-driven cavity-spin system in Ref.~\cite{Nathan_2019}, and by our collaborators to study readout of topological qubits in Ref.~\cite{Munk_2020}.
The universal Lindblad equation was also implemented in numerical simulations in Ref.~\cite{Kirsanskas_2018}, in order to support the hypothesized  master equation there (see Sec.~\ref{sec:ady_section}).

The principle underlying our derivation of the ULE is that there does not exist a unique Markov approximation in the Markovian regime $\Gamma\tau \ll 1$. 
Rather there exists  an infinite family of Markov approximations yielding distinct time-local master equations for the system that each are valid  on an equivalent level of approximation.
From this family of equivalent master equations, we identified a master equation in the Lindblad form, the ULE.

An interesting avenue of future studies is the  mathematical exploration of this equivalence class of Markov approximations; in particular, it will be interesting to investigate whether the above freedom of choice can be exploited further, to obtain master equations that are even more efficient or accurate, or perhaps explicitly respect desired symmetries or conservation laws. 
Another relevant question  along this direction of research is whether higher-order bath correlations and non-Markovian corrections can also be incorporated in the framework we develop here, and yield efficient and accurate master equations for the system.

As stimulus for another direction of future work, we speculate that the correction bounds we obtained  can be improved further. 
In particular, while we do not show it here, for Ohmic baths,
 the energy scale $\Gamma$ scales linearly with the {high-energy} cutoff of the bath (see Sec.~\ref{sec:bath_time_scales}). 
 However, this divergence arises from ultra-short (i.e., effectively time-local) correlations and reflects a divergent renormalization of the Hamiltonian through the Lamb shift.
Hence, adding a correction to the bare system Hamiltonian to compensate the divergent terms, we speculate that much better bounds can be obtained for the correction the ULE. 
Often, such a correcting counterterm is physically well-motivated. 
We believe further analysis of the problem using this principle can lead to significant improvement of the error  bounds  for the ULE.

As  important secondary results, in this work, we  obtained rigorous error bounds for the Bloch-Redfield equation, and established a ``quantum speed limit'' for the rate of bath-induced evolution of open quantum systems. 
These results may also be relevant for future work.
The results were established using a  perturbative approach in Appendix~\ref{app:br_correction}, in which the time-derivative of the reduced density matrix of the system is systematically expanded in orders of the dimensionless number $\Gamma \tau$. 
We speculate that this approach may be used in the future  to obtain master equations that are valid at higher orders in $\Gamma \tau$.

In summary, we have rigorously derived a Lindblad-form master equation for open quantum systems that offers several advantages over previously existing methods.
We expect that the efficiency, wide applicability, and simplicity of our method opens up new possibilities for future studies of   open quantum systems.

{\it Acknowledgements ---} 
We thank Ivar Martin, Gil Refael, Karsten Flensberg, Martin Leijnse,  Morten I. K. Munk,  Gediminas Kirsanskas, Evgeny Mozgunov, Tatsuhiko Ikeda, and Archak Purkayastha for helpful comments and useful discussions. FN and MR gratefully acknowledge the support of Villum Foundation, the European Research Council (ERC) under the European Union Horizon 2020 Research and Innovation Programme (Grant Agreement No.678862),  and CRC 183 of the Deutsche Forschungsgemeinschaft.
%
 \bibliographystyle{apsrev}
\bibliography{../bibliography_master_equation,../footnotes_master_equation}

\begin{thebibliography}{46}
\expandafter\ifx\csname natexlab\endcsname\relax\def\natexlab#1{#1}\fi
\expandafter\ifx\csname bibnamefont\endcsname\relax
  \def\bibnamefont#1{#1}\fi
\expandafter\ifx\csname bibfnamefont\endcsname\relax
  \def\bibfnamefont#1{#1}\fi
\expandafter\ifx\csname citenamefont\endcsname\relax
  \def\citenamefont#1{#1}\fi
\expandafter\ifx\csname url\endcsname\relax
  \def\url#1{\texttt{#1}}\fi
\expandafter\ifx\csname urlprefix\endcsname\relax\def\urlprefix{URL }\fi
\providecommand{\bibinfo}[2]{#2}
\providecommand{\eprint}[2][]{\url{#2}}

\bibitem[{\citenamefont{Scully and Zubairy}(1996)}]{Scully_book}
\bibinfo{author}{\bibfnamefont{M.}~\bibnamefont{Scully}} \bibnamefont{and}
  \bibinfo{author}{\bibfnamefont{M.~S.} \bibnamefont{Zubairy}},
  \emph{\bibinfo{title}{Quantum Optics}} (\bibinfo{publisher}{Akademie Verlag},
  \bibinfo{year}{1996}).

\bibitem[{\citenamefont{Van~Kampen}(2007)}]{VanKampen_book}
\bibinfo{author}{\bibfnamefont{N.~G.} \bibnamefont{Van~Kampen}},
  \emph{\bibinfo{title}{Stochastic Processes in Physics and Chemistry}}
  (\bibinfo{publisher}{North Holland}, \bibinfo{year}{2007}).

\bibitem[{\citenamefont{Nielsen and Chuang}(2010)}]{Nielsen_chuang}
\bibinfo{author}{\bibfnamefont{M.~A.} \bibnamefont{Nielsen}} \bibnamefont{and}
  \bibinfo{author}{\bibfnamefont{I.~L.} \bibnamefont{Chuang}},
  \emph{\bibinfo{title}{Quantum Computation and Quantum Information: 10th
  Anniversary Edition}} (\bibinfo{publisher}{Cambridge University Press},
  \bibinfo{year}{2010}).

\bibitem[{\citenamefont{Feshbach}(1958)}]{Feshbach_1958}
\bibinfo{author}{\bibfnamefont{H.}~\bibnamefont{Feshbach}},
  \bibinfo{journal}{Annals of Physics} \textbf{\bibinfo{volume}{5}},
  \bibinfo{pages}{357} (\bibinfo{year}{1958}).

\bibitem[{\citenamefont{Zanardi and Rasetti}(1997)}]{Zanardi_1997}
\bibinfo{author}{\bibfnamefont{P.}~\bibnamefont{Zanardi}} \bibnamefont{and}
  \bibinfo{author}{\bibfnamefont{M.}~\bibnamefont{Rasetti}},
  \bibinfo{journal}{Phys. Rev. Lett.} \textbf{\bibinfo{volume}{79}},
  \bibinfo{pages}{3306} (\bibinfo{year}{1997}).

\bibitem[{\citenamefont{Bourennane et~al.}(2004)\citenamefont{Bourennane, Eibl,
  Gaertner, Kurtsiefer, Cabello, and Weinfurter}}]{Bourennane_2004}
\bibinfo{author}{\bibfnamefont{M.}~\bibnamefont{Bourennane}},
  \bibinfo{author}{\bibfnamefont{M.}~\bibnamefont{Eibl}},
  \bibinfo{author}{\bibfnamefont{S.}~\bibnamefont{Gaertner}},
  \bibinfo{author}{\bibfnamefont{C.}~\bibnamefont{Kurtsiefer}},
  \bibinfo{author}{\bibfnamefont{A.}~\bibnamefont{Cabello}}, \bibnamefont{and}
  \bibinfo{author}{\bibfnamefont{H.}~\bibnamefont{Weinfurter}},
  \bibinfo{journal}{Phys. Rev. Lett.} \textbf{\bibinfo{volume}{92}},
  \bibinfo{pages}{107901} (\bibinfo{year}{2004}).

\bibitem[{\citenamefont{Verstraete et~al.}(2009)\citenamefont{Verstraete, Wolf,
  and Ignacio~Cirac}}]{Verstraete_2009}
\bibinfo{author}{\bibfnamefont{F.}~\bibnamefont{Verstraete}},
  \bibinfo{author}{\bibfnamefont{M.~M.} \bibnamefont{Wolf}}, \bibnamefont{and}
  \bibinfo{author}{\bibfnamefont{J.}~\bibnamefont{Ignacio~Cirac}},
  \bibinfo{journal}{Nature Physics} \textbf{\bibinfo{volume}{5}},
  \bibinfo{pages}{633} (\bibinfo{year}{2009}).

\bibitem[{\citenamefont{Diehl et~al.}(2011)\citenamefont{Diehl, Rico, Baranov,
  and Zoller}}]{Diehl_2011}
\bibinfo{author}{\bibfnamefont{S.}~\bibnamefont{Diehl}},
  \bibinfo{author}{\bibfnamefont{E.}~\bibnamefont{Rico}},
  \bibinfo{author}{\bibfnamefont{M.~A.} \bibnamefont{Baranov}},
  \bibnamefont{and} \bibinfo{author}{\bibfnamefont{P.}~\bibnamefont{Zoller}},
  \bibinfo{journal}{Nature Physics} \textbf{\bibinfo{volume}{7}},
  \bibinfo{pages}{971} (\bibinfo{year}{2011}).

\bibitem[{\citenamefont{Breuer and Petruccione}(2002)}]{Breuer}
\bibinfo{author}{\bibfnamefont{H.~P.} \bibnamefont{Breuer}} \bibnamefont{and}
  \bibinfo{author}{\bibfnamefont{F.}~\bibnamefont{Petruccione}},
  \emph{\bibinfo{title}{The theory of open quantum systems}}
  (\bibinfo{publisher}{Oxford University Press}, \bibinfo{address}{Great
  Clarendon Street}, \bibinfo{year}{2002}).

\bibitem[{\citenamefont{Gardiner and Zoller}(2004)}]{GardinerZoller}
\bibinfo{author}{\bibfnamefont{C.}~\bibnamefont{Gardiner}} \bibnamefont{and}
  \bibinfo{author}{\bibfnamefont{P.}~\bibnamefont{Zoller}},
  \emph{\bibinfo{title}{Quantum Noise}} (\bibinfo{publisher}{Springer-Verlag
  Berlin Heidelberg}, \bibinfo{year}{2004}).

\bibitem[{\citenamefont{Nakajima}(1958)}]{Nakajima_1958}
\bibinfo{author}{\bibfnamefont{S.}~\bibnamefont{Nakajima}},
  \bibinfo{journal}{Progress of Theoretical Physics}
  \textbf{\bibinfo{volume}{20}}, \bibinfo{pages}{948} (\bibinfo{year}{1958}).

\bibitem[{\citenamefont{Zwanzig}(1960)}]{Zwanzig_1960}
\bibinfo{author}{\bibfnamefont{R.}~\bibnamefont{Zwanzig}},
  \bibinfo{journal}{The Journal of Chemical Physics}
  \textbf{\bibinfo{volume}{33}}, \bibinfo{pages}{1338} (\bibinfo{year}{1960}).

\bibitem[{\citenamefont{Wangsness and Bloch}(1953)}]{Wangsness_1953}
\bibinfo{author}{\bibfnamefont{R.~K.} \bibnamefont{Wangsness}}
  \bibnamefont{and} \bibinfo{author}{\bibfnamefont{F.}~\bibnamefont{Bloch}},
  \bibinfo{journal}{Phys. Rev.} \textbf{\bibinfo{volume}{89}},
  \bibinfo{pages}{728} (\bibinfo{year}{1953}).

\bibitem[{\citenamefont{Redfield}(1965)}]{Redfield_1965}
\bibinfo{author}{\bibfnamefont{A.~G.} \bibnamefont{Redfield}},
  \bibinfo{journal}{Advances in Magnetic and Optical Resonance}
  \textbf{\bibinfo{volume}{1}}, \bibinfo{pages}{1 } (\bibinfo{year}{1965}).

\bibitem[{\citenamefont{Davies}(1974)}]{Davies_1974}
\bibinfo{author}{\bibfnamefont{E.~B.} \bibnamefont{Davies}},
  \bibinfo{journal}{Comm. Math. Phys.} \textbf{\bibinfo{volume}{39}},
  \bibinfo{pages}{91} (\bibinfo{year}{1974}).

\bibitem[{\citenamefont{Majenz et~al.}(2013)\citenamefont{Majenz, Albash,
  Breuer, and Lidar}}]{Majenz_2013}
\bibinfo{author}{\bibfnamefont{C.}~\bibnamefont{Majenz}},
  \bibinfo{author}{\bibfnamefont{T.}~\bibnamefont{Albash}},
  \bibinfo{author}{\bibfnamefont{H.-P.} \bibnamefont{Breuer}},
  \bibnamefont{and} \bibinfo{author}{\bibfnamefont{D.~A.} \bibnamefont{Lidar}},
  \bibinfo{journal}{Phys. Rev. A} \textbf{\bibinfo{volume}{88}},
  \bibinfo{pages}{012103} (\bibinfo{year}{2013}).

\bibitem[{\citenamefont{Mozgunov and Lidar}(2020)}]{Mozgunov_2020}
\bibinfo{author}{\bibfnamefont{E.}~\bibnamefont{Mozgunov}} \bibnamefont{and}
  \bibinfo{author}{\bibfnamefont{D.}~\bibnamefont{Lidar}},
  \bibinfo{journal}{{Quantum}} \textbf{\bibinfo{volume}{4}},
  \bibinfo{pages}{227} (\bibinfo{year}{2020}).

\bibitem[{\citenamefont{Lindblad}(1976)}]{Lindblad_1976}
\bibinfo{author}{\bibfnamefont{G.}~\bibnamefont{Lindblad}},
  \bibinfo{journal}{Communications in Mathematical Physics}
  \textbf{\bibinfo{volume}{48}}, \bibinfo{pages}{119} (\bibinfo{year}{1976}).

\bibitem[{\citenamefont{Gorini et~al.}(1976)\citenamefont{Gorini, Kossakowski,
  and Sudarshan}}]{Gorini_1976}
\bibinfo{author}{\bibfnamefont{V.}~\bibnamefont{Gorini}},
  \bibinfo{author}{\bibfnamefont{A.}~\bibnamefont{Kossakowski}},
  \bibnamefont{and} \bibinfo{author}{\bibfnamefont{E.~C.~G.}
  \bibnamefont{Sudarshan}}, \bibinfo{journal}{Journal of Mathematical Physics}
  \textbf{\bibinfo{volume}{17}}, \bibinfo{pages}{821} (\bibinfo{year}{1976}).

\bibitem[{\citenamefont{Dalibard et~al.}(1992)\citenamefont{Dalibard, Castin,
  and M\o{}lmer}}]{Dalibard_1992}
\bibinfo{author}{\bibfnamefont{J.}~\bibnamefont{Dalibard}},
  \bibinfo{author}{\bibfnamefont{Y.}~\bibnamefont{Castin}}, \bibnamefont{and}
  \bibinfo{author}{\bibfnamefont{K.}~\bibnamefont{M\o{}lmer}},
  \bibinfo{journal}{Phys. Rev. Lett.} \textbf{\bibinfo{volume}{68}},
  \bibinfo{pages}{580} (\bibinfo{year}{1992}).

\bibitem[{\citenamefont{Dum et~al.}(1992)\citenamefont{Dum, Zoller, and
  Ritsch}}]{Dum_1992}
\bibinfo{author}{\bibfnamefont{R.}~\bibnamefont{Dum}},
  \bibinfo{author}{\bibfnamefont{P.}~\bibnamefont{Zoller}}, \bibnamefont{and}
  \bibinfo{author}{\bibfnamefont{H.}~\bibnamefont{Ritsch}},
  \bibinfo{journal}{Phys. Rev. A} \textbf{\bibinfo{volume}{45}},
  \bibinfo{pages}{4879} (\bibinfo{year}{1992}).

\bibitem[{\citenamefont{Carmichael}(1993)}]{Carmichael_1993}
\bibinfo{author}{\bibfnamefont{H.}~\bibnamefont{Carmichael}},
  \emph{\bibinfo{title}{An open systems approach to quantum optics}}
  (\bibinfo{publisher}{Springer}, \bibinfo{year}{1993}).

\bibitem[{\citenamefont{Kir\ifmmode~\check{s}\else \v{s}\fi{}anskas
  et~al.}(2018)\citenamefont{Kir\ifmmode~\check{s}\else \v{s}\fi{}anskas,
  Francki\'e, and Wacker}}]{Kirsanskas_2018}
\bibinfo{author}{\bibfnamefont{G.}~\bibnamefont{Kir\ifmmode~\check{s}\else
  \v{s}\fi{}anskas}},
  \bibinfo{author}{\bibfnamefont{M.}~\bibnamefont{Francki\'e}},
  \bibnamefont{and} \bibinfo{author}{\bibfnamefont{A.}~\bibnamefont{Wacker}},
  \bibinfo{journal}{Phys. Rev. B} \textbf{\bibinfo{volume}{97}},
  \bibinfo{pages}{035432} (\bibinfo{year}{2018}).

\bibitem[{\citenamefont{Kleinherbers et~al.}(2020)\citenamefont{Kleinherbers,
  Szpak, K\"onig, and Sch\"utzhold}}]{Kleinherbers_2020}
\bibinfo{author}{\bibfnamefont{E.}~\bibnamefont{Kleinherbers}},
  \bibinfo{author}{\bibfnamefont{N.}~\bibnamefont{Szpak}},
  \bibinfo{author}{\bibfnamefont{J.}~\bibnamefont{K\"onig}}, \bibnamefont{and}
  \bibinfo{author}{\bibfnamefont{R.}~\bibnamefont{Sch\"utzhold}},
  \bibinfo{journal}{Phys. Rev. B} \textbf{\bibinfo{volume}{101}},
  \bibinfo{pages}{125131} (\bibinfo{year}{2020}).

\bibitem[{\citenamefont{Nathan}(2018)}]{Phd_thesis}
\bibinfo{author}{\bibfnamefont{F.}~\bibnamefont{Nathan}}, Ph.D. thesis,
  \bibinfo{school}{University of Copenhagen} (\bibinfo{year}{2018}).

\bibitem[{fn:({\natexlab{a}})}]{fn:spectral_norm_def}
\bibinfo{note}{Here the spectral norm is defined as the maximum singular
  value-norm: $\norm{X}=\sup_{\psi,\phi}|\langle\psi|X|\phi\rangle|$, where the
  supremum is taken over all normalized states. We note that, to apply our
  framework to a system where $X$ may be unbounded, some additional physically
  justfied truncation of the Hilbert space is needed.}

\bibitem[{fn:({\natexlab{b}})}]{fn:sb_hamiltonian}
\bibinfo{note}{The square root is introduced for convenience, since the
  ``bare'' system-bath coupling $\sqrt{\gamma }$ only appears in even powers in
  the master equations we obtain. As a result of this parameterization, $B$ has
  dimensions of $[{\rm Energy}]^{1/2}$. These units of $B$ are a natural choice
  when the bath has a continuous energy spectrum~\cite{Breuer}, such as is the
  case for the Ohmic bath in Sec.~\ref{sec:bath_time_scales}.}

\bibitem[{fn:({\natexlab{c}})}]{fn:general_lindblad_form}
\bibinfo{note}{To be precise, a general Lindblad form allows the
  time-derivative of $\rho$ to be given by a sum of multiple terms on the form
  in Eq.~(\ref{eq:lindblad_1}), where the Lamb shift and each jump operator may
  be time-dependent.}

\bibitem[{fn:({\natexlab{d}})}]{fn:correction_bounds}
\bibinfo{note}{In Appendix~\ref{app:br_correction}, we obtain a stricter bound,
  namely $\norm{\xi(t)}\leq 2\Gamma_0 \tau_0$, where $\Gamma_0^{-1}$ and
  $\tau_0$ are distinct, but typically comparable, measures for the
  characteristic timescales of bath-induced evolution and bath correlations.
  These quantities are defined in Appendix~\ref{app:time_scale_relationship},
  where we also show that $2\Gamma_0^2 \tau _0\leq \Gamma^2 \tau$. The
  timescales $\Gamma_0^{-1}$ and $\tau_0$ were also identified in
  Ref.~\cite{Mozgunov_2020}, where analogous bounds for the trace norm of the
  correction $\xi(t)$ were derived in terms of these time-scales. While we
  could have used the timescales $\Gamma_0^{-1}$ and $\tau_0$ to express the
  bound for $\norm{\xi(t)}$ in the BR equation, the steps leading to the ULE
  induce errors whose bounds we can only express in terms of $\Gamma$ and
  $\tau$. To simplify the discussion, in the main text we therefore use the
  (looser) bound $\Gamma^2 \tau$ in Eq.~(\ref{eq:br_bound}).}

\bibitem[{\citenamefont{del Campo et~al.}(2013)\citenamefont{del Campo,
  Egusquiza, Plenio, and Huelga}}]{DelCampo_2013}
\bibinfo{author}{\bibfnamefont{A.}~\bibnamefont{del Campo}},
  \bibinfo{author}{\bibfnamefont{I.~L.} \bibnamefont{Egusquiza}},
  \bibinfo{author}{\bibfnamefont{M.~B.} \bibnamefont{Plenio}},
  \bibnamefont{and} \bibinfo{author}{\bibfnamefont{S.~F.}
  \bibnamefont{Huelga}}, \bibinfo{journal}{Phys. Rev. Lett.}
  \textbf{\bibinfo{volume}{110}}, \bibinfo{pages}{050403}
  (\bibinfo{year}{2013}).

\bibitem[{\citenamefont{Deffner and Campbell}(2017)}]{Deffner_2017}
\bibinfo{author}{\bibfnamefont{S.}~\bibnamefont{Deffner}} \bibnamefont{and}
  \bibinfo{author}{\bibfnamefont{S.}~\bibnamefont{Campbell}},
  \bibinfo{journal}{Journal of Physics A: Mathematical and Theoretical}
  \textbf{\bibinfo{volume}{50}}, \bibinfo{pages}{453001}
  (\bibinfo{year}{2017}).

\bibitem[{fn:({\natexlab{e}})}]{fn:g_decay}
\bibinfo{note}{Specifically, the relative weight of the jump correlator
  $|g(t)|$ beyond a particular time $t$, $\intop_t^\infty {\rm d}t' |g(t')|/C$,
  is bounded by $\tau/t$ [this is straightforward to verify from
  Eq.~(\ref{eq:time_scales})]. Note also that for many physically relevant
  cases, such as for the Ohmic bath discussed below, the jump correlator decays
  much faster than by this power law (often exponentially). In particular, if
  the bath spectral function is smooth in a way such that, for some $n$, $I_n
  \equiv \frac{1}{\sqrt{2\pi}}\intop_{-\infty}^\infty {\rm d}\omega |(\partial
  _\omega)^n\sqrt{J(\omega)}|$ is a finite number, one can show that $|g(t)|$
  is always bounded by $I_n/|t|^n$. This can be straightforwardly shown by
  using the definition of $g(t)$ from Eq.~(\ref{eq:jump_correlator}), along
  with the triangle inequality.}

\bibitem[{fn:({\natexlab{f}})}]{fn:error_accumulation}
\bibinfo{note}{In principle, the error induced by neglecting $\xi'(t)$ in the
  ULE may accumulate over time and result in inaccurate values of $\rho'(t)$
  for $t\gtrsim(\Gamma^2 \tau )^{-1}$. While the bound we obtained for
  $\norm{\xi'(t)}$ can be used to infer rigorous results for the error to
  $\rho'(t)$ (using e.g., the spectral gap of the Liouvillian), such a
  discussion is beyond the scope of this paper. We expect it is often a good
  strategy to simply compare of the correction bound $2\Gamma^2\tau$ to the
  other relevant energy scales of the physical model and from this comparison
  determine whether the correction $\xi'(t)$ can safely be neglected using
  physical arguments. We expect this approach will include a much wider range
  of models than those allowed by rigorous mathematical results.}

\bibitem[{\citenamefont{Ding et~al.}(2006)\citenamefont{Ding, Zhou, He, and
  Zha}}]{Ding_2006}
\bibinfo{author}{\bibfnamefont{C.}~\bibnamefont{Ding}},
  \bibinfo{author}{\bibfnamefont{D.}~\bibnamefont{Zhou}},
  \bibinfo{author}{\bibfnamefont{X.}~\bibnamefont{He}}, \bibnamefont{and}
  \bibinfo{author}{\bibfnamefont{H.}~\bibnamefont{Zha}}, in
  \emph{\bibinfo{booktitle}{Proceedings of the 23rd International Conference on
  Machine Learning}} (\bibinfo{publisher}{Association for Computing Machinery},
  \bibinfo{address}{New York, NY, USA}, \bibinfo{year}{2006}), pp.
  \bibinfo{pages}{281--288}.

\bibitem[{fn:({\natexlab{g}})}]{fn:floquet_intro}
\bibinfo{note}{Here the Floquet states and quasienergies define the unique set
  of stationary solutions to the Schrodinger equation of the form
  $|\psi(t)\left.\hspace{-1mm}\right> = e^{-i\varepsilon _n
  t}|\phi_n(t)\left.\hspace{-1mm}\right>$, see Ref.~\cite{Shirley_1965} for
  more details}.

\bibitem[{\citenamefont{Nathan et~al.}(2019)\citenamefont{Nathan, Martin, and
  Refael}}]{Nathan_2019}
\bibinfo{author}{\bibfnamefont{F.}~\bibnamefont{Nathan}},
  \bibinfo{author}{\bibfnamefont{I.}~\bibnamefont{Martin}}, \bibnamefont{and}
  \bibinfo{author}{\bibfnamefont{G.}~\bibnamefont{Refael}},
  \bibinfo{journal}{Physical Review B} \textbf{\bibinfo{volume}{99}},
  \bibinfo{pages}{094311} (\bibinfo{year}{2019}).

\bibitem[{\citenamefont{Bl\"umel et~al.}(1991)\citenamefont{Bl\"umel,
  Buchleitner, Graham, Sirko, Smilansky, and Walther}}]{Blumel_1991}
\bibinfo{author}{\bibfnamefont{R.}~\bibnamefont{Bl\"umel}},
  \bibinfo{author}{\bibfnamefont{A.}~\bibnamefont{Buchleitner}},
  \bibinfo{author}{\bibfnamefont{R.}~\bibnamefont{Graham}},
  \bibinfo{author}{\bibfnamefont{L.}~\bibnamefont{Sirko}},
  \bibinfo{author}{\bibfnamefont{U.}~\bibnamefont{Smilansky}},
  \bibnamefont{and} \bibinfo{author}{\bibfnamefont{H.}~\bibnamefont{Walther}},
  \bibinfo{journal}{Phys. Rev. A} \textbf{\bibinfo{volume}{44}},
  \bibinfo{pages}{4521} (\bibinfo{year}{1991}).

\bibitem[{\citenamefont{Kohler et~al.}(2005)\citenamefont{Kohler, Lehmann, and
  H\"anggi}}]{Kohler_2004}
\bibinfo{author}{\bibfnamefont{S.}~\bibnamefont{Kohler}},
  \bibinfo{author}{\bibfnamefont{J.}~\bibnamefont{Lehmann}}, \bibnamefont{and}
  \bibinfo{author}{\bibfnamefont{P.}~\bibnamefont{H\"anggi}},
  \bibinfo{journal}{Physics Reports} \textbf{\bibinfo{volume}{406}},
  \bibinfo{pages}{379 } (\bibinfo{year}{2005}).

\bibitem[{\citenamefont{Hone et~al.}(2009)\citenamefont{Hone, Ketzmerick, and
  Kohn}}]{Hone_2009}
\bibinfo{author}{\bibfnamefont{D.~W.} \bibnamefont{Hone}},
  \bibinfo{author}{\bibfnamefont{R.}~\bibnamefont{Ketzmerick}},
  \bibnamefont{and} \bibinfo{author}{\bibfnamefont{W.}~\bibnamefont{Kohn}},
  \bibinfo{journal}{Phys. Rev. E} \textbf{\bibinfo{volume}{79}},
  \bibinfo{pages}{051129} (\bibinfo{year}{2009}).

\bibitem[{fn:({\natexlab{h}})}]{fn:gamma_finite}
\bibinfo{note}{Here we assume that $\Gamma $ is finite; this condition is
  already required for the universal Lindblad equation to be valid.}

\bibitem[{\citenamefont{Munk et~al.}(2020)\citenamefont{Munk, Schulenborg,
  Egger, and Flensberg}}]{Munk_2020}
\bibinfo{author}{\bibfnamefont{M.~I.~K.} \bibnamefont{Munk}},
  \bibinfo{author}{\bibfnamefont{J.}~\bibnamefont{Schulenborg}},
  \bibinfo{author}{\bibfnamefont{R.}~\bibnamefont{Egger}}, \bibnamefont{and}
  \bibinfo{author}{\bibfnamefont{K.}~\bibnamefont{Flensberg}},
  \bibinfo{journal}{Phys. Rev. Research} \textbf{\bibinfo{volume}{2}},
  \bibinfo{pages}{033254} (\bibinfo{year}{2020}).

\bibitem[{\citenamefont{Shirley}(1965)}]{Shirley_1965}
\bibinfo{author}{\bibfnamefont{J.~H.} \bibnamefont{Shirley}},
  \bibinfo{journal}{Phys. Rev.} \textbf{\bibinfo{volume}{138}},
  \bibinfo{pages}{B979} (\bibinfo{year}{1965}).

\bibitem[{\citenamefont{Albert et~al.}(2016)\citenamefont{Albert, Bradlyn,
  Fraas, and Jiang}}]{Albert_2016}
\bibinfo{author}{\bibfnamefont{V.~V.} \bibnamefont{Albert}},
  \bibinfo{author}{\bibfnamefont{B.}~\bibnamefont{Bradlyn}},
  \bibinfo{author}{\bibfnamefont{M.}~\bibnamefont{Fraas}}, \bibnamefont{and}
  \bibinfo{author}{\bibfnamefont{L.}~\bibnamefont{Jiang}},
  \bibinfo{journal}{Phys. Rev. X} \textbf{\bibinfo{volume}{6}},
  \bibinfo{pages}{041031} (\bibinfo{year}{2016}).

\bibitem[{\citenamefont{Evans et~al.}(1985)\citenamefont{Evans, Grigolini, and
  Parravicini}}]{Evans_book}
\bibinfo{author}{\bibfnamefont{M.~W.} \bibnamefont{Evans}},
  \bibinfo{author}{\bibfnamefont{P.~P.} \bibnamefont{Grigolini}},
  \bibnamefont{and} \bibinfo{author}{\bibfnamefont{G.~P.}
  \bibnamefont{Parravicini}}, \emph{\bibinfo{title}{Memory Function Approaches
  to Stochastic Problems in Condensed Matter}}, vol.~\bibinfo{volume}{62} of
  \emph{\bibinfo{series}{Advances in Chemical Physics}}
  (\bibinfo{publisher}{John Wiley \& Sons}, \bibinfo{year}{1985}).

\bibitem[{fn:({\natexlab{i}})}]{fn:superoperator_ref}
\bibinfo{note}{See Appendix~\ref{sec:superoperators} for a general discussion
  of superoperators.}

\bibitem[{fn:({\natexlab{j}})}]{fn:h_static_approx}
\bibinfo{note}{This can be proven by going to the interaction picture with
  respect to $H_\S(t)$, and using $\norm{H_\S (s)-H_\S (t)} \leq |s-t|v$.}

\end{thebibliography}

\appendix

\section{Correction to the Bloch-Redfield equation}
\label{app:br_correction}
Here we derive the rigorous upper bounds for the correction to the Bloch-Redfield (BR) equation that were quoted in Sec.~\ref{sec:bloch_redfield_conditions} in the main text. 
We derive the bounds for the
  general case where multiple 
 noise channels  connect the system and the bath. 
As a part of our derivation, in Sec.~\ref{app:speed_limit} below we  establish the upper bound for the rate of bath-induced evolution in the system, $\norm{\partial _t \tilde \rho}$, that we quoted in Sec.~\ref{sec:bath_time_scales} of  the main text. 


In this Appendix we work exclusively in the interaction picture (see Sec.~\ref{sec:born_markov}). 
To avoid cumbersome notation, we therefore use different notation here than in the main text, and neglect the  $\tilde {\cdot}$ accent on all interaction picture operators.
Thus, throughout this Appendix, $\rho(t)$, $ \rho_{\S\B} (t)$, $ H(t)$, $X_\alpha (t)$, and $B_\alpha (t)$ denote the interaction picture operators  $\tilde \rho(t)$, $\tilde  \rho_{\S\B} (t)$, $ \tilde H(t)$, $\tilde X_\alpha (t)$, and $\tilde B_\alpha (t)$  from the main text, respectively.

\subsection{Superoperator formalism}
\label{sec:superoperators}
Our derivation of  error bounds for the BR equation exploits  the fact that linear
operators on a Hilbert space (such as  density matrices) can themselves be seen as   vectors, or ``kets''.
To make this vector  nature of operators  explicit, in the following we use double brackets $|\cdot\rrangle$ to indicate operators acting on the Hilbert spaces  $\mathbb  H_\S, \mathbb  H_\B,$ and $\mathbb  H_{\S\B}$ of the system $\S$, bath $\B$, or the combined system $\S\B$. 
In this way, any operator which is denoted by $\mathcal O$ in standard notation is denoted by the ket $|\mathcal O\rrangle$ in the derivation below. 
The vector space of operator kets that act on Hilbert space $\mathbb  H_i$ (with $i = \{\S, \B, \S\B\}$)  defines an operator Hilbert space $\mathbb H_{i}^{2}$, defined with the  inner product  $\llangle R|S\rrangle \equiv \Tr (R^\dagger S)$.
This notation is commonly used in the literature, see, e.g.,  Ref.~\cite{Albert_2016} for a recent example. 
Note that the operator  space of the combined system $\S\B$, $\mathbb H_{\S\B}^{2}$, inherits the tensor product structure of the standard Hilbert space of $\S\B$, $\mathbb  H_{\S\B}$: $\mathbb  H^2_{\S\B}\cong \mathbb  H_\S^2 \otimes \mathbb  H_{\B}^2$, where $\mathbb  H_\S^2$ and $\mathbb  H^2_\B$ denote the  operator spaces of the system $\S$ and bath $\B$, respectively.

In the superoperator notation above, the  von-Neumann equation for the density  matrix of the combined system  (in the interaction picture), $\partial _t \rho_{\S\B}(t) = - i[H(t),\rho_{\S\B}(t)]$ translates to a linear  Schr\"odinger-type equation: 
\be 
  \partial _t | \rho_{\S\B}(t)\rrangle =   -i\, \hat {\mathcal H}(t)| \rho_{\S\B}(t)\rrangle,
\label{eq:DensityMatrixSchrodingerEq}
\ee
where  $\hat{\mathcal H}(t)$ denotes the commutator with 
 $H(t)$:
$  \hat {{\mathcal H}}(t) | \mathcal O\rrangle = |\, [H(t), \mathcal O]\,\rrangle$. 
Note that  $ \hat{{\mathcal H}}(t)$ acts linearly on $|\rho_{\S\B}\rrangle$, and hence it can be represented as a matrix  acting on the operator space $\mathbb H_{\S\B}^2$.
Below we furthermore show that $\hat{\mathcal H}(t)$ is Hermitian, and hence can be seen as a ``Hamiltonian'' acting on 
 $\mathbb{H}^2_{\S\B}$.
We refer to $\hat{\mathcal H}(t)$, and other 
linear transformations on operator kets, as superoperators. 
To make notation unambiguous, in the following we  use the ``hat''  accent  ($\, \hat{}\, $) to indicate superoperators.

A useful class of superoperators which we employ extensively in the  following is  left and right  multiplication  by some given operator: 
for any  operator $|A\rrangle$ in $\mathbb H_{i}^2$ (for $i=\{\S,\B,\S\B\}$), we define the left- and right-multiplication superoperators $\hat{A}^{l}$ and $\hat{A}^{r}$  as 
\be 
\hat{A} ^l | \mathcal O\rrangle = |A  \mathcal O\rrangle, \quad \hat{A}^r| \mathcal O\rrangle = | \mathcal O A \rrangle.
\label{eq:o_r_l_def}
\ee 
From the above definition, one can verify that the superoperator $\hat {\mathcal H}(t)$ in Eq.~\eqref{eq:DensityMatrixSchrodingerEq} is given by $\hat{H}^l(t) - \hat  H^r(t)$,
where, for any time-dependent operator $A(t)$, and for $m=\{l,r\}$, we use $\hat A^m(t)$ as shorthand for $\widehat{A(t)}^m$ to avoid cumbersome notation.
The right- and left-multiplication superoperators have a few useful properties that we use below: 
firstly, we note that, by associativity, $\hat{A}^l \hat{B}^l = (\widehat{AB})^l$, while $\hat{A}^r \hat{B}^r = (\widehat{BA})^r$.
Moreover, 
  the Hermitian conjugate of the superoperator $\hat{A}^{m}$ [i.e., $(\hat{A}^{m})^\dagger$] is given by $ \hat{(A^\dagger)}{}^{m}$. 
  This follows from Eq.~\eqref{eq:o_r_l_def}, along with the definition of the inner product $\llangle \cdot|\cdot\rrangle$: $\llangle \mathcal{O}_1| {A}^l|\mathcal{O}_2\rrangle = \llangle \mathcal{O}_1| {A}\mathcal{O}_2\rrangle = \llangle {A}^\dagger\mathcal{O}_1| \mathcal{O}_2\rrangle$.  
For this reason, in the following, we let $\hat A^m{}^{\dagger}$ simply refer to  $(\hat A^m)^\dagger = (\hat A^\dagger)^m$.  
From these results, it follows that 
$\hat{\mathcal H}(t)$ is Hermitian: $\hat{\mathcal H}(t) =\hat{\mathcal H}^\dagger(t)$, as we claimed above.

In deriving the error bounds for the Bloch-Redfield equation, we will make use of the norms of superoperators. 
We define the norm of the superoperator $\hat{A}$ acting on $\mathbb H^2_i$ as 
\be
\norm{\hat{A}} \equiv \sup_{{|\mathcal O\rrangle\in \mathbb H_{i}^2}}\frac{\norm{\hat{A} |\mathcal O\rrangle}}{\norm{|\mathcal O\rrangle}},
\label{eq:superoperator_norm}
\ee
where, here and in the following, $\norm{|\mathcal O\rrangle}$ 
 denotes the  spectral  norm of the operator $|\mathcal{O}\rrangle$. 
 Note that $\norm{|\mathcal O\rrangle}$ is {\it not} identical to $\sqrt{\llangle \mathcal O|\mathcal O\rrangle}$; rather, $\sqrt{\llangle \mathcal O|\mathcal O\rrangle}$   gives the Frobenius norm of $|\mathcal O\rrangle$. 
From the  definition above, it follows that the superoperator norm is submultiplicative: $\norm{\hat A  \hat B} \leq \norm{\hat A}\norm{\hat B}$. 
Moreover, using the submultiplicativity of the spectral norm along with the definitions in Eq.~\eqref{eq:o_r_l_def}, we conclude that, for any operator $|\mathcal{O}\rrangle$  and for  $m=\{l,r\}$, $\norm{\hat{\mathcal O}^{m}} = \norm{|\mathcal O\rrangle}$.

Using the superoperator notation above, we now consider the evolution  of the reduced density matrix of the system in the interaction picture,   $|\rho(t)\rrangle$.
Recalling that the superoperator  $\hat{\mathcal H}(t)$ in Eq.~\eqref{eq:DensityMatrixSchrodingerEq} is Hermitian, the ``Schr\"odinger equation'' for the density matrix of the combined system, 
 Eq.~\eqref{eq:DensityMatrixSchrodingerEq}, has the well-known solution 
\be 
|\rho_{\S\B}(t)\rrangle = \hat {\mathcal U}(t,s)|\rho_{\S\B}(s)\rrangle,
\label{eq:SuperOperatorSolution}
\ee
where   $ \hat {\mathcal U}(t,s)$ denotes the unitary evolution superoperator of the combined system, given by  
$ 
\hat {\mathcal U}(t,s) = \mathcal T e^{-i \int_{s}^t {\rm d}t'\,   \hat {\mathcal H}(t') }.
$
By taking the time-derivative, one can  verify that $\hat{\mathcal U} (t,s) =  {\hat U_{\S\B}^l(t,s) \hat U^{r\dagger}_{\S\B}(t,s)}$, where $U_{\S\B}(t,s) = \mathcal T e^{-i \int_{s}^t {\rm d}t'\, H(t')}$ denotes the (ordinary) 
time-evolution operator of the combined system. 
Thus $|\rho_{\S\B}(t)\rrangle = |\,U_{\S\B}(t,s) \rho_{\S\B}(s) U_{\S\B}^\dagger(t,s)\,\rrangle$.
Using the properties of the superoperator norm below Eq.~\eqref{eq:superoperator_norm}, we conclude that
 that $\norm{\hat {\mathcal U}(t,s)} = 1$.

In the superoperator notation, the partial trace $\Tr_\B$ over the bath degrees of freedom can be expressed as the dual vector (bra) of the bath identity operator $|I_\B\rrangle$. 
Here, as for ordinary bra-ket notation,   $\llangle Y_\B|$ is understood as the linear mapping $\mathbb H^2_{\S\B} \to \mathbb H ^2 _\S$, such that, for $|M_{\S\B}\rrangle = \sum_{a,b}M_{ab}|a_\S\rrangle |b_\B\rrangle$, 
$ 
\llangle Y_\B| M_{\S\B} \rrangle  = \sum_{a ,b} M_{ab}|a_\S\rrangle\llangle Y_\B|b_\B\rrangle
$, where $a$ and $b$ label orthonormal bases for the operator spaces on $\S$ and $\B$, respectively.
(Recall that the operator space $\mathcal H_{\S\B}^2$ inherits the tensor product structure of $\mathcal H_{\S\B}$.)
As a result, we may write the reduced density matrix of the system $\S$ as  $|\rho(t) \rrangle =\llangle I_\B|\rho_{\S\B}(t)\rrangle$. 
Inserting the above result into Eq.~\eqref{eq:SuperOperatorSolution}, and  using our assumption that  $|\rho_{\S\B}(t_0)\rrangle=|\rho_{0}\rrangle  |\rho_\B\rrangle$  for some time $t_0$ in the remote past (see Sec.~\ref{sec:model} in the main text), we find 
\be 
|\rho(t)\rrangle  =  \llangle I_\B|\Us(t,t_0) |\rho_\B\rrangle|\rho_0\rrangle. \label{eq:reduced_u}
\ee

To obtain a master equation for $|\rho(t)\rrangle$, we explicitly take the time-derivative in Eq.~\eqref{eq:reduced_u}, obtaining 
\be
\partial _t |\rho(t)\rrangle  = -i\llangle I_\B|\Hs(t)\Us(t,t_0) |\rho_\B\rrangle|\rho_0\rrangle.
\label{eqa:rho_eom_0}
\ee
Using the decomposition $H(t) = \sqrt{\gamma}\sum_\alpha X_\alpha(t) B_\alpha(t)$
 [Eq.~\eqref{eq:h_sb_def} in the main text, translated to the interaction picture], we find   $\Hs(t) = \sqrt{\gamma } \sum_{m,\alpha }\nu_m\hat X^m_\alpha (t)\hat B^m_\alpha (t)  $,   where $m = \{l, r\}$, with  $\nu_l = 1$ and $\nu_r= -1$.
 Recalling that  (for each $\alpha $)  $X_\alpha (t)$ acts trivially on the bath degrees of freedom,
  we obtain  
\be 
\partial _t |\rho(t)\rrangle = -i \sqrt{\gamma } \sum_{m, \alpha}\! \nu_m \hat X_{\alpha }^m(t)\llangle I_\B|\hat B_{\alpha }^m(t) \hat {\mathcal U}(t,t_0) |\rho_\B\rrangle|\rho_0\rrangle.
\label{eq:du_r_dt_expression_1}
\ee
\subsection{Statistical properties of the bath}
To obtain a convenient expression for the bath expectation value $\llangle I_\B|\hat B_{\alpha }^m(t) \hat{\mathcal U}(t,t_0)|\rho_\B \rrangle$ in Eq.~\eqref{eq:du_r_dt_expression_1}, 
we make use of  our assumption  that the bath is Gaussian. 
For simplicity, in this section we assume that all bath operators are bosonic. 
Similar considerations can be applied for fermionic bath operators.

For  a Gaussian bath, the expectation value of any bath operator can be computed from the two-point correlation function using Wick's theorem. 
In the superoperator notation we use, with  $\hat B _j\equiv \hat B_{\alpha _j}^{m_j}(t_j)$ (where $m_j =\{l,r\}$, while $\alpha _j$ refers to the noise channel index), 
Wick's theorem applied to a product of $k$ bath operators takes the form: 
\be
\llangle \hat B_1 \ldots \hat B_k\rrangle = \sum_{j=2}^k\llangle \hat B_1 \hat B_j\rrangle \llangle \hat A_{2,j-1}\hat A_{j+1,k}\rrangle,
\label{eq:app:continuous_wick}
\ee
where $\hat A_{i,j}=\prod_{n=i}^j \hat B_n$ for $j\geq i$,  $\hat A_{i,j}=1$ for $j<i$,  and we introduced the shorthand $\llangle \hat{ \mathcal O} \rrangle \equiv \llangle I_B|\hat{\mathcal O} |\rho_B\rrangle$ to simplify notation.
Wick's theorem for superoperators, as stated in Eq.~\eqref{eq:app:continuous_wick}, can be proven by direct computation using the definitions of the superoperators $\{\hat B_j\}$, along with Wick's theorem for the (non-super) operators $\{B_{\alpha _j}(t_j)\}$~\cite{Breuer,GardinerZoller}. 

By iteration of Wick's theorem [Eq.~\eqref{eq:app:continuous_wick}], it is straightforward to show that the expectation value $\llangle \cdot\rrangle$ of any polynomial functional of the bath superoperators $\{\hat B_\alpha(t)\}$ can be expressed  
 fully in terms of the (two-point) bath superoperator correlation functions
\be
J_{\alpha \beta }^{mn}(t-t')\equiv \llangle \hat B_{\alpha }^{m} (t)\hat B_{\beta  }^{n}(t')\rrangle.
\label{eqa:superoperator_correlator}
\ee
The  bath superoperator correlation functions hold the same information as the ordinary bath correlation function $\bs J(t-s)$ [see Eq.~\eqref{eq:correlation_function_m} in the main text]: 
letting $\bs J^{mn}(t)$ denote the  matrix with elements $\{J_{\alpha \beta }^{mn}(t)\}$, and using the cyclic property of the trace, one can verify that, for $m=\{l,r\}$, $\boldsymbol J^{ml}(t) = \boldsymbol J(t)$, while $\boldsymbol J^{mr}(t) = \ \boldsymbol J^\dagger(t)$.

\begin{widetext}
Importantly, the unitary evolution superoperator of the combined system $\S\B$, $\hat {\mathcal U}(t,s)$,  is  analytic, and hence
can be  expanded as a polynomial  of the bath superoperators $\{\hat B^m_\alpha(t)\}$.
By using this expansion along with Wick's theorem [Eq.~\eqref{eq:app:continuous_wick}], one can then verify that 
\be  
 \llangle  \hat B_\alpha ^{m}(t) \hat {\mathcal U}(t,s)\rrangle  = \int_{-\infty}^\infty \!\!\!{\rm d} t' \sum_{\beta ,n}  J_{\alpha \beta }^{mn}(t-t') \left<\hspace{-2mm}\left< \frac{\delta\hat {\mathcal U}(t,s)}{\delta \hat B_\beta^n(t')}\right>\hspace{-2mm}\right>,
\label{eq:functional_derivative}
\ee
where ${\delta}/{\delta \hat B_\beta^n(t')}$ denotes the functional derivative  with respect to $\hat B_\beta ^n(t')$.
Specifically, $\delta \hat B_\alpha ^m(t)/{\delta \hat B_\beta^n(t')} = \delta_{\alpha \beta }\delta_{mn}\delta(t-t')$, where $\delta_{ij}$ denotes the Kronecker symbol, and $\delta(t)$ the Dirac delta function. 

Using the Trotter decomposition of  $\hat {\mathcal U}(t,s)$ along with $\hat {\mathcal H}(t) = -i\sqrt{\gamma } \sum_{m,\alpha  } \nu_m \hat X_\alpha  ^m (t)\hat B_\alpha  ^m(t)$, one can verify  that, for   $t'$ in the interval between $s$ and $t$,
\be 
\frac{\delta \hat{\mathcal U}(t,s)}{\delta \hat B^n_\beta (t')} = - i \sqrt{\gamma }\nu_n{\mathcal U}(t,t')\hat X^n_\beta (t') \hat{\mathcal U}(t',s),
\label{eq:app:u_func_derivative}
\ee 
 while ${\delta \hat {\mathcal U}(t,s)}/{\delta \hat B_\beta ^n(t')} = 0$ when $t'$ is outside the interval between $s$ and $t$.
Inserting Eqs.~\eqref{eq:functional_derivative} and \eqref{eq:app:u_func_derivative} into Eq.~\eqref{eq:du_r_dt_expression_1} gives
\begin{align}
\partial _t |\rho(t)\rrangle 
 = - \gamma \sum_{m,n;\alpha, \beta } \nu_{m}\nu_n \hat X_{\alpha}^m (t) \int_{t_0}^t \!{\rm d}s\,  J_{\alpha \beta }^{mn}(t-s) 
\llangle I_\B|\hat {\mathcal U}(t,s) \hat X_{\beta }^n (s)\hat{\mathcal U}(s,t_0)|\rho_\B\rrangle\, |\rho_0\rrangle .
 \label{eq:dur_dt_exact_expression}
\end{align}
\end{widetext}

Eq.~\eqref{eq:dur_dt_exact_expression}  is a crucial result, and forms the basis for the derivation below.
Importantly, the result is  {exact} for Gaussian baths, and does not rely on any other approximations or assumptions. 
Eq.~\eqref{eq:dur_dt_exact_expression} can be generalized to  non-Gaussian baths by expanding the left hand side of  Eq.~\eqref{eq:functional_derivative} in terms of the (nonvanishing) higher-order correlation functions of the bath.
While  such an extension to non-Gaussian baths is  in principle straightforward, 
in this Appendix we restrict ourselves for simplicity to the case of Gaussian baths.

\subsection{Upper bound for rate of bath-induced evolution} 
\label{app:speed_limit}
While Eq.~\eqref{eq:dur_dt_exact_expression}  looks somewhat complicated, we may already use it in its present form to infer important facts about 
the evolution of the system. 
Specifically, in this subsection, using Eq.~\eqref{eq:dur_dt_exact_expression}, we    identify an upper limit  for the rate of bath-induced evolution in the system, $\norm{\partial _t|\rho\rrangle}$. 
This result was quoted in Sec.~\ref{sec:bath_time_scales} of the main text (recall that $|\rho(t)\rrangle$ in this Appendix is identical to $\tilde \rho(t)$ in the main text). 
The arguments and concepts we use here will  also be used in  the following subsections, when we derive error bounds for the Bloch-Redfield equation. 

To derive an upper bound for $\norm{\partial _t| \rho\rrangle}$, we  take the (spectral) norm on both sides in
Eq.~\eqref{eq:dur_dt_exact_expression}. 
Using the triangle inequality  along with 
$\norm{\hat{X}^m_\alpha (t)|\mathcal O\rrangle }\leq \norm{|{\mathcal O}\rrangle}$ (this follows from the properties of the superoperator norm listed in Sec.~\ref{sec:superoperators} and the fact that the operators $X_\alpha$ are assumed to have unit spectral norm), we thereby obtain
\begin{align}
&\norm{\partial _t |\rho(t)\rrangle }\leq  
\gamma \!\!\!\sum_{m,n;\alpha, \beta  }  \int_{t_0}^t \!{\rm d}s\,|J_{\alpha \beta }^{mn}(t-s)| 
k_\beta ^n(t,s),\!
 \label{eq:app:dt_rho_inequality}
\end{align}
where $k_\beta ^n(t,s) \equiv \norm{ \llangle  I_\B|\hat {\mathcal U}(t,s) \hat X_{\beta }^n (s)\hat{\mathcal U}(s,t_0)|\rho_\B\rrangle |\rho_0\rrangle}$. 
We now prove that 
$k_\beta ^n(t,s) \leq 1$. 
To establish this bound,  it is simplest to consider the cases $n=l$ and $n=r$ separately. 
Specifically, below we prove that $k^l_\beta (t,s)\leq 1$. 
The proof for $k^r_\beta (t,s)\leq 1$  proceeds along the same lines.  

To establish that $k_\beta ^l(t,s)\leq 1$,  we write $k_\beta ^l(t,s) = \norm{|Q\rrangle}$, where 
\be 
|Q\rrangle\equiv\llangle  I_\B|\hat {\mathcal U}(t,s) \hat X_{\beta }^{l} (s)\hat{\mathcal U}(s,t_0)|\rho_\B\rrangle |\rho_0\rrangle
\label{eqa:q_def}
\ee 
represents an operator on the system $\S$.
We now  note that  $\hat{\mathcal U}(s,t_0)|\rho_\B\rrangle|\rho_0\rrangle = |\rho_{\S\B}(s)\rrangle$, since $|\rho_{\S\B}(t_0)\rrangle = |\rho_\B \rrangle|\rho_0\rrangle$. 
Thus 
$
|Q\rrangle= \llangle I_\B|\hat{\mathcal{U}}(t,s) \hat X_{\beta }^l (s)|\rho_{\S\B} (s)\rrangle.
$

We now convert the above expression for $|Q\rrangle$ into standard (non-superoperator) notation for the corresponding operator $Q$ that acts on system $\S$: 
\be
Q= \Tr_\B \big[U_{\S\B}(t,s)X_\beta (s)\rho_{\S\B}(s)U_{\S\B}^\dagger(t,s)\big],
\label{eqa:q_expr}
\ee
where $U_{\S\B}(t,s)=\mathcal Te^{-i\int_{s}^t\!{\rm d}t' H(t')}$ denotes the standard (i.e, non-super) unitary evolution operator of the combined system in the interaction picture [see  discussion below Eq.~(\ref{eq:SuperOperatorSolution})]. 
We recall that the spectral norm of $Q$, also denoted $\norm{|Q\rrangle}$,  is given by the maximal value of $|\langle \phi|Q|\psi\rangle| $ for any two normalized states $|\psi\rangle$, $|\phi\rangle$ in the system Hilbert space $\mathbb H_\S$.
To bound this number, we exploit the cyclic property of the trace along with Eq.~\eqref{eqa:q_expr} to write 
\begin{align}
\langle \phi|Q|\psi\rangle= \Tr_{\S\B} \left[C\rho_{\S\B}(s)\right],
\label{eq:q_expval}
\end{align}
where $C\equiv U_{\S\B}^\dagger(t,s)( |\psi\rangle\langle \phi|\otimes I_\B) U_{\S\B}(t,s) X_\beta (s)$, with $I_\B$ denoting the identify operator on the bath Hilbert space, $\mathbb H_\B$.
Next, we use the  spectral decomposition  of $\rho_{\S\B}(s)$, $\rho_{\S\B}(s) = \sum_i |n_i\rangle\langle n_i| p_i$, where $\{|n_i\rangle\}$ form an orthonormal basis for the Hilbert space of the combined system, $\mathbb H_{\S\B}$, and the eigenvalues $\{p_i\}$ are non-negative and have unit sum. 
Inserting this 
 into Eq.~\eqref{eq:q_expval}, we  find 
$
\langle \phi|Q|\psi\rangle = \sum_i \langle n_i|C|n_i\rangle p_i.
$
Using the triangle inequality, along with $|\langle n_i|C|n_i\rangle|\leq \norm{C}$, where $\norm{\cdot}$ denotes the spectral norm, we find
\be 
|\langle \phi|Q|\psi\rangle| \leq \norm{C},
\ee
 where we also exploited the non-negativity and unit sum of the eigenvalues $\{p_i\}$. 
Using the submultiplicativity of the spectral norm and the fact that 
$\norm{|\psi\rangle\langle \phi|\otimes I_\B}\leq 1$ when $|\psi\rangle$ and $|\phi\rangle$ are normalized, one can verify that $\norm{C}\leq 1 $.
Thus, for any normalized states $|\psi\rangle$ and $|\phi\rangle$, $|\langle \psi|Q|\phi\rangle| \leq 1$. 
We thus  conclude that $\norm{|Q\rrangle}=k_\beta ^l (t,s)$, must be smaller than or equal to $1$. 
The same line of arguments shows that $k_\beta ^r(t,s)\leq 1$.
Recalling that $k_\beta ^n(t,s)$ by construction cannot be negative, we thus conclude 
\be 
0\leq k_\beta ^n(t,s)\leq 1.
\label{eqa:k_result}
\ee

We now use Eq.~(\ref{eqa:k_result}) 
 in Eq.~\eqref{eq:app:dt_rho_inequality} to obtain
\begin{align}
\norm{\partial _t |\rho(t)\rrangle}\leq \gamma \!\!  \sum_{m,n;\alpha, \beta }  \int_{t_0}^t \!{\rm d}s\, | J^{mn}_{\alpha \beta }(t-s)| .
\end{align}
Evaluating the sum, using the results below Eq.~\eqref{eqa:superoperator_correlator}, we obtain 
\begin{align}
\norm{\partial _t |\rho(t)\rrangle}\leq 4\gamma    \int_{t_0}^t \!{\rm d}s\, \norm{\boldsymbol J(t-s)}_1, \label{eq:app:dt_rho_inequality_2}
\end{align}
where $\norm{\cdot}_1$ denotes  the entrywise matrix $1$-norm, such that for any matrix $\boldsymbol M$ with elements $\{M_{\alpha \beta }\}$, $\norm{\boldsymbol M}_1 \equiv \sum_{\alpha \beta }|M_{\alpha \beta }|$.
To obtain  Eq.~\eqref{eq:app:dt_rho_inequality_2}, we used the relation $\norm{\boldsymbol M}_1 = \norm{\boldsymbol M^\dagger}_1$, which follows from the definition above.

Extending the lower limit of integration in Eq.~\eqref{eq:app:dt_rho_inequality_2} to $-\infty$ and changing integration variables, we finally obtain  
\be
\norm{\partial _t |\rho(t)\rrangle}\leq \vmax,
\quad 
\vmax \equiv 4\gamma  \int_0^\infty \!\!\! {\rm d} t\, \norm{\boldsymbol J(t)}_1.
\label{eq:speed_limit}
\ee
Thus, the energy scale  $\vmax$  sets an upper bound for the rate of bath-induced evolution  in the system, $\norm{|\partial _t \rho(t)\rrangle}$.
This timescale was also identified in Ref.~\cite{Mozgunov_2020} (see main text and Appendix~\ref{app:time_scale_relationship} for further discussion).
In Appendix~\ref{app:time_scale_relationship} we further show that $\Gamma_0 \leq \Gamma/2$, where  $\Gamma$ was given in Eq.~\eqref{eq:time_scales_m} in the main text [see Eq.~\eqref{eq:time_scales} for   the special case of a single noise channel].
Thus, recalling that  $|\rho(t)\rrangle$ corresponds to $ \tilde \rho(t)$ in the main text, we have shown that 
\be 
\norm{\partial _t \tilde \rho(t)} \leq \Gamma/2.
\ee
This was the result quoted in Sec.~\ref{sec:bath_time_scales} of the main text.\\

\subsection{Error induced by the Born approximation}
\label{sec:born_approximation}
Until now, our derivation has been exact, with our only assumptions  being that the bath is   Gaussian, and that the  system and bath were decoupled  at some point $t_0$ in the remote past (see Sec.~\ref{sec:born_markov} in the main text). 
At this point,  exact manipulations cannot take us further, and  we thus need  make our first approximation: the Born approximation. 

To make the Born approximation, we integrate  the equation of motion for the evolution superoperator of the combined system, $\partial _t \hat{ \mathcal U}(t,s)= -i\hat{\mathcal H}(t)\hat{\mathcal U}(t,s)$:  
\be 
\hat {\mathcal U}(t,s) = 1 - i \int_{s}^t \!{\rm d}t'\, \hat {\mathcal H}(t')\Us(t',s).
\label{eq:u_iterative_solution}
\ee
By directly substituting this expression in for the factor of $\Us(t,s)$ in  Eq.~\eqref{eq:dur_dt_exact_expression}, we obtain 
\begin{widetext}
\begin{align}
\partial _t| \rho(t)\rrangle
 = - \gamma\!\!  \sum_{m,n;\alpha ,\beta } \! \!\nu_{m}\nu_n \int_{t_0}^t \!{\rm d}s\,
 \hat X_{\alpha}^m (t)   J_{\alpha \beta }^{mn}(t-s)\left(
 \hat X_{\beta }^n (s) 
|\rho({s}) \rrangle
 - i  \int_{s}^t \!{\rm d}t'\, \llangle I_\B| \Hs(t')\Us(t',s)\hat X^n_\beta (s)\Us(s,t_0)|\rho_\B\rrangle|\rho_0\rrangle
 \right).
 \label{eq:dur_dt_exact_expression_2}
\end{align}
For the first term in the parentheses above we used that $\hat X_\beta ^n (t)$ acts trivially on the bath, along with $\Us(s,t_0)|\rho_0\rrangle|\rho_\B\rrangle = |\rho_{\S\B}(s)\rrangle$ and $\llangle I_\B|\rho_{\S\B}(s)\rrangle = |\rho(s)\rrangle$,
 such that $\llangle I_\B|\hat X^n_\beta (s) \Us(s,t_0)|\rho_\B\rrangle|\rho_0\rrangle = \hat X^n_\beta (s) |\rho(s)\rrangle$.

Next, we separate the two terms in the parentheses in Eq.~\eqref{eq:dur_dt_exact_expression_2}. 
Referring to the second term in the resulting expression  as $| \xiB(t)\rrangle$ (we discuss this term in further detail below), we find 
\be 
\partial _t |\rho(t)\rrangle = - \gamma  \int_{t_0}^t \!{\rm d}s\,
  \DeltaB(t,s) |\rho(s)\rrangle  + | \xiB(t)\rrangle,
\quad\ {\rm where}\ \quad
  \DeltaB(t,s) \equiv-\gamma\!\!\sum_{m,n;\alpha ,\beta }\!\!\nu_m\nu_n  \hat X^m_\alpha  (t)\hat X^n_\beta (s) J^{mn}_{\alpha \beta }(t-s).
\label{eq:dissipator_definition}
 \ee
By applying the definitions of the quantities $\hat X^m_\alpha (t)$, $\nu_m$ and $J^{mn}_{\alpha \beta }(t)$, one can verify that the first term in the right-hand side of Eq.~(\ref{eq:dissipator_definition}) is identical to the 
master equation for $|\rho(t)\rrangle$ in the Born approximation~\cite{Breuer} [see text above Eq.~\eqref{eq:correlation_function_def} in the main text for the single-channel case].
Hence, the Born approximation is equivalent  to neglecting the  term $|\xiB(t)\rrangle$ in the  above, and we  identify $|\xiB(t)\rrangle$ as the error induced by the Born approximation. 
Note that the Born-approximated master equation for $|\rho(t)\rrangle$ [Eq.~\eqref{eq:dissipator_definition} with the correction $|\xiB(t)\rrangle$ neglected] can also be obtained through other  approaches than the one we use here.
For example, this result may also be obtained using the Nakajima-Zwanzig equation (see, e.g., Refs.~\cite{Breuer,Evans_book}).
 
We  now seek a  bound for the norm of $|\xiB(t)\rrangle$, i.e.,   the {norm of the} error in $\partial_t| \rho(t)\rrangle$ induced by the Born-approximation. 
Matching Eqs.~(\ref{eq:dur_dt_exact_expression_2}) and (\ref{eq:dissipator_definition}), we see that 
\be
|\xiB(t)\rrangle   = i \gamma\!\!\!\sum_{m,n;\alpha ,\beta }\!\!\!\nu_m\nu_n\int^t_{t_0} {\rm d}s\,   J^{mn}_{\alpha \beta }(t-s) \hat X^m_\alpha (t)
  \int_{s}^t \!{\rm d}t'\, \llangle I_\B| \Hs(t')\Us(t',s)\hat X^n_\beta (s)\Us(s,t_0) |\rho_\B\rrangle|\rho_0\rrangle.
\label{eqa:xi_b_def}
\ee
%
 \end{widetext}
 
To obtain a bound for $\norm{|\xiB(t)\rrangle}$, we take the norm on both sides of Eq.~(\ref{eqa:xi_b_def}) above. 
Using  the triangle inequality and submultiplicativity of the superoperator norm, along with $\norm{\hat X^m_\alpha (t)}=1$, 
we find 
\begin{align}
\norm{ | \xiB(t)\rrangle}  \leq \gamma\!\!\sum_{m,n;\alpha ,\beta }\int^t_{t_0} {\rm d}s\,   |J^{mn}_{\alpha \beta }(t-s)|\int_{s}^t \!{\rm d}t'\,  q^n_\beta (t,t',s),
\label{eq:app:xi_norm_ineq}
 \end{align}
 where 
\begin{align}
&q^n_\beta (t,t',s) \equiv \label{eqa:q_def}
\norm{ \llangle I_\B| \Hs(t')\Us(t',s)\hat X^n_\beta (s)\Us(s,t_0) |\rho_\B\rrangle|\rho_0\rrangle}.\notag 
\end{align}
Following the same line of arguments that {showed that the number $k^n_\beta (t,s)$ in Sec.~\ref{app:speed_limit} was bounded by $1$}, 
one can  verify that
\be
0\leq  q^n_\beta (t,t',s)\leq \Gamma_0,
\ee
where $\vmax$ was defined in Eq.~\eqref{eq:speed_limit}.
Substituting this result into Eq.~(\ref{eq:app:xi_norm_ineq}) and evaluating the integral over $t'$, we find 
\begin{align}
\norm{| \xiB(t)\rrangle}   \leq   4 \gamma \Gamma_0 \int^{t}_{t_0} \!{\rm d}s\, \norm{\boldsymbol J(t-s)}_1 \cdot|t-s| ,
\end{align}
where the matrix norm $\norm{\cdot}_1$ was defined in Sec.~\ref{app:speed_limit}.
Extending the lower limit of integration to $-\infty$ {and using the definition of $\Gamma_0$ in Eq.~\eqref{eq:speed_limit}},  we  obtain  
\begin{align}
\norm{| \xiB(t)\rrangle}   \leq \Gamma_0^2 \tau_0, 
\quad 
\tau_0 \equiv \frac{\int_0^\infty\! {\rm d}t\,  t \norm{\boldsymbol J(t)}_1 }{\int_0^\infty\!{\rm d}t\,  \norm{\boldsymbol J(t)}_1 }.
\label{eq:tau_c_def}
\end{align}
The timescale $\tau_0$, which was also identified in Ref.~\cite{Mozgunov_2020},  can be seen as a measure for the characteristic  decay  timescale of correlations in the bath, and we expect it to typically be comparable to the timescale $\tau$ from the main text (see Appendix~\ref{app:time_scale_relationship} and Sec.~\ref{sec:bath_time_scales} in the main text for further discussion). 
Importantly, in Appendix~\ref{app:time_scale_relationship} we show that $\Gamma_0 \tau_0 \leq\Gamma \tau$.

Using the above results, along with $\Gamma _0 \leq \Gamma/2$ (see Sec.~\ref{app:speed_limit}), we conclude that  
\be
\norm{|\xiB(t)\rrangle}\leq \Gamma^2 \tau/2.
\ee
%
Recalling that $|\xiB(t)\rrangle$ gives the correction to the Born-approximated master equation for the system [Eq.~\eqref{eq:dissipator_definition}],  we conclude that  the Born approximation induces an error in the expression for $\partial _t \tilde \rho(t)$ whose spectral norm is no greater than $\Gamma^2 \tau/2$.

\subsection{Error induced by the Markov approximation}
We now implement the Markov approximation, which is the second approximation necessary to derive  the Bloch-Redfield equation.
Below, we show that  Markov approximation induces an error in the expression for $|\partial _t \rho(t)\rrangle$ (i.e., $\partial _t\tilde \rho(t)$ in the main text) whose spectral norm is bounded by $\Gamma^2 \tau/2$. 
This bound is  identical to the error bound we obtained for  the Born approximation in  Sec.~\ref{sec:born_approximation}.
In this sense, the Markov approximation is valid on an equivalent level of approximation as the Born approximation: the  validity of one approximation  by our arguments implies the validity of the other.

To implement the Markov approximation, we insert  $|\rho(s)\rrangle =|\rho(t)\rrangle +(|\rho(s)\rrangle-|\rho(t)\rrangle)$ into
 Eq.~\eqref{eq:dissipator_definition},  thereby obtaining
\be 
\partial _t| \rho(t)\rrangle=\int_{t_0}^t\! {\rm d}s\, \DeltaB(t,s)|\rho(t)\rrangle  +  |\xiM(t)\rrangle+ |\xiB(t)\rrangle, 
\label{eq:markov_first_step}
\ee
where  
\be 
| \xiM(t)\rrangle =\int_{t_0}^t\! {\rm d}s\,\DeltaB(t,s)\big(|\rho(s)\rrangle-|\rho (t)\rrangle\big). 
\label{eqa:xi_m_def}
\ee

We note that  neglecting the terms $|\xiB(t)\rrangle$ and $|\xiM(t)\rrangle$  in Eq.~\eqref{eq:markov_first_step}  results in a Markovian master equation for the system.
Recalling that $|\xiB(t)\rrangle$ arises from the Born approximation,  we hence identify $|\xiM(t)\rrangle$ as the error induced by the Markov approximation.
The Bloch-Redfield equation~\cite{Breuer} [see Eq.~\eqref{eq:redfield_dissipator} in the main text for the single-channel case] is obtained by neglecting these two terms, and subsequently  taking the limit $t_0 \to -\infty$, i.e., using our assumption that $t_0$ was in the remote past.
In Sec.~\ref{seca:transient_correction} we discuss the physical justification for this assumption, and provide a bound for the correction that arises when this limit is not taken. 

To obtain an upper bound for the error induced by the Markov approximation, $|\xiM(t)\rrangle$, we  take the norm on both sides in Eq.~\eqref{eqa:xi_m_def} and use the triangle inequality. 
Recalling from Sec.~\ref{app:speed_limit} that $\norm{\partial _t|\rho(t)\rrangle}\leq \Gamma_{0}$, we have  $\norm{|\rho(s)\rrangle -|\rho(t)\rrangle}\leq \vmax  |t-s|$, where $\Gamma_0$ was defined in Eq.~\eqref{eq:speed_limit}.
Moreover, we note $\norm{ \DeltaB(t,s)} \leq 4\gamma  \norm{\boldsymbol J(t-s)}_1$; this can be shown using the triangle inequality in Eq.~\eqref{eq:dissipator_definition}. 
Combining these inequalities, we find 
\be 
\norm{|\xiM(t)\rrangle} \leq 4\vmax \gamma\int^t_{t_0} \!{\rm d}s\,    \norm{\boldsymbol J(t-s)}_1  |t-s|.
\label{eq:delta_v_inequality_1}
\ee
Extending the lower limit of integration to $-\infty$, and using the definitions of $\Gamma_0$ and $\tau_0$ in Eqs.~\eqref{eq:speed_limit}~and~\eqref{eq:tau_c_def}, we  conclude that 
$
\norm{|\xiM(t)\rrangle} \leq \vmax^2 \tau_0. 
$
Recalling that $\Gamma^2 _0\tau_0 \leq \Gamma^2 \tau/2$ (see Sec.~\ref{sec:born_approximation} and Appendix~\ref{app:time_scale_relationship}), we thus find 
\be
\norm{|\xiM(t)\rrangle} \leq \Gamma^2 \tau/2. \label{eq:xi_M_bound}
\ee
The result in Eq.~(\ref{eq:xi_M_bound}) shows that the error in the expression for $\partial _t |\rho(t)\rrangle$ (corresponding to $\partial _t \tilde \rho(t)$ in the main text) induced by the Markov approximation, $|\xiM(t)\rrangle$, has spectral norm no greater than $\Gamma^2 \tau/2$, as we claimed. 


Based on the derivation above, we conclude that the density matrix of the system evolves according to the Markovian master equation
\be 
\partial _t |\rho(t) \rrangle = \int_{t_0}^t\! {\rm d}s\, \DeltaB(t,s) |\rho(t)\rrangle + |\xi(t)\rrangle ,
\label{eqa:br_equation_m}
\ee
where  $\DeltaB(t,s)$ is given in Eq.~\eqref{eq:dissipator_definition}, and $|\xi(t)\rrangle \equiv |\xiB(t)\rrangle + |\xiM(t)\rrangle$ denotes the error induced by the Born-Markov approximation. 
From our results above that $|\xiB(t)\rrangle, |\xiM(t)\rrangle \leq \Gamma^2 \tau/2$, we hence conclude that   the total error induced by the Markov and Born approximations is bounded by $\Gamma^2 \tau$, as we claimed in the main text.

\subsection{Transient correction from initialization at $t_0$}
\label{seca:transient_correction} 

As a final step in our derivation, here we show that when $t_0$ is in the remote past, the error induced by extending $t_0$ to $-\infty$ in Eq.~\eqref{eqa:br_equation_m} is negligible compared to the error induced by the Born-Markov approximation,  $|\xi(t)\rrangle \sim \mathcal O(\Gamma^2 \tau)$.
Specifically, we show that the spectral norm of this error is bounded by $\Gamma \tau/(t-t_0)$, and hence is negligible when $t-t_0\gg \Gamma^{-1}$.    
By setting $t_0\to -\infty$ in Eq.~\eqref{eqa:br_equation_m}, we obtain 
\be 
\partial _t |\rho(t) \rrangle = \hat {\mathcal D}_{\rm R}(t) |\rho(t)\rrangle + |\xi(t)\rrangle ,
\label{eqa:br_equation_1}
\ee
where $\hat{\mathcal D}_{\hspace{0.5pt}\rm R}(t)  \equiv \int_{-\infty}^t{\rm d}s\, \DeltaB(t,s)$. 
This is the Bloch-Redfield equation~\cite{Breuer} [including the error induced by the Born-Markov approximation, see Eq.~\eqref{eq:redfield_dissipator} in the main text for the single-channel case].

To establish a bound for the error induced by setting $t_0\to -\infty$ in Eq.~\eqref{eqa:br_equation_m}, we  rewrite  Eq.~\eqref{eqa:br_equation_m} as follows:
\be
\partial _t|\rho(t)\rrangle = \left[\hat {\mathcal D}_{\hspace{0.5pt} \rm R}(t)+ \hat {\mathcal D}_{\hspace{0.5pt} \rm T} (t)\right]|\rho(t)\rrangle + |\xi (t)\rrangle,
\label{eqa:br_with_transient}
\ee
where
$
\hat {\mathcal D}_{\hspace{0.5pt} \rm T} (t) \equiv  - \int_{-\infty}^{t_0}\! {\rm d}s\, \DeltaB(t,s).
$
This term can  be seen as  the transient correction to the BR equation induced by the absence of system-bath correlations in our assumed initial state at time $t_0$, $|\rho_{\S\B}(t_0)\rrangle = |\rho_0\rrangle|\rho_\B\rrangle$.
This ``correction'' is thus an artifact of our choice of initial state (see Sec.~\ref{sec:born_markov}). 
Below, we show that  $\norm{\hat {\mathcal D}_{\hspace{0.5pt} \rm T} (t)|\rho(t)\rrangle}\leq\Gamma \tau/(t-t_0)$. 
Thus, when $t-t_0 \gg \Gamma^{-1}$, i.e., after a time long enough for weak correlations to be established between the system and the bath, the transient correction $\hat {\mathcal D}_{\hspace{0.5pt} \rm T} (t)$ is negligible compared to the bound we obtained for the error induced by the Born-Markov approximation, $\Gamma^2 \tau$. 
As a result, the BR equation [Eq.~\eqref{eqa:br_equation_1}] accurately describes the system's evolution in this limit.

To show that $\norm{\hat{\mathcal D}_{\hspace{0.5pt} \rm T}(t)|\rho(t)\rrangle} \leq\Gamma \tau/(t-t_0)$,  we consider the superoperator norm of $\hat{\mathcal D}_{\hspace{0.5pt} \rm T}(t)$ [see Eq.~\eqref{eq:superoperator_norm}].
Noting that $\norm{\DeltaB (t,s)}\leq 4\gamma  \norm{\boldsymbol  J(t-s)}_1$ [this can be shown using the triangle inequality in Eq.~\eqref{eq:dissipator_definition}], we find 
\be 
\norm{\hat {\mathcal D}_{\hspace{0.5pt} \rm T} (t) } \leq 4\gamma \int_{-\infty}^{t_0}{\rm d}s\, \norm{\bs J(t-s)}_1.
\label{eqa:xi0_ineq}\ee 
Using the fact that $t>t_0$,  we have that $|t-s|\geq |t-t_0|$ {for all $s\leq t_0$.} 
Thus, 
\be
\norm{\hat {\mathcal D}_{\hspace{0.5pt} \rm T} (t) }\leq 4 \gamma \int_{-\infty}^{t}\!\!\!{\rm d}s\, \norm{\bs J(t-s)}_1\frac{|t-s|}{|t-t_0|}.
 \ee 
Changing variables of integration and using the definitions of $\Gamma_0$ and $\tau_0$ in Eqs.~\eqref{eq:speed_limit}~and~\eqref{eq:tau_c_def}, we conclude
\be
\norm{\hat {\mathcal D}_{\hspace{0.5pt} \rm T} (t) } \leq \frac{\tau_0 \Gamma_0}{t-t_0}.
 \ee  
Using the fact that that $\Gamma_0 \tau_0 \leq \Gamma \tau$ (see Appendix~\ref{app:time_scale_relationship}) along with the definition of the superoperator norm, we conclude that
$
\norm{\hat{\mathcal D}_{\hspace{0.5pt} \rm T}(t)|\rho(t)\rrangle } \leq \frac{\Gamma \tau}{t-t_0},
$
as we claimed. 

%

\section{Relationship between bath timescales}
\label{app:time_scale_relationship}

In this Appendix we discuss the relationship  
 between the bath timescales $\Gamma$ and $\tau$ introduced in Eq.~\eqref{eq:time_scales_m} of the main text, and the 
 timescales $\Gamma_0^{-1}, \tau_0$ identified in Eqs.~\eqref{eq:speed_limit}~and~\eqref{eq:tau_c_def} of Appendix~\ref{app:br_correction}:
  \be 
\Gamma_0 = 4 \gamma   \int_{0}^\infty \!\!\!{\rm d}t\, \norm{\boldsymbol J(t)}_{1},\quad \tau_0 = \frac{\int_{0}^\infty \!{\rm d}t\,t\norm{\boldsymbol J(t) }_{1}}{\int_{0}^\infty\! {\rm d}t\, \norm{\boldsymbol J(t)}_{1}},
\label{eqa:alternative_time_scales}
\ee
where $\bs J(t)$ denotes the matrix-valued bath correlation function (see Sec.~\ref{sec:multiple_noise_channels}), and  $\norm{\bs M }_1\equiv \sum_{\alpha \beta }|M_{\alpha \beta }|$ refers to the entrywise matrix $1$-norm of a matrix $\bs M$ with elements $\{M_{\alpha \beta }\}$ (see Appendix~\ref{app:br_correction}). 
The above timescales $\Gamma_0^{-1}$ and $\tau_0$ were also identified in  Ref.~\cite{Mozgunov_2020}. 

Like the timescales $\Gamma^{-1}$ and $\tau$,  $\Gamma_0^{-1}$ and $\tau_0$ serve as measures for the characteristic timescales for bath-induced evolution, and the decay 
bath correlations, respectively. 
In contrast to $\Gamma^{-1} $ and $\tau$, which are defined in terms of the ``jump correlator'' $\bs g(t)$ [see Eqs.~(\ref{eq:jump_correlator})~and~\eqref{eq:jump_correlator_m} of the main text], the timescales $\Gamma_0^{-1}$ and $\tau_0^{-1}$ above are defined directly from the bath correlation function $\bs J(t)$.
However, as discussed in Sec.~\ref{sec:bath_time_scales} and demonstrated in Fig.~\ref{fig:jump_correlator}, {we expect these two distinct ways of characterizing the timescales 
 of the bath} 
 to give comparable results in most cases. 
Further supporting this point, in this Appendix, we rigorously prove the following inequalities between the timescales $\{\Gamma_0^{-1},\tau_0\}$ and $\{\Gamma,\tau\}$:
\be 
\Gamma_ 0 \leq \Gamma/2  \quad {\rm and}\quad \Gamma_0 \tau_0 \leq \Gamma\tau.
\label{eqa:timescale_inequalities}
\ee
These inequalities were  used in Appendix~\ref{app:br_correction}.

We first show that $\Gamma_0\leq \Gamma/2$. 
We note from the definition of $\bs J(t)$ in Eq.~\eqref{eq:correlation_function_m} that 
$\boldsymbol J(t) = \boldsymbol J^\dagger(-t)$.
Using $\norm{\boldsymbol M}_1 = \norm{\boldsymbol M^\dagger}_1$, we thus have
 $\norm{\bs J(t)}_1 = \norm{\bs J(-t)}_1$. 
Using this result in 
 Eq.~\eqref{eqa:alternative_time_scales}, we find
\be 
\Gamma_0 = 2\gamma \int_{-\infty}^\infty \!{\rm d}t\, \norm{\boldsymbol J(t)}_1.
\label{eq:app:g0_def_1}
\ee
To obtain a bound for $\norm{\bs J(t)}_1$ we note that  $\boldsymbol J(t)$ is related to the jump correlator $\boldsymbol g(t)$ through the convolution
$
\boldsymbol J(t) = \int_{-\infty}^\infty {\rm d}s\, \boldsymbol g(t-s)\boldsymbol g(s).
$
This result follows from the definition of $\bs g(t)$ in Sec.~\ref{sec:multiple_noise_channels}, and is the multi-channel generalization of the result quoted above Eq.~\eqref{eq:f_superoperator} in the main text.
Using the triangle inequality, we obtain
\be
\norm{\boldsymbol J(t)}_1 \leq  \int_{-\infty}^\infty \!\!\! {\rm d}s\, \norm{\boldsymbol g(t-s)\boldsymbol g(s)}_1.
\label{eqa:jc_ineq_1}
\ee

To rewrite the integrand above, we now prove that, for any two matrices $\bs A$ and $\bs B$, 
\be
 \norm{\boldsymbol A \boldsymbol B}_1\leq \norm{\boldsymbol A^\dagger}_{2,1}\norm{\boldsymbol B}_{2,1},
 \label{eqa:m_norm_ineq}
\ee
where the matrix norm $\norm{\cdot}_{2,1}$ was defined below Eq.~\eqref{eq:time_scales_m} in the main text: 
$\norm{\boldsymbol M}_{2,1}  \equiv \sum_\beta  (\sum_{\alpha }|M_{\alpha \beta }|^2)^{1/2}$ for a matrix $\boldsymbol M$ with  elements $\{M_{\alpha \beta }\}$. 
To prove Eq.~\eqref{eqa:m_norm_ineq}, we recall that  
$
\norm{\boldsymbol A \boldsymbol B}_1 = \sum_{\alpha \beta \gamma } |A_{\alpha \beta }B_{\beta \gamma }|.
$
We consider the sum over the index $\beta $ first.
Using the Cauchy-Schwartz inequality, we find 
\be 
\sum_\beta  |A_{\alpha \beta }B_{\beta \gamma }| \leq \big(\sum_\beta |A_{\alpha \beta }|^2 \big)^{1/2} \big(\sum_{\beta '}|B_{\beta' \gamma }|^2\big)^{1/2}.
\ee
Using this inequality in the expression for $\norm{\bs A\bs B}_1$ in Eq.~(\ref{eqa:m_norm_ineq}), we conclude 
$ 
\norm{\boldsymbol A \boldsymbol B}_1 \leq c_A c_B,
$
where $c_A = \sum_\alpha \big(\sum_\beta |A_{\alpha \beta }|^2 \big){}^{1/2}$ and $c_B = \sum_\gamma  \big(\sum_{\beta '}|B_{\beta '\gamma }|^2\big){}^{1/2}$.
Recalling the definition of the  norm $\norm{\cdot}_{2,1}$, we  identify $c_A=\norm{\boldsymbol A^\dagger}_{2,1}$ and $c_B = \norm{\boldsymbol B}_{2,1}$.
Thus Eq.~\eqref{eqa:m_norm_ineq} holds.

Combining  Eqs.~\eqref{eq:app:g0_def_1}-\eqref{eqa:m_norm_ineq}, we obtain
\be
\Gamma_0 \leq  2\gamma \int_{-\infty}^\infty \!\!\! {\rm d}s\int_{-\infty}^\infty \!\!\!{\rm d}t\, \norm{\boldsymbol g^\dagger(t-s)}_{2,1}\norm{\boldsymbol g(s)}_{2,1}.
\ee
The hermiticity  of $\bs g(\omega)$  implies that $\bs g(t)  = \bs g^\dagger(-t)$ (see Sec.~\ref{sec:multiple_noise_channels}).
Using this result in the above and shifting the variables of integration, we obtain 
\be 
\Gamma_0 \leq 2\gamma \left[\int_{-\infty}^\infty \!\!\!{ \rm d}t\, \norm{\bs g(t)}_{2,1}\right]^2.
\label{eqa:g0_result}
 \ee
Comparing this result with the definition of $\Gamma$ in Eq.~\eqref{eq:time_scales_m} in the main text, we conclude that $\Gamma_0 \leq \Gamma /2$. 

We now prove the second inequality in Eq.~\eqref{eqa:timescale_inequalities},  $\Gamma_0\tau_0 \leq \Gamma \tau$. 
Using the fact that $\norm{\boldsymbol J(t)}_1=\norm{\boldsymbol J(-t)}_1$ [see text above Eq.~\eqref{eq:app:g0_def_1}], along with the definitions of $\Gamma_0$ and $\tau_0$ in Eq.~\eqref{eqa:alternative_time_scales}, we have 
\be 
\Gamma_0 \tau_0 = 2\gamma  \int_{-\infty}^\infty \!\!\!{\rm d}t \, |t| \cdot \norm{\boldsymbol J(t) }_1.
\ee
Using   Eqs.~\eqref{eqa:jc_ineq_1}-\eqref{eqa:m_norm_ineq} along with $\bs g(t) = \bs g^\dagger(-t)$ [see text above Eq.~\eqref{eqa:g0_result}], we obtain
\be 
\Gamma_0 \tau_0 \leq 2\gamma \int_{-\infty}^\infty \!\!\!{\rm d}t \int_{-\infty}^\infty \!\!\!{\rm d}s \, |t| k(s-t) k(s),
\ee
where we introduced the shorthand $k(t)\equiv \norm{\bs g(t)}_{2,1}$. 
Using that  $|t| \leq |s-t|+|s|$ and  shifting variables of integration, 
 one can then verify that 
$ 
\Gamma_0 \tau_0 \leq 2\gamma \int_{-\infty}^\infty {\rm d}t'\int_{-\infty}^\infty {\rm d}s \, (|t'|+|s|) k(t') k(s).
$
Exploiting the symmetry of this expression under exchange of $t'$ and $s$, we find
\be
\Gamma_0 \tau_0 \leq 4\gamma  \int_{-\infty}^\infty \!\!\!{\rm d}t'\int_{-\infty}^\infty \!\!\!{\rm d}s \, |t'| k(t') k(s).
\ee
Recalling that $k(t)\equiv \norm{\bs g(t)}_{2,1}$, and comparing with the definitions of $\Gamma$ and $\tau$ in Eq.~\eqref{eq:time_scales_m}, we identify the right-hand side above as $\Gamma \tau$. 
Thus,  $\Gamma _0\tau_0 \leq \Gamma\tau$, and Eq.~\eqref{eqa:timescale_inequalities} holds. 
This was what we wanted to prove and concludes this Appendix.

\section{Derivation of the ULE} 
\label{app:ule_derivation}

In this Appendix,  we rigorously derive the universal Lindblad equation (ULE) in the interaction picture 
[Eq.~\eqref{eq:general_ip_lindblad} of the main text]. 

As in Appendices~\ref{app:br_correction}~and~\ref{app:time_scale_relationship}, we consider  here the case of arbitrary system-bath coupling $H_{\rm int}$, 
 such that the system and bath are connected through multiple noise channels (see Sec.~\ref{sec:model} in the main text). 
In the main text, we heuristically derived the ULE  for the  case of a single noise channel [Eq.~\eqref{eq:lindblad_form_1}]. 
This result is a special case of the more general result that we rigorously prove here, and hence this Appendix also serves as a  proof of Eq.~\eqref{eq:lindblad_form_1}.

As discussed in the main text,  the ULE   [Eq.~\eqref{eq:general_ip_lindblad}] 
 holds for a modified density matrix $\rho'(t)$ whose spectral norm-distance to the exact density matrix $\tilde \rho(t)$ (in the interaction picture)  remains bounded by  $\Gamma \tau$ at all times.
Here the bath timescales $\Gamma^{-1}$ and $\tau$ were defined in Eq.~\eqref{eq:time_scales_m} in the main text. 
In the Markovian limit, $\Gamma\tau \ll 1$, which is required for  the ULE to be valid (see Secs.~\ref{sec:bloch_redfield_conditions}~and~\ref{sec:single_channel} in the main text), the modified density matrix $\rho'$ is thus nearly identical to the true density matrix $\tilde \rho$, and accurately describes the state of the system. 

Our derivation below proceeds in three steps. 
In Sec.~\ref{app:modified_density_matrix}, we define the modified density matrix $\rho'(t)$ [see Eq.~\eqref{eq:app:rho'_def}] and prove that its spectral norm-distance to  $\tilde \rho(t)$ remains bounded by $\Gamma\tau$ at all times.
Subsequently, in Sec.~\ref{seca:rho'_master_equation}, we show that $\rho'(t)$  evolves according to the master equation 
 \begin{widetext}
\be
\partial _t \rho'(t) = \mathcal L(t)[\rho'(t)]+\xi'(t), 
\label{eqa:rho'_result}
\quad \quad \mathcal L(t)   \equiv \!\int_{-\infty}^\infty \!\!\! {\rm d}s \int_{-\infty}^\infty \!\!\!{\rm d}s'\, \mathcal F(s,t,s'),
 \ee
where the spectral norm of $\xi'(t)$ is bounded by $ 2\Gamma^2 \tau$, and, for any operator $A$, we have defined
\be 
\mathcal F(s,t,s')[A] =  \gamma \!\!\sum_{\alpha ,\beta ,\lambda}\!\!  \theta(s-s') \left(g_{\alpha \lambda}(s-t)g_{\lambda \beta }(t-s') [\tilde X_\alpha (s),A \tilde X_\beta (s')]  +g^*_{\alpha \lambda}(s-t)g^*_{\lambda \beta }(t-s') [\tilde X_\beta (s')A ,\tilde X_\alpha (s)] \right).
\label{eqa:f_def_m}
\ee
\end{widetext}
Note that the definitions above
 generalize the  superoperators  $\mathcal L(t)$ and  $\mathcal F(s,t,s')$ in Sec.~\ref{sec:single_channel}  to   cases with multiple noise channels~\cite{fn:superoperator_ref}. 
As the third and final step of our derivation, in Sec.~\ref{sec:lindblad_form} we show that the superoperator $\mathcal L(t)$ takes the Lindblad form in Eq.~\eqref{eq:general_ip_lindblad}.
Thereby we reach the goal of this Appendix,  proving that $\rho'(t)$ evolves according to the Lindblad-form master equation in Eq.~\eqref{eq:general_ip_lindblad} of the main text.

\subsection{Modified density matrix}
\label{app:modified_density_matrix}
Here we define the modified density matrix $\rho'(t)$, and prove that $\norm{\rho'(t)-\tilde \rho(t)}\leq \Gamma \tau$ at all times. 
Our approach is to identify a transformation $\rho'(t) \equiv [1 + \mathcal{M}(t)]\tilde{\rho}(t)$ such that, if $\tilde{\rho}(t)$ satisfies the Bloch-Redfield equation [Eq.~\eqref{eqa:br_equation_m}], then, up to an error of order $\Gamma^2 \tau$, $\rho'(t)$ evolves according to Eq.~(\ref{eqa:rho'_result}) (which can then be expressed in Lindblad form).
We will bound the norm-distance between $\rho'(t)$ and $\tilde{\rho}(t)$ using the explicit form of this transformation.

To motivate our definition of $\rho'(t)$,  we note that the multi-channel Bloch-Redfield (BR)  equation [Eq.~\eqref{eqa:br_equation_m} in Appendix A] can be written as 
\be 
\partial _t \tilde \rho (t) =   \int_{-\infty}^\infty \!\!\! {\rm d}s'\int_{-\infty}^\infty\!\!\!{\rm d}s\, \mathcal F(t,s,s')[ \tilde \rho(t) ] +\xi(t), \label{eqa:dt_rho_f_decomp}
\ee
where $\xi(t)$ denotes the error induced by the Born-Markov approximation, with norm bounded by $\Gamma^2 \tau$. 
The expression above generalizes the single-channel result in Eq.~\eqref{eq:f_superoperator} in the main text to the case of multiple noise channels.
It is convenient to rewrite the right-hand side above in terms of the superoperator 
\be 
\mathcal G(t,s)\equiv \int_{-\infty}^\infty\!\! {\rm d}s' \,\mathcal F(t,s,s').
\label{eq:app:g_def}
\ee
Specifically, we express the BR equation [Eq.~\eqref{eqa:dt_rho_f_decomp}] as 
\be 
\partial _t \tilde \rho(t) = \int_{-\infty}^\infty \!\! {\rm d}s \, \mathcal G(t,s)[\tilde \rho(t)]  + \xi(t). 
\label{eqa:br_g_expression}
\ee
Similarly,  we may rewrite Eq.~\eqref{eqa:rho'_result} (our target equation of motion for the modified density matrix $\rho'$) as 
\be 
\partial _t \rho' (t)  = \int_{-\infty}^\infty \!\!\! {\rm d}s \, \mathcal G(s,t)[ \rho'(t)] + \xi'(t). 
\label{eqa:ule_g_expression}
\ee 
We will identify the precise form of the correction $\xi'(t)$ in the derivation below.

Note that, when neglecting the corrections $\xi(t)$ and $\xi'(t)$, the only difference between Eqs.~\eqref{eqa:br_g_expression} and  
\eqref{eqa:ule_g_expression} is the order of the arguments in the superoperator $\mathcal G$.
The modified density matrix $\rho'(t)$ is  obtained from a (time-local) linear operation on $\tilde \rho(t)$ that transforms Eq.~\eqref{eqa:br_g_expression} into  Eq.~\eqref{eqa:ule_g_expression}. 
As we show in Sec.~\ref{seca:rho'_master_equation} below, such a linear transformation is generated by the superoperator $[1+\mathcal M(t)]$, where

\be 
\mathcal M(t) \equiv 
\int_t^\infty\!\!{\rm d}s\!\int^t_{-\infty} \!\!\!{\rm d}s'\, [ \mathcal G(s,s') - \mathcal G(s',s)]. %
\label{eq:m_def}
\ee
Specifically, we define $\rho'(t)$ as follows: 
\be 
\rho'(t) = \left[1+\mathcal M (t)\right][\tilde \rho(t)].
\label{eq:app:rho'_def}
\ee
In Sec~\ref{seca:rho'_master_equation}, we  show that $\rho'(t)$, as defined above, evolves according to Eq.~\eqref{eqa:ule_g_expression}. 
Before proving this, we show here that $\rho'$ deviates from $\tilde \rho$ by a correction whose spectral norm is bounded by $\Gamma \tau$ at all times:  $\norm{\rho'(t)- \tilde \rho(t)}\leq \Gamma \tau$.

To show that $\norm{\rho'(t)- \tilde \rho(t)}\leq \Gamma \tau$, we prove below that, for any operator $A$, 
\be
\norm{\mathcal M(t) [A]} \leq    \Gamma \tau \norm{A}.
\label{eqa:m_inequality}
\ee
By the definition of $\rho'(t)$ in Eq.~\eqref{eq:app:rho'_def}, this result in particular implies that $\norm{\rho'(t)- \tilde \rho(t)}\leq \Gamma \tau$, since $\norm{\tilde \rho(t)}\leq 1$.
We will also use Eq.~\eqref{eqa:m_inequality} for other purposes in Sec.~\ref{seca:rho'_master_equation}.

To prove Eq.~\eqref{eqa:m_inequality}, we use the triangle inequality in 
 Eq.~\eqref{eq:m_def} to obtain
\be 
\norm{\mathcal M(t)[A]} \leq
\int_t^\infty\!\!{\rm d}s\!\int^t_{-\infty} \!\!\!\!{\rm d}s' \Big( \norm{\mathcal G(s,s')[A]} + \norm{\mathcal G(s',s)[A]}\Big). %
\label{eqa:m_ineq_11}
\ee
Using the triangle inequality  
 in Eq.~\eqref{eq:app:g_def}, we have 
\be
\norm{\mathcal G(t,s)[A]} \leq  
  \int_{-\infty}^\infty\!\!\! {\rm d}s' \,\norm{\mathcal F(t,s,s')[A]}.
 \label{eqa:g_inequality_0}
\ee
From the definition of $\mathcal F$ in Eq.~\eqref{eqa:f_def_m}, using the triangle inequality and the submultiplicativity of the spectral norm, one can verify that 
$
\norm{\mathcal F(t,s,s')[A]} \leq  4 \gamma   \norm{\bs g(t-s) \bs g(s-s')}_1 \norm{A} \theta(t-s),
$
where the $1$-matrix norm $\norm{\cdot}_1$ is defined in Sec.~\ref{app:speed_limit}, and  we used that $\norm{\tilde X_\alpha (t) } \leq 1$. 
In  Appendix~\ref{app:time_scale_relationship}, we  established that $\norm{\bs g(t)\bs g(s)}_1 \leq \norm{\bs g(-t)}_{2,1} \norm{\bs g(s)}_{2,1}$, where the matrix norm $\norm{\cdot}_{2,1}$ is defined in Sec.~\ref{sec:multiple_noise_channels}. 
Thus,
\be 
\norm{\mathcal F(t,s,s')[A]} \leq  4 \gamma  \norm{\bs g(s-t)}_{2,1}\norm{\bs g(s-s')}_{2,1}\theta(t-s')\norm{A}.
\notag \ee
Using this result in  Eq.~\eqref{eqa:g_inequality_0},  we find  
\be
\norm{\mathcal G(t,s)[A]}\leq G(t-s)\norm{A},
\label{eq:app:g_inequality}
\ee
where
\be 
G(t) \equiv 
4\gamma \norm{\bs g(-t)}_{2,1} \int_{-\infty}^t \!\!\!{\rm d}s\, \norm{\bs g(-s)}_{2,1}.
\label{eqa:g_func_def}\ee
Using Eq.~\eqref{eq:app:g_inequality} in Eq.~\eqref{eqa:m_ineq_11}, we obtain
\be 
\norm{ \mathcal M(t)[A] } \leq   \int_t^\infty\!\!\!{\rm d}s\!\int^t_{-\infty} \!\!\!{\rm d}s' [ G(s-s') + G(s'-s)]\norm{A}. 
\label{eqa:g_result_1}
\ee

To rewrite Eq.~\eqref{eqa:g_result_1}, we note that, for any function $f(s)$,
$
\int_t^\infty{\rm d}s\!\int^t_{-\infty} {\rm d}s'  f(s-s') =\int_0^\infty{\rm d}s \, s  f(s) 
$
(this can be verified by change of integration variables). 
Using this result in  Eq.~\eqref{eqa:g_result_1}, we obtain
\be
\norm{ \mathcal M(t)[A] } \leq   \int_{-\infty}^\infty\!\!\!{\rm d t} \, |t| G(t)\norm{A}.
\label{eqa:m_ineq_16}
\ee 
We now seek a convenient bound for $G(t)$. 
Extending the upper limit of integration  in Eq.~\eqref{eqa:g_func_def} to $\infty$, and using the definition of $\Gamma$ in Eq.~\eqref{eq:time_scales_m}, we obtain 
\be 
G(t)\leq \sqrt{4\gamma \Gamma} \norm{\bs g(-t)}_{2,1}. 
\label{eqa:g_ineq_3}
\ee
Using this result in Eq.~\eqref{eqa:m_ineq_16} along with   the definitions of $\Gamma$ and $\tau$ in  Eq.~\eqref{eq:time_scales_m},  we   conclude that the right-hand side of Eq.~\eqref{eqa:m_ineq_16} is bounded by $\Gamma \tau \norm{A}$. 
Thus, Eq.~\eqref{eqa:m_inequality} holds. 
By the arguments below Eq.~\eqref{eqa:m_inequality}, we hence conclude  $\norm{\rho'(t)-\tilde \rho(t)}\leq \Gamma \tau$. 
This was what we wanted to show. 

\subsection{Master equation for  modified density matrix} 
\label{seca:rho'_master_equation}
We now show that $\rho'(t)$, as defined in Eq.~\eqref{eq:app:rho'_def}, evolves according to the master equation in Eq.~\eqref{eqa:rho'_result}. 
To  establish this result, we explicitly take the time-derivative of $\rho'(t)$ in Eq.~\eqref{eq:app:rho'_def}, obtaining 
\be 
\partial _t \rho' (t) = \partial _t \tilde \rho(t)  + \partial _t \mathcal M(t)[\tilde \rho(t)]+ \mathcal M(t)[\partial _t \tilde \rho(t)], 
\label{eqa:chain_rule}\ee
where we exploited the linear dependence of   $\mathcal M(t)[\tilde \rho]$ on $\tilde \rho$. 
We consider the second term first in the above. 
Using the definition of  $\mathcal M(t)$ in Eq.~\eqref{eq:m_def}, one can verify by explicit computation that
\be
\partial _t\mathcal M(t) = 
\int_{-\infty}^{\infty}\!\!\! {\rm d}s\,  \mathcal G(s,t) -\int^\infty_{-\infty} \!\!\!{\rm d}s\,  \mathcal G(t,s) ,
\label{eqa:m_result_2}
\ee
Inserting this result into   Eq.~\eqref{eqa:chain_rule}, and using   Eq.~\eqref{eqa:br_g_expression} along with $\mathcal L(t)  =\int_{-\infty}^{\infty}\! {\rm d}s\,  \mathcal G(s,t)$ [see Eqs.~\eqref{eqa:rho'_result} and \eqref{eq:app:g_def}], we obtain
\be 
\partial _t \rho'(t) =  
\mathcal L(t) [\tilde \rho(t)] +\xi_1'(t) 
+\xi(t),
\label{eqa:rho'_me_1}
\ee
where $\xi'_1(t) \equiv \mathcal M(t) [\partial _t \tilde \rho(t)] $.
Noting that $\norm{\partial _t \tilde \rho(t)}\leq \Gamma /2$ (see Sec.~\ref{sec:bath_time_scales}), and that $\norm{\mathcal M(t)[A]} \leq \Gamma\tau \norm{A}$ [Eq.~\eqref{eqa:m_inequality}], we conclude that $\norm{\xi_1'(t)}\leq \Gamma^2 \tau/2$. 

As the final step in our derivation, 
we show  that we may replace the argument $\tilde \rho(t)$ of $\mathcal L(t)$ in Eq.~\eqref{eqa:rho'_me_1} by $\rho'(t)$, at the cost of a correction $\xi'_2(t)$ whose spectral norm is bounded by $\Gamma ^2\tau/2$. 
To show this, we  exploit the linearity of $\mathcal L(t)$ to write 
\be 
\mathcal L(t) [\tilde \rho(t)] =   \mathcal L(t) [ \rho'(t)] + \mathcal L(t) [\Delta \rho(t)], 
\label{eqa:l_arg_subs}
\ee
where $ \Delta \rho(t) \equiv \tilde\rho(t)-\rho'(t)$. 
We now show that 
$
\norm{\mathcal L(t) [A]} \leq \Gamma \norm{A}/2,
\label{eqa:l_ineq_1}
$ such that the second term in Eq.~\eqref{eqa:l_arg_subs} is bounded by $\Gamma^2 \tau/2$ (recall that $\norm{\Delta \rho(t)}\leq \Gamma \tau$, see Sec. \ref{app:modified_density_matrix}). 
To prove this result, 
we use $\mathcal L(t)  =\int_{-\infty}^{\infty}\! {\rm d}s\,  \mathcal G(s,t)$ along with Eq.~\eqref{eq:app:g_inequality}  to obtain 
\be 
\norm{\mathcal L(t)[A]}\leq \int_{-\infty}^{\infty} \!\!\!{\rm d}t\,G(t)\norm{A},
\ee
where $G(t)$ was defined in Eq.~\eqref{eqa:g_func_def}.
By explicit computation, using the definition of $\Gamma$ in Eq.~\eqref{eq:time_scales_m}, one can verify that $\int_{-\infty}^\infty\!{\rm d}t \,G(t) \leq \Gamma /2$.
Thus $\norm{\mathcal L(t)[A]}\leq \Gamma \norm{A}/2$. 
We conclude that 
\be
\mathcal L(t) [\tilde \rho(t)] =   \mathcal L(t) [ \rho'(t)] +\xi'_2(t) , 
\label{eqa:c22}
\ee
 where $\norm{\xi'_2(t)} \leq \Gamma^2 \tau /2$. 

Using the relation in Eq.~\eqref{eqa:c22} in Eq.~(\ref{eqa:rho'_me_1}),   we conclude that $\rho'(t)$, as defined in Eq.~(\ref{eq:app:rho'_def}), evolves according to Eq.~\eqref{eqa:rho'_result}, 
with $\xi'(t) = \xi_1'(t) + \xi_2'(t) + \xi(t)$.   
Since the spectral norm of $\xi(t)$ is bounded by $\Gamma^2 \tau$, while the spectral norms of $\xi_1'(t)$ and $\xi_2'(t)$ are both bounded by $\Gamma^2 \tau/2$,  we conclude that $\norm{\xi'(t)}\leq 2\Gamma^2 \tau$.
Note that the bound for the error $\xi_1'(t)+\xi_2'(t)$ induced by the modified Markov approximation described above
is identical to the bound for the error induced by the Born-Markov approximation, $\xi(t)$.

\subsection{Lindblad form of master equation}
\label{sec:lindblad_form}

As the final step in our derivation, we now show  that the right-hand side of the  master equation for $\rho'$ in Eq.~\eqref{eqa:rho'_result} is identical to the right-hand side of the universal Lindblad equation [Eq.~\eqref{eq:general_ip_lindblad} in the main text].
To prove this result, we first modify the expression for  the superoperator $\mathcal L(t)$ that was defined in  Eq.~\eqref{eqa:rho'_result}-\eqref{eqa:f_def_m}.
By decomposing the step function in  Eq.~\eqref{eqa:f_def_m} into its symmetric and antisymmetric components: $\theta(s-s') = \frac{1}{2}(1+\sgn(s-s'))$,  we find 
\begin{widetext}
\be
\mathcal L(t) = \mathcal L_{\hspace{0.5pt} \rm S}(t) + \mathcal L_{\rm A}(t), \quad\quad  \mathcal L_{i}(t) \equiv \!\int_{-\infty}^\infty \!\!\! {\rm d}s'\int_{-\infty}^\infty \!\!\!{\rm d}s\, \mathcal F_{i}(s,t,s'), \quad i =\{{\rm S},{\rm A}\}.
\label{eqa:lf_l_decomp}
\ee
For any density matrix $\rho$, we have defined
\begin{align}
&\mathcal F_{\hspace{0.5pt} \rm S} (s,t,s') [\rho]  =   
 -\frac{\gamma}{2} \sum_{\alpha ,\beta ,\lambda} [g_{\alpha \lambda}(s-t)\tilde X_{\alpha }(s), g_{ \lambda\beta }(t-s')\tilde X_{\beta }(s')  \rho]  +H.c.,
 \label{eqa:f_s_def}
\\
&\mathcal F_{\hspace{-0.5pt}\rm A} (s,t,s')[  \rho]  = -  \frac{\gamma}{2}\sum_{\alpha ,\beta }\phi_{\alpha \beta }(s-t,s'-t) [ \tilde X_{\alpha }(s),\tilde X_{\beta }(s')  \rho]  +H.c.,
\label{eq:f_def_2}
\end{align}
\end{widetext}
where $\{\phi_{\alpha \beta }(s,t)\}$ denote the matrix elements of the $N\times N$ matrix $\bs \phi(t,s) \equiv \bs g(t)\bs g(-s) \sgn(t-s)$   that was defined  below Eq.~\eqref{eq:lamb_shift_multi} the main text.
Below, we show that the superoperator $\mathcal L_{\rm S}$ in Eq.~\eqref{eqa:lf_l_decomp}  generates the dissipative component of the ULE, while  $\mathcal L_{\rm A}$ generates the Lamb shift.

We consider the term $\mathcal L_{\hspace{0.5pt} \rm S}$ first. 
By direct computation, one can  verify that 
\be 
\mathcal L_{\hspace{0.5pt} \rm S} (t)[\rho]=  - \frac{1}{2}  \sum_{\lambda} [\tilde L^\dagger_\lambda(t), \tilde L_\lambda(t)   \rho]  + H.c.,
\label{eq:l_e_def}
\ee
where $\tilde L_\lambda(t)$ denotes the interaction picture jump operator  defined in  Eq.~\eqref{eq:jump_operator_m} in the main text.
Here we used that
$ 
\tilde L_\lambda ^\dagger(t)=\sqrt{\gamma} \int_{-\infty}^\infty \! {\rm d}s'\, g_{\alpha \lambda}(s-t)\tilde X_{\alpha }(s),
$
which follows from  the relation $\bs g(t)= \bs g^\dagger(-t)$, along with the definition of $\tilde L_\lambda(t)$. 
Writing out all terms in Eq.~\eqref{eq:l_e_def}, we obtain
\be 
\mathcal L_{\hspace{0.75pt} \rm S} (t)[\rho] = \sum_\lambda \Big[\tilde L_\lambda(t)  \rho \tilde L^\dagger_\lambda(t)   -\frac{1}{2}  \{\tilde L^\dagger_\lambda (t)\tilde L_\lambda(t),   \rho\}\Big].
\label{eqa:ls_result}
\ee
Hence 
$\mathcal L_{\hspace{0.5pt} \rm S}(t)$  is in the Lindblad form and generates the dissipative part of the  ULE. 

Next, we consider the term 
 $\mathcal L_{\rm A}$ in Eq.~\eqref{eqa:lf_l_decomp}.
By expanding the commutator in Eq.~\eqref{eq:f_def_2}, we obtain $\mathcal F_{\rm A} (s,t,s') [  \rho]= T_1(s,t,s')-T_2(s,t,s')  + H.c. $, where 
\begin{eqnarray}
T_1(s,t,s')&\equiv  \frac{\gamma}{2}\sum_{\alpha ,\beta }\phi_{\alpha \beta }(s-t,s'-t) \tilde X_{\beta }(s')  \rho \tilde X_{\alpha }(s), 
\notag
 \\
T_2(s,t,s') &\equiv  \frac{\gamma}{2}\sum_{\alpha ,\beta }\phi_{\alpha \beta }(s-t,s'-t) \tilde X_{\alpha }(s) \tilde X_{\beta }(s')  \rho. 
\notag
\end{eqnarray}

We now  show that 
 $T_1(s,t,s') = -T_1^\dagger(s',t,s)$.
This implies that the net contribution to $\mathcal L_{\rm A}(t)$ from $T_1$ and its Hermitian conjugate vanishes: $\int_{-\infty}^\infty \! {\rm d}s'\int_{-\infty}^\infty \!{\rm d}s\, [T_1(s,t,s') +T_1^\dagger(s,t,s')]=0$, and hence [see  Eq.~\eqref{eqa:lf_l_decomp}]
\be
\mathcal L_{\rm A}(t)[\rho] = -\int_{-\infty}^\infty \!\!\! {\rm d}s'{\rm d}s\, [T_2(s,t,s') +T_2^\dagger(s,t,s')].
\label{eqa:la_result1}\ee
To prove  that $T_1(s,t,s') = -T_1^\dagger(s',t,s)$, we note that $\bs \phi(t,s)=-\bs \phi^\dagger(s,t)$ [this follows from the definition of $\bs \phi$ below Eq.~\eqref{eq:f_def_2} along with $\bs g(t) = \bs g^\dagger(-t)$].
Using this identity in the definition of $T_1$ above, we find, after a relabelling of summation variables,
\be
T_1(s,t,s') =  -\frac{\gamma}{2}\sum_{\alpha ,\beta }\!\phi^*_{\alpha \beta   }(s'-t,s-t) \tilde X_{\alpha  }(s')  \rho \tilde X_{\beta  }(s)  .
\notag\ee
We identify the right-hand side as $-T_1^\dagger(s',t,s)$ (see definition of $T_1$ above). 
Thus, $T_1(s,t,s') = -T_1^\dagger(s',t,s)$, and hence Eq.~\eqref{eqa:la_result1} holds.

We finally note that 
$
\int_{-\infty}^\infty  {\rm d}s'\int_{-\infty}^\infty{\rm d}s\, T_2(s,t,s') = i\tilde \Lambda (t) \rho,
$
where 
\be
\tilde \Lambda(t)  =   \frac{\gamma}{2i}  \int_{-\infty}^\infty\!\!\! {\rm d}s\int_{-\infty}^\infty\!\!\!{\rm d}s'  \sum_{\alpha \beta } \tilde X_\alpha (s)\tilde X_\beta (s')\phi_{\alpha \beta}(s-t,s'-t)
\ee
 denotes the Lamb shift from Eq.~\eqref{eq:lamb_shift} in the main text.
Hence the anti-symmetric component $\mathcal L_{\rm A}$ generates the Lamb shift  in the ULE, as we claimed:
\be
\mathcal L_{\rm A}(t)[\rho]= -i [\tilde \Lambda(t),\rho].
\label{eqa:la_result}
\ee

Combining Eqs.~\eqref{eqa:lf_l_decomp},~\eqref{eqa:ls_result},~and~\eqref{eqa:la_result}, we obtain
\begin{align}
\mathcal L(t) [\rho ]    =& - i[\tilde\Lambda(t),\tilde \rho(t)] \\
&+ \sum_{\lambda}\Big[ \tilde  L_\lambda(t)\tilde \rho  (t)\tilde L^\dagger_\lambda(t)-\frac{1}{2}\{\tilde L^\dagger_\lambda(t)\tilde L_\lambda(t), \tilde \rho (t)\}\Big].\notag
\end{align}
Thus,  the superoperator $\mathcal L(t)$ is in the Lindblad form.
Using this result in Eq.~\eqref{eqa:rho'_result}, we conclude that the modified density matrix $\rho'(t)$, as defined in Eq.~\eqref{eq:app:rho'_def}, evolves according to the ULE in Eq.~\eqref{eq:general_ip_lindblad}, with the correction term $\xi'(t)$ being bounded by $2\Gamma^2 \tau$. 
Proving this  was the goal of this appendix.

\section{Lamb shift for static Hamiltonians}
\label{app:lamb_shift_static}
In this appendix  we derive the expression for the  Lamb shift in Eq.~\eqref{eq:lamb_shift_static} of the main text, which 
 holds for  cases where the system Hamiltonian $H_\S$ is  time-independent.

Eq.~\eqref{eq:lamb_shift_static} is most conveniently derived in the interaction picture. 
We recall from Eq.~\eqref{eq:lamb_shift} that, in the interaction picture, the Lamb shift  
  is given by 
\be
\tilde \Lambda(t)  =   \frac{\gamma}{2i}  \infint\!\! {\rm d}s'\!\!\infint\!\!{\rm d}s  \sum_{\alpha \beta } \tilde X_\alpha (s)\tilde X_\beta (s')\phi_{\alpha \beta }(s-t,s'-t),
\label{eq:lamb_shift_a1}
\ee
where
$
\{\phi_{\alpha \beta }(s,s')\}$ denote the  elements of the 
 matrix 
$
\bs \phi(t,s)\equiv \bs g(t)\bs g(-s) \sgn(t-s),
$
and $\boldsymbol g(t)$ denotes the matrix-valued jump correlator defined in  Eq.~\eqref{eq:jump_correlator_m} in the main text.

As a first step in our derivation, we   decompose the time-evolved system operator
 $\tilde X_\alpha (t)$ in terms of the eigenstates  $\{|n\rangle\}$ and energies $\{E_n\}$ of the system Hamiltonian $H_\S$:
\be
\tilde X_\alpha (t) = \sum_{m,n} X_{mn}^{(\alpha) }e^{-iE_{nm}t} |m\rangle\langle n|,
\label{eqa:evolved_x}
\ee
where, as in the main text, $X_{mn}^{(\alpha )}\equiv \langle m|X_\alpha |n\rangle$, while $E_{nm}\equiv E_n-E_m$. 
Inserting Eq.~\eqref{eqa:evolved_x} into Eq.~\eqref{eq:lamb_shift_a1}, shifting  variables of integration, and using $E_{lm}+E_{nl}=E_{nm}$ along with the definition of $\bs \phi_{\alpha \beta }(t,s)$, we obtain 
\be 
\tilde \Lambda(t) =   \!\!\!\!\sum_{mnl;\alpha\beta }  \!\!\! X^{(\alpha)}_{ml} X^{(\beta )}_{ln}\!   {f}_{\alpha \beta }(E_{lm},E_{nl})e^{-iE_{nm}t}|m\rangle\langle n|,
\label{eqa:ls_f_expr}
\ee
where $\{f_{\alpha \beta }(p,q)\}$ denote the  elements of the  matrix 
\be
%
\boldsymbol f(p,q)  = \frac{\gamma }{2i}\int_{-\infty}^\infty\!\!\! {\rm d} t\int_{-\infty}^\infty\!\!\!{\rm d}s \,  \sgn(s-t) e^{-i(pt+qs)}\boldsymbol g(t)\boldsymbol  g(-s).
\ee
Note that $\boldsymbol{f}(p,q)$  is the Fourier transform of $\boldsymbol \phi(t,s)$, up to a constant prefactor. 

We now express the jump correlator in terms of its Fourier transform: $\bs g(t) = \int _{-\infty}^\infty \! {\rm d}\omega\, e^{-i\omega t } \bs g(\omega)$.
After factoring out the integrals over $t$ and $s$, we obtain
\be
\bs f (p,q)  = \frac{\gamma }{2i }\int_{-\infty}^\infty\!\!\!{\rm d}\omega\!\int_{-\infty}^\infty\!\!\!{\rm d}\omega'\, \bs g(\omega)\bs g(\omega') k(p+\omega,q-\omega'),
\label{eq:f_q1_q2_expression}\ee
where 
$k(p,q)  \equiv\int_{-\infty}^\infty\! {\rm d} s'\int_{-\infty}^\infty\!{\rm d}s \,  \sgn(s-s') e^{-i(ps+qs')}.
$
By explicit computation, one can verify that
\be 
k(p,q) = - 4 \pi i   \delta(p+q) \Re\left(\frac{1}{p-i0^+}\right),
\ee
where $\delta(x)$ denotes the Dirac delta function.
Using this  result  in Eq.~\eqref{eq:f_q1_q2_expression}, integrating out $\omega'$, and subsequently shifting variables of integration, we find 
\begin{align}
 \bs f(p,q) = -2\pi\gamma  \int_{-\infty}^\infty\!\!\!{\rm d}\omega\, \bs g(\omega-p)\bs g(\omega+q)  \Re\left(\frac{1}{\omega-i0^+}\right). \notag
\end{align}
We can rewrite this to the following: 
\begin{align}
\bs  f(p,q) = -2\pi\gamma  \, \mathcal P \int_{-\infty}^\infty\!\!\!{\rm d}\omega \, \frac{\bs g(\omega-p)\bs g(\omega+q)}{\omega} ,
\label{eqa:ls_st_f_result}\end{align}
where $\mathcal P $ denotes the Cauchy principal value.

As a final step in our derivation, we use the expression for $\tilde \Lambda$ in  Eqs.~\eqref{eqa:ls_st_f_result}~and~\eqref{eqa:ls_f_expr}  to compute the Lamb shift  in the Schrodinger picture, $\Lambda$. 
We recall that  $\Lambda = U(t)\tilde \Lambda(t) U^\dagger(t)$ [see below Eq.~\eqref{eq:sp_dissipator_def} in the main text], where $U(t) = e^{-iH_\S t}$ denotes the unitary evolution operator generated by the system Hamiltonian $H_\S$.
Noting that for time-independent system Hamiltonians,  $U(t) |m\rangle\langle n|U^\dagger(t) = e^{iE_{nm}t}|m\rangle\langle n|$,  this implies that 
\be 
 \Lambda =  \sum_{m,n,l}\,  \sum_{\alpha \beta }  X^{(\alpha)}_{mn}  X^{(\beta)}_{nl}  f_{\alpha \beta }(E_{mn},E_{nl}) |m\rangle\langle l|,
\ee
where the matrix 
$\bs f(p,q)$ is defined in Eq.~\eqref{eqa:ls_st_f_result}. 
This was the result quoted in the main text.

We note that the above line of arguments  can be  generalized   to periodically driven systems with a few modifications.
However, for the sake of brevity, we do not provide such a derivation here.




\section{Conditions for slow time-dependence}
\label{app:time_independent_approx}

In this Appendix we identify the conditions on the time-dependence of the  system Hamiltonian, $H_\S (t)$,  under which the Schr\"odinger picture jump operators $\{L_\lambda(t)\}$ and Lamb shift $\Lambda(t)$ can be computed from the eigenstates and energies of the instantaneous Hamiltonian $ H_\S(t)$, using Eq.~\eqref{eq:static_jump_operators} of the main text.

To show this explicitly for the jump operator $L_\lambda(t)$, we note that 
 $U(t,s) = e^{-i (t-s) H_\S(t)} + \mathcal O (v (t-s)^2)$, where $v=\sup_{s\leq t'\leq t}\norm{\partial _t H_\S(t')}$ denotes the maximal rate of change of $H_\S$~\cite{fn:h_static_approx}.
Using this form of $U(t,s)$ in Eq.~\eqref{eq:sp_l} of the main text, along with the results from Sec.~\ref{sec:time_independent}, we conclude that $L_\lambda(t)$  can be computed from the spectrum and eigenstates of $H_\S(t)$ through Eq.~\eqref{eq:lamb_shift_static}, up to  a correction of order $\sqrt{\Gamma } v (\tau_2)^2 $ [note  from Eq.~\eqref{eq:lindblad_1} that the jump operators have units of $({\rm Energy})^{1/2}$]. 
Here $(\tau_2)^2 \equiv \int_{-\infty}^\infty\!\!{\rm d}t\,  \norm{\boldsymbol g(t) t^2}_{2,1}/\mathcal N$, where $\mathcal N\equiv \sqrt{\Gamma/4\gamma }$ (see Sec.~\ref{sec:multiple_noise_channels} in the main text for the definition of the matrix norm $\norm{\cdot}_{2,1}$).
The timescale $\tau_2$  gives the square root of the second moment of the normalized distribution $\norm{\boldsymbol{g}(t)}_{2,1}/\mathcal N$ [see definition of $\Gamma$ in Eq.~\eqref{eq:time_scales_m}], and we expect it to typically be comparable to the first moment $\tau$. 
Thus,  when $H_\S (t)$ changes slowly on the  correlation  timescale of the bath $\tau$, i.e., $\partial_t H_\S(t) (\tau_2)^2 \ll 1$,   the jump operators of the system $\{L_\lambda(t)\}$  
 can be  computed from the instantaneous Hamiltonian $H_\S(t)$ using Eq.~\eqref{eq:static_jump_operators}.
 A similar result holds for the Lamb shift $\Lambda(t)$.

\section{Calculation of transport properties}
\label{app:transport}
Here we define the heat and magnetization currents computed for the non-equilibrium spin chain in Sec.~\ref{sec:numerics}.
The average heat current $\bar I_E$ can be identified from the equation of motion for the energy in the spin chain: $ \partial _t \langle E(t)\rangle = \Tr [H \partial _t\rho]$. 
Using the universal Lindblad equation [Eqs.~\eqref{eq:sp_lindblad}-\eqref{eq:sp_dissipator_def}], along with $[H,H]=0$, we find $\partial _t \langle E(t)\rangle = \sum_\lambda \langle I^{(\lambda)}_E\rangle$, where $ I_E^{(\lambda)} =  L^\dagger_\lambda H L_\lambda - \frac{1}{2}\{L^\dagger_\lambda L_\lambda,H\}$.
For $\lambda=1,2$, we identify $I^{(\lambda)}_E$ as  the heat current flowing into the system from bath $\lambda$. 
Since   the energy  of  the chain is bounded, the  time-averaged heat current  from bath $1$ must exactly compensate the average heat current from bath $2$. 
Hence, we identify $\bar I_E $ as the time-averaged expectation value of $ -I_E^{(1)}$.
The magnetization current $\bar I_M$ can be obtained similarly from the equation of motion for the magnetization $M$,  using $[H,M]=0$.

\end{document}